\documentclass{SciPost}
\usepackage{graphicx}
\usepackage{color}
\usepackage{dcolumn}
\usepackage{bm}
\usepackage{dutchcal}
\usepackage{hyperref}
\usepackage{physics}
\usepackage{amsmath}
\usepackage{color}
\usepackage{empheq}
\usepackage[toc,page]{appendix}
\usepackage{caption}
\usepackage{subcaption}
\usepackage{comment}
\usepackage{microtype}
\usepackage{eucal}
\usepackage[cal=cm, scr=rsfso]{mathalfa}
\DeclareMathAlphabet{\mathpzc}{OT1}{mdbch}{m}{it}

\newcommand{\VR}[1]{\textrm{\textcolor{blue}{#1}}}

\newcommand{\grafe}[1]{\left\{ #1 \right\}}
\newcommand{\tonde}[1]{\left( #1 \right)}
\newcommand{\quadre}[1]{\left[ #1 \right]}

\allowdisplaybreaks  

\newcommand{\parhead}[1]{%
  \vspace{0.8\baselineskip}%
  \noindent\textbf{#1 }\!
}

\hypersetup{
  colorlinks,
  linkcolor={red!50!black},
  citecolor={blue!50!black},
  urlcolor={blue!80!black}
}

\usepackage[bitstream-charter]{mathdesign}
\DeclareSymbolFont{usualmathcal}{OMS}{cmsy}{m}{n}
\DeclareSymbolFontAlphabet{\mathcal}{usualmathcal}

\fancypagestyle{SPstyle}{
  \fancyhf{}
  \lhead{\colorbox{scipostblue}{\bf \color{white} ~SciPost Physics }}
  \rhead{{\bf \color{scipostdeepblue} ~Submission }}
  
  \fancyfoot[C]{\textbf{\thepage}}
}

\begin{document}

\pagestyle{SPstyle}

\begin{center}{\Large \textbf{\color{scipostdeepblue}{
        Fragile \emph{vs} robust Multiple Equilibria phases in generalized Lotka-Volterra model with non-reciprocal interactions
  }}}
\end{center}

\begin{center}\textbf{
    Thomas Louis-Sarrola\textsuperscript{1$\star$},
    Valentina Ros\textsuperscript{1},
  }
\end{center}

\begin{center}
  {\bf 1} Université Paris-Saclay, CNRS, LPTMS, 91405, Orsay, France
  \\

  $\star$ \href{mailto:email1}{\small thomas.louis-sarrola@universite-paris-saclay.fr}\,,\quad
\end{center}

\section*{\color{scipostdeepblue}{Abstract}}
\textbf{\boldmath{%
    We investigate the Multiple Equilibria phase of generalized Lotka–Volterra dynamics with random, non-reciprocal interactions. We compute the topological complexity of equilibria, which quantifies how rapidly the number of equilibria of the dynamical equations grows with the total number of species. We perform the calculation for arbitrary degree of non-reciprocity in the interactions, distinguishing between configurations that are dynamically stable to invasions by species absent from the equilibrium, and those that are not. We characterize the properties of typical (i.e., most numerous) equilibria at a given diversity, including their average abundance, mutual similarity, and internal stability. This analysis reveals the existence of two distinct ME phases, which differ in how internally stable equilibria behave under invasions by absent species. We discuss the implications of this finding for the system’s dynamical behavior.
}}

\vspace{\baselineskip}

\noindent\textcolor{white!90!black}{%
  \fbox{\parbox{0.975\linewidth}{%
      \textcolor{white!40!black}{
        \begin{tabular}{lr}%
          \begin{minipage}{0.6\textwidth}%
            {\small Copyright attribution to authors. \newline
              This work is a submission to SciPost Physics. \newline
              License information to appear upon publication. \newline
            Publication information to appear upon publication.}
          \end{minipage} &
          \begin{minipage}{0.4\textwidth}
            {\small Received Date \newline Accepted Date \newline Published Date}%
          \end{minipage}
      \end{tabular}}
  }}
}


\vspace{10pt}
\noindent\rule{\textwidth}{1pt}
\tableofcontents
\noindent\rule{\textwidth}{1pt}
\vspace{10pt}


\section{Introduction}
Nonreciprocal interactions drive a broad class of out-of-equilibrium dynamics and are nowadays at the core of an established line of research. Systems with a \emph{large} number of heterogeneous components interacting asymmetrically are a particularly interesting realization of non-reciprocity, since their high dimensionality provides a natural framework for analytical approaches. This setting is well justified for modeling biological neural networks~\cite{sompolinsky1988statistical, sompolinsky1986temporal, crisanti1987dynamics, sompolinsky1988chaos} and ecosystems composed of many coexisting species such as the microbiota, tropical rainforests, or plankton communities~\cite{faustMicrobialInteractionsNetworks2012,hutchinsonParadoxPlankton1961,stompLargescaleBiodiversityPatterns2011}. The models often incorporate randomness to represent interactions among  neurons or species, with different levels of structural organization. The fully unstructured case in which couplings are drawn independently at random was, for instance, famously exploited by R. May in his seminal analysis of the diversity–stability problem in ecology~\cite{mayWillLargeComplex1972a}.

In ecosystems modeling, the non-reciprocal interaction terms enter the dynamical equations governing the time evolution of species abundances. These (random) interactions compete with single-species terms that encode the intrinsic growth or suppression of the abundances in the absence of other species, as determined by environmental resources and intra-species dynamics~\cite{akjouj2024complex}. This competition naturally leads to distinct dynamical regimes: at weak randomness (henceforth, variability), the abundances relax to time-independent values, while at stronger variability they persistently fluctuate in time. In presence of non-reciprocity, these fluctuations display signatures of chaotic behavior~\cite{fournier2025high, fournier2025non, fournier2026chaos, royNumericalImplementationDynamical2019, pearce2020stabilization, blumenthal2024phase, sanders2018prevalence}. While the first type of behavior is  simple to characterize analytically, the second poses a quite significant theoretical challenge.

We focus here on a prototypical model exhibiting such a transition, the generalized Lotka–Volterra equations with random interactions (rGLV). The existence of a dynamical transition in this (and equivalent) models is known since the early studies~\cite{rieger1989solvable,diederich1989replicators, opper1992phase}. In the weak variability phase, the dynamical equations admit a unique fixed point (or equilibrium) configuration that is stable with respect to perturbations of the species abundances, both for the species coexisting at the fixed point and also for those that are absent, and can potentially invade.
When the variability reaches a critical value, this equilibrium loses its stability, and the system correspondingly displays a qualitatively different dynamics. It is expected that this complex dynamical phase appears concomitantly with the emergence of a multitude of fixed points of the dynamical equations~\cite{opper1999replicator}. In the limiting case of reciprocal interactions, this expectation is reinforced by the similarity between the rGLV model and spin-glass models~\cite{opper1992phase, biscari1995replica, tokita2004species}; in this limit indeed the rGLV equations describe a dynamical descent into an energy landscape (conservative dynamics), and the stable fixed points are local minima of such landscape.
As in other high-dimensional optimization problems with randomness~\cite{franz2016simplest, franz2019critical, folena2022marginal}, one expects a multitude of marginally stable local minima~\cite{biroliMarginallyStableEquilibria2018} and a dynamical descent that exhibits slowing down and aging due to marginality~\cite{bouchaud1998out}. The existence of exponentially many fixed points has also been shown in the opposite limit of strong non-reciprocity~\cite{rosQuenchedComplexityEquilibria2023,rosGeneralizedLotkaVolterraEquations2023}, and dynamical studies in this setting~\cite{arnoulxdepireyManySpeciesEcologicalFluctuations2024, kessler2015generalized} hint that certain fixed points may exert an influence on the (chaotic) dynamics in this case, too. More broadly, investigating to what extent the out-of-equilibrium dynamics of non-reciprocal systems can be understood through the fixed points of the dynamics is an open challenge that is attracting increasing interest~\cite{yang2025relationship, stubenrauch2025fixed, wang2024high, fyodorovNonlinearAnalogueMayWigner2016}. In this work, we provide a statistical characterization of the fixed points of the gLVE for arbitrary degrees of non-reciprocity, thereby establishing the basis for assessing their impact on dynamics.

\begin{figure}
  \centering
  \includegraphics[width=0.6\textwidth]{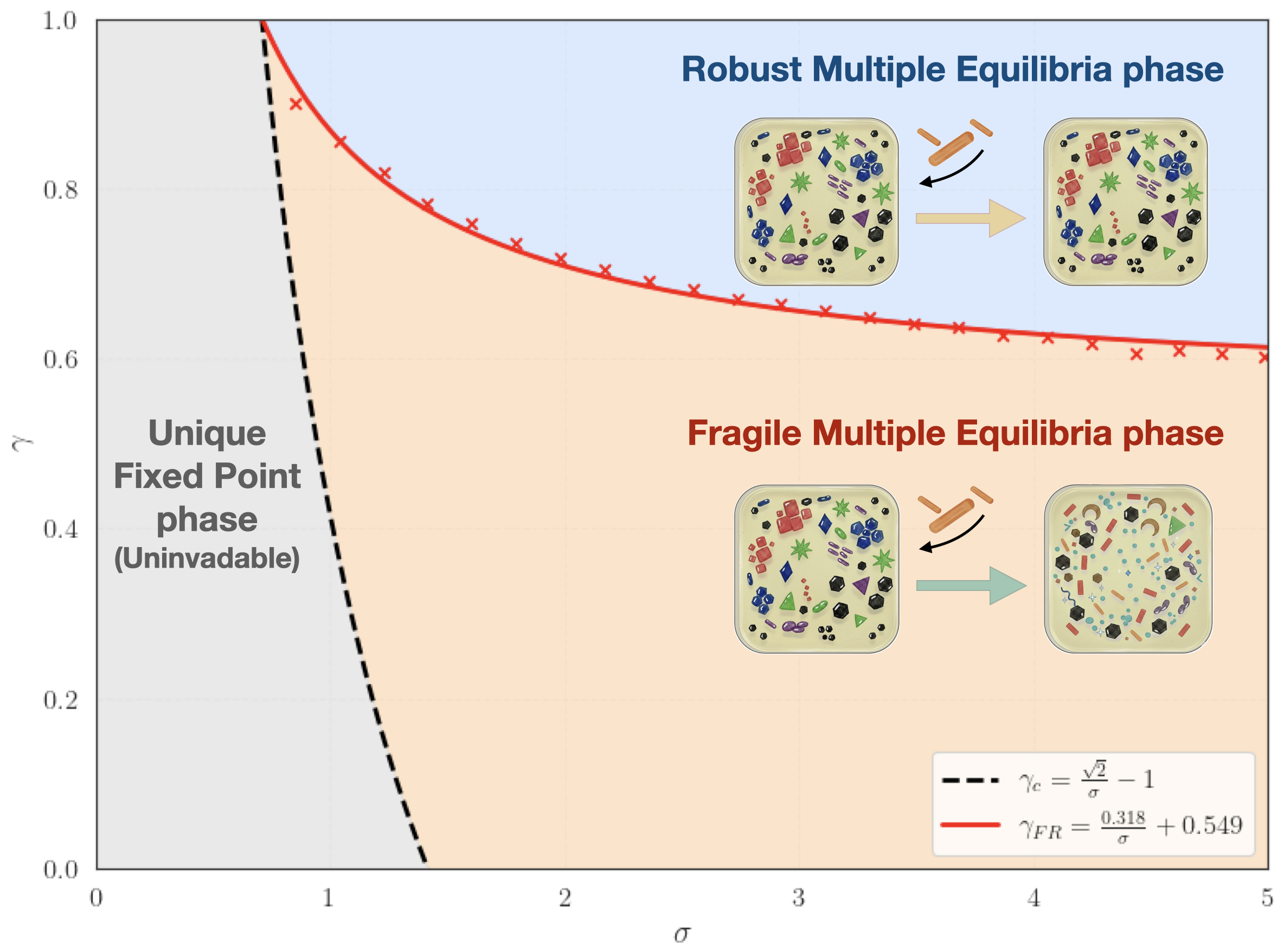}
  \caption{Phase diagram in the variability-reciprocity space $(\sigma,\gamma)$, highlighting the existence of three distinct phases. The black dotted line marks the boundary between the Unique Fixed Point (UFP) phase (gray), where a unique uninvadable and internally stable equilibrium exists, and the Multiple Equilibria (ME) phase. The latter is split into a Fragile (orange) phase, where all internally stable equilibria are unstable to invasions, and a Robust (blue) phase, where uninvadable internally stable equilibria exist in exponential number. The red crosses identify the critical line $\gamma_{FR}$, the red full line is the hyperbolic fit $\gamma_{FR} \approx 0.318/\sigma + 0.549$.}
  \label{fig:gamma_crit}
\end{figure}

\section{The multiple equilibria phase}
\paragraph*{Fixed points and stability.} We consider the rGLV equations
\begin{equation}\label{eq:LVmulti}
  \frac{dN_i}{dt}
  = N_i F_i(\vec N), \; \;  F_i(\vec N)=\Bigl(\kappa_i - N_i -  \sum_{j=1}^S \alpha_{ij}\,N_j\Bigr)
\end{equation}
where $N_i$ is the properly scaled abundance of species $i=1, \cdots, S$, $F_i(\vec{N})$ is the effective growth rate associated to species $i$, and $\kappa_i$ denote the species carrying capacities. The interaction couplings between species
\begin{equation}\label{eq:IntMat}
  \alpha_{ij} = \frac{\mu}{S} + \frac{\sigma}{\sqrt{S}}\,a_{ij},  \quad  \bigl\langle a_{ij}a_{kl}\bigr\rangle
  =\delta_{ik}\,\delta_{jl} + \gamma\,\delta_{il}\,\delta_{jk}
\end{equation}
are Gaussian variables with mean \(\mu/S\), variance \(\sigma^2/S\) and reciprocity parameter $\gamma \in [0,1]$: they are entries of random matrices belonging to the Gaussian elliptic ensemble~\cite{lehmann1991eigenvalue, sommers1988spectrum}.
We set $\kappa_i=1$ and assume $S \gg 1$.
A fixed point (or equilibrium) $\vec{N}^*$ of the rGLV equations is a configuration satisfying
\begin{equation}\label{eq:EqDef}
  N_i^*\left(  1 - N_i^* -\sum_{j} \alpha_{ij}N^*_j\right)=0, \quad N^*_i \geq 0
\end{equation}
for all $i=1,\dots, S$. Any solution of \eqref{eq:EqDef} can have a certain number of vanishing entries $N_i^*=0$, and a fraction $\phi(\vec{N}^*) =  |\grafe{i: N_i^*>0}|/S$ of species with positive abundance, defining the fixed point \emph{diversity} (that is, the number of coexisting species in the equilibrium).

An important aspect in assessing the influence of fixed points on the dynamics is their stability. We distinguish two complementary notions of stability: \emph{internal stability} denotes the linear stability with respect to fluctuations in the abundances of the species that are present at the fixed point ($N^*_i>0$); this is controlled by the eigenvalues of the Jacobian matrix restricted to the subspace of coexisting species~\cite{stone2018feasibility}, defined as
\begin{equation}\label{eq:Jacobian}
  H_{ij}(\vec{N}^*) = \frac{\partial F_i(\vec{N}^*)}{\partial N_j}  \quad \text{for } i,j: N^*_i, N^*_j>0,
\end{equation}
and requires that all its eigenvalues have negative real parts. \emph{Uninvadability} concerns instead the stability with respect to the invasion by species that are absent in the fixed point ($N^*_i=0$), and it is satisfied whenever
\begin{equation}\label{eq:UnInv}
  F_i(\vec{N}^*) < 0  \quad \text{for } i: N^*_i=0.
\end{equation}
For large $S$, the spectral properties of the matrices \eqref{eq:Jacobian} at a fixed point $\vec{N}^*$ can be characterized using random matrix theory, as we recall below. A generalization of the argument in~\cite{mayWillLargeComplex1972a} implies that an equilibrium is internally stable provided that its diversity is bounded by $\phi(\vec{N}^*)\leq \phi_{\text{May}} \equiv [{\sigma^2(1+\gamma)^2}]^{-1}$. When this inequality is saturated, the equilibrium $\vec{N}^*$ is marginally stable.

\paragraph*{Topological complexity.} When $\sigma$ is small and $\mu$ is positive, there is a unique solution of \eqref{eq:EqDef}, which is both internally stable and uninvadable: the rGLV system is in the Unique Fixed Point (UFP) phase (gray area in Fig.~\ref{fig:gamma_crit}). Dynamical trajectories of the abundances, initialized arbitrarily, converge to this configuration. Properties of this unique stable equilibrium, such as its diversity, are determined as solutions of self-consistent equations derived in the limit of large $S$~\cite{diederich1989replicators, opper1992phase}. The internal stability of this unique uninvadable equilibrium breaks down at a critical value $\sigma_c (\gamma) = \sqrt{2}/(\gamma+1)$ (black dashed curve in Fig.~\ref{fig:gamma_crit}), as first shown in~\cite{rieger1989solvable}: at this point, the diversity of the UFP reaches $\phi=1/2$ and the equilibrium becomes marginally stable, while remaining uninvadable. For $\gamma=0$, it is known that at $\sigma_c$ the uninvadable equilibrium also loses its uniqueness: the typical value of the number $ \mathcal{N}^{(u)}_{S}(\phi)$ of uninvadable equilibria of given $\phi$ starts growing exponentially in $S$ and the system enters into a Multiple Equilibria (ME) phase, which is the focus of this work. In particular, the \emph{topological complexity of uninvadable} fixed points,
\begin{equation}\label{eq:CompUn}
  \Sigma^{(u)}_{\sigma, \gamma} (\phi) = \lim_{S\to\infty} \frac{1}{S}\left \langle \log \mathcal{N}^{(u)}_S(\phi) \right \rangle,
\end{equation}
has been computed explicitly in~\cite{rosGeneralizedLotkaVolterraEquations2023, rosQuenchedComplexityEquilibria2023} for $\gamma=0$. It is shown that, within the exponentially large family of fixed points that are stable against invasions, none is also internally stable. This feature is expected to break down when increasing the reciprocity $\gamma$: in particular, for $\gamma=1$ one knows that uninvadable internally stable fixed points exist in exponential number (at least, their average number is exponentially large~\cite{opper1999replicator, rosQuenchedComplexityEquilibria2023}). In parallel, the equilibria that actually matter the most for the out-of-equilibrium dynamics for $\gamma=0$ might be the invadable ones~\cite{arnoulxdepireyManySpeciesEcologicalFluctuations2024, kessler2015generalized}, which are not considered in~\cite{rosGeneralizedLotkaVolterraEquations2023, rosQuenchedComplexityEquilibria2023}. The latter are even more numerous than the uninvadable: denoting with $\mathcal{N}_{S}^{(t)}(\phi)$ the total number of equilibria, which includes invadable ones, we define the \emph{total topological complexity}
\begin{equation}\label{eq:CompBoth}
  \Sigma_{\sigma, \gamma}^{(t)} (\phi) = \lim_{S\to\infty} \frac{1}{S}\left\langle \log \mathcal{N}_S^{(t)}(\phi) \right \rangle \geq  \Sigma^{(u)}_{\sigma, \gamma} (\phi).
\end{equation}
Here, we discuss the behavior of  $\Sigma_{\sigma, \gamma}^{(\alpha)} (\phi)$ for both $\alpha \in \{t, u\}$ and for arbitrary reciprocity degree $\gamma \in [0,1]$.

\section{The replicated Kac-Rice computation}

\paragraph*{Outline of the computation.} The calculation of $\Sigma_{\sigma, \gamma}^{(\alpha)} (\phi)$ is performed using the replicated Kac–Rice formalism introduced in~\cite{rosComplexEnergyLandscapes2019}. This technique relies on the Kac–Rice formula for the average number of solutions of a system of random equations, in our case~\eqref{eq:EqDef}. This formula expresses $\left\langle \mathcal{N}_S^{(\alpha)}(\phi) \right \rangle$  as the integral of a density~\cite{fyodorov2015high, ros2023high}, and it can be straightforwardly extended to compute higher $n$-th moments:
\begin{equation}\label{eq:AvMeasure}
  \left\langle [\mathcal{N}^{(\alpha)}_S(\phi)]^n \right \rangle = \hspace{-.1cm}\int\hspace{-.1cm}\prod_{a=1}^n\prod_{i=1}^S dN_i^a \,  \rho_{S,n}^{(\alpha)} \tonde{\{\vec{N}^a\}_{a=1}^n; \phi}.
\end{equation}
The density $\rho^{(\alpha)}_{S,n}$ enforces that the vectors $\vec{N}^a$ satisfy \eqref{eq:EqDef}, i.e., that they are equilibria. For $\alpha=u$, it also enforces the condition~\eqref{eq:UnInv}.
Its explicit form is given in the next section. To obtain the complexity, this must be combined with the replica limit~\cite{mezardSpinGlassTheory1986}
\begin{equation}\label{eq:Reptrick}
  \Sigma_{\sigma, \gamma}^{(\alpha)} (\phi) = \lim_{S\to\infty} \lim_{n \to 0} \frac{\left\langle  [\mathcal{N}^{(\alpha)}_S(\phi)]^n \right \rangle  -1}{S n},
\end{equation}
in analogy with the standard procedure used to compute the free energy of conservative disordered systems at equilibrium. The calculation follows closely~\cite{ rosQuenchedComplexityEquilibria2023} (see \cite{rosHighdimensionalRandomLandscapes2025b} for a pedagogical review of these techniques): we report it in the Appendix for completeness. The key step in this procedure is a dimensionality reduction, which allows one to rewrite~\eqref{eq:AvMeasure} as an integral over a smaller set of variables (or \emph{order parameters}) $\mathbf{y} \in \mathbb{R}^d$ with $d (=15) \ll S$,
\begin{equation}\label{eq:SPINT}
  \left\langle [\mathcal{N}^{(\alpha)}_S(\phi)]^n \right \rangle
  =  \int \hspace{-.1cm}d \mathbf{y} \, e^{S n \, \bar{\mathcal{A}}^{(\alpha)}(\mathbf{y},\phi) + o(S n)}.
\end{equation}
The entries of $\mathbf{y}$ are collective functions~\footnote{By collective we mean that they are defined as sums over all components of the $\vec{N}^a$, and therefore involve all species.} of the equilibria $\vec{N}^a$, some of which are easily interpretable: they include the average abundance $m:=m(\vec{N}^a)=S^{-1}\sum_{i=1}^S N_i$, their self-similarity $q_1:=q_1(\vec{N}^a)= S^{-1} \sum_{i=1}^S (N_i^a)^2$ and the mutual similarity between pairs of distinct equilibria, $q_0:=q_0(\vec{N}^a, \vec{N}^b)= S^{-1} \sum_{i=1}^S N_i^a  N_i^b$, as well as the average effective growth rate $p:=p(\vec{N}^a)= S^{-1} \sum_{i=1}^S F_i(\vec{N}^a)$. For $S \gg 1$, the moments~\eqref{eq:SPINT} are dominated by the family of fixed points having the properties $\mathbf{y}_*^{(\alpha)}$ that maximize the function at the exponent, and the integral can be evaluated using a saddle-point approximation. Combined with \eqref{eq:Reptrick}, this yields $\Sigma_{\sigma, \gamma}^{(\alpha)}(\phi)= \bar{\mathcal{A}}^{(\alpha)}(\mathbf{y}_*^{(\alpha)},\phi)$ where $\mathbf{y}_*^{(\alpha)}$ are solutions of the saddle point equations $\nabla_{{\bf y}}\bar{\mathcal{A}}^{(\alpha)}(\mathbf{y}_*^{(\alpha)},\phi)=0$. The vector $\mathbf{y}_*^{(\alpha)}$ directly encodes the properties (average abundance, self-similarity, similarity, average growth rate) of the \emph{typical} fixed points (i.e., those that are most numerous) at given $\phi$; fixed points with values of the order parameters that differ from $\mathbf{y}_*^{(\alpha)}$ are exponentially less numerous. The Kac-Rice approach thus generalizes the result for the UFP phase, where the unique equilibrium is characterized by a value of the parameters $m, q_1$ and $p$ that can be computed exploiting the assumption of uniqueness~\cite{opper1992phase, buninEcologicalCommunitiesLotkaVolterra2017}. Notice that our notation for the order parameters $m, q_1, q_0, p$, without explicit indices $a,b$ labeling the specific equilibria where these functions are evaluated, implicitly assumes that for $S \gg 1$ these quantities take the same value across typical equilibria at fixed $\phi$. In the language of disordered systems, this is an assumption of Replica Symmetry.

\paragraph{The Kac-Rice Density.} We provide the explicit form of the density appearing in the replicated Kac-Rice formula for the higher moments, Eq.~\eqref{eq:AvMeasure}, and of the function $\bar{\mathcal{A}}^{(\alpha)}(\mathbf{y},\phi)$ in \eqref{eq:SPINT}. From these formulas, the complexities can be derived directly through a saddle point method. We denote with $\mathbf{N} = (\vec{N}^1,...,\vec{N}^n)$ and $\mathbf{F}(\mathbf{N}) = (\vec{F}^1,\dots,\vec{F}^n)$ the concatenations of $n$ vectors of dimension $S$, where $\vec{F}^a := F(\vec{N}^a)$ is a shorthand notation for the effective growth rate associated to the configuration $\vec{N}^a$. $I=\left\{i_1, \cdots, i_{\lfloor S\phi \rfloor}\right\}$ are  index sets with $\lfloor S \phi \rfloor$ distinct components $i_k \in \left\{1, \cdots, S \right\}$. With this notation, the density in \eqref{eq:AvMeasure} reads:
\begin{equation}\label{eq:DensityEM}
  \rho_{S,n}^{(\alpha)} \tonde{\bf{N}; \phi}=\hspace{-.3 cm}\sum_{\substack{(I_1, \cdots, I_n)\\  |I_a| = S\phi}}  \int \prod_{i=1}^S \prod_{a=1}^n  d {f}^a_i  \Theta^{(\alpha)}_{I_a}(\vec{N}^a, \vec{f}^a)  \mathcal{W}_{\mathbf{N}}^{(n)}(\mathbf{f}).
\end{equation}
In this formula, the $I_a$ are index sets that identify the species that are present in the configuration $\vec{N}^a$, i.e., those with $N^a_i >0$, and
\begin{equation}\label{eq:Enforce}
  \Theta^{(\alpha)}_{I_a}(\vec{N}^a, \vec{f}^a) =\prod_{i\in I^a} \theta(N_i^a) \delta(f_i^a) \prod_{i \notin I^a} \delta(N_i^a) \chi^{(\alpha)}(f_i^a)
\end{equation}
with
\begin{equation}
  \chi^{(\alpha)}(f)=
  \begin{cases}
    \theta(-f) \quad \text{if} \quad \alpha=u\\
    1 \quad \text{if} \quad \alpha=t.
  \end{cases}
\end{equation}
The integrand
\begin{equation}
  \mathcal{W}_{\mathbf{N}}^{(n)}(\mathbf{f}) =  \mathcal{P}_{\mathbf{N}}^{(n)}(\mathbf{f})\,  \mathcal{D}_{\mathbf{N}}^{(n)}(\mathbf{f})
\end{equation}
contains the  probability that the effective growth rates $\vec{F}^a$ take a given value $\vec{f}^a$,
\begin{equation}\label{eq:Dist}
  \mathcal{P}_{\mathbf{N}}^{(n)}({\bf f}) = \left\langle \prod_{a=1}^n \delta\left(  \vec{F}(\vec{N}^a) - \vec{f}^a\right) \right\rangle,
\end{equation}
where $\langle \cdot \rangle$ denotes the average over the random interaction couplings, and the expectation of the product of Jacobian matrices~\eqref{eq:Jacobian} evaluated at the different $\vec{N}^a$,
\begin{equation}\label{eq:FanCondDet}
  \mathcal{D}_{\mathbf{N}}^{(n)} (\mathbf{f}) = \left\langle \prod_{a=1}^n \left| \det\left(\frac{\partial F_i^{a}}{\partial N_j^a} \right)_{i,j \in I^a}\right| \Big | \mathbf{F}(\mathbf{N}) = \mathbf{f}\right\rangle,
\end{equation}
conditional to $\vec{F}^a=\vec{f}^a$. Therefore, \eqref{eq:Enforce} enforces that the configurations $\vec{N}^a$ are (uninvadable) fixed points of the dynamical equations, and \eqref{eq:FanCondDet} accounts for the multiplicity of these solutions.
The representation~\eqref{eq:SPINT} is obtained exploiting the fact that the statistics of the Gaussian variables $F_i^a$ and of their derivatives is isotropic, and depends on the $\vec{N}^a, \vec{f}^a$ only trough the rotationally-invariant functions
\begin{equation}
  \begin{split}
    &q(\vec{N}^a, \vec{N}^b) = \vec{N}^a \cdot \vec{N}^b/S, \quad
    m(\vec{N}^a)  = \vec{N}^a \cdot \vec{1}/S, \\
    &\xi(\vec{f}^a, \vec{f}^b)  = \vec{f}^a \cdot \vec{f}^b/S, \quad p(\vec{N}^a)  = \vec{f}^a \cdot \vec{1}/S,\\
    &\quad z(\vec{N}^a, \vec{f}^b)  = (1-\delta_{ab}) \vec{N}^a \cdot \vec{f}^b/S,
  \end{split}
\end{equation}
where $ \vec{1}$ is the $S$-dimensional vector with all components equal to 1. We introduce the quantities $q_{ab}=q(\vec{N}^a, \vec{N}^b)$ (and similarly for the others) representing the values taken by these functions.
A change of variables allows to rewrite the integrals \eqref{eq:DensityEM} over ${\bf N}, {\bf F}$ as an integral over these quantities only. As it is standard in these type of calculations, the Jacobian of this change of variables is parametrized in terms of a set of conjugate parameters (Lagrange multipliers) $ \hat{q}_{ab}, \hat{\xi}_{ab}, \hat{z}_{ab},\hat{m}_a,\hat{p}_a, \hat{\phi}_a$, where $\hat{\phi}_a$ additionally enforces that the configurations $\vec{N}^a$ have diversity $\phi$. Under the assumption that the order parameters associated to fixed points at given $\phi$ concentrate when $S \to \infty$ to values that are constant among the fixed points (Replica Symmetry), we can set
\begin{equation*}
  \begin{split}
    &q_{ab} = \delta_{ab} q_1 + (1-\delta_{ab}) q_0, \quad \xi_{ab} = \delta_{ab} \xi_1 + (1-\delta_{ab}) \xi_{0}\\
    &z_{ab} =  (1-\delta_{ab}) z, \quad m_a = m, \quad p_a = p,
  \end{split}
\end{equation*}
and similarly for the conjugate parameters. Introducing
$${\bf x}=(q_1, q_0, \xi_1, \xi_0, z, m,p)
\qquad \hat{\bf x}=(\hat{q}_1, \hat{q}_0, \hat{\xi}_1, \hat{\xi}_0, \hat z, \hat m, \hat p, \hat \phi)
\qquad {\bf y}=({\bf x}, \hat{\bf x})
$$ and performing an expansion to leading order in large $S$ and small $n$, we get~\eqref{eq:SPINT}
with
\begin{equation}\label{eq:Action}
  \begin{split}
    &\bar{\mathcal{A}}^{(\alpha)}=  \bar{\mathcal{p}}(\mathbf{x}) + n\,\mathcal{d}(\phi) - \frac{1}{2} \left( \hat{q}_0 q_0 + \hat{\xi}_0 \xi_0 + 2 \hat{z} z \right)\\
    &+ \hat{q}_1 q_1 + \hat{\xi}_1 \xi_1 + \hat{m} m + \hat{p} p + \hat{\phi} \phi + \bar{\mathcal{J}}^{(\alpha)}(\hat{\mathbf{x}}).
  \end{split}
\end{equation}
In this function,
\begin{equation*}
  \begin{split}
    \bar{p}(\mathbf{x}) &=
    \frac{(1- \mu m)}{\sigma^2 (q_1 - q_0)} \!\quadre{\frac{m (q_1 - q_0 + \gamma z)}{ (1 + \gamma)(q_1 - q_0)} +p - \frac{(1 - \mu m)}{2}}+\\
    &\quad\frac{q_0}{2\sigma^2 (q_1 - q_0)^2} \!\tonde{\!\xi_1\! - \!\xi_0\!- \!\frac{2 z}{1 + \gamma}\!}\!- \frac{1}{2\sigma^2} \!\tonde{\!\frac{\xi_1}{q_1 - q_0} \!+\!\frac{1}{1 + \gamma}\!}\!-\\
    &\quad\frac{\gamma}{2\sigma^2 (1 + \gamma)} \!\frac{z^2 (q_1 + q_0)}{(q_1 - q_0)^3}  - \frac{\log\left[2\pi\sigma^2(q_1 - q_0)\right]}{2}- \frac{q_0}{2(q_1 - q_0)}
  \end{split}
\end{equation*}
is the contribution coming from the large-$S$ expansion of \eqref{eq:Dist}, while
\begin{equation}
  \mathpzc{d}(\phi) =
  \begin{cases}
    \displaystyle
    \frac{1 - s(\phi)}{4 \gamma \sigma^2}  + \phi \quadre{\log \tonde{\frac{1 + s(\phi)}{2}}  - \frac{1}{2} }& \text{if } a(\phi) < 1 \\
    \displaystyle
    \frac{1}{2 \sigma^2} \frac{1}{1 + \gamma} - \frac{\phi}{2} + \frac{\phi}{2} \log(\sigma^2 \phi) & \text{if } a(\phi) > 1
  \end{cases}
\end{equation}
with $s(\phi)=\sqrt{1 - 4 \gamma \sigma^2 \phi}$ and $a(\phi)=\sigma \sqrt{\phi}(1 + \gamma)$ derives from the conditional expectation \eqref{eq:FanCondDet}. The remaining terms are the contribution of the Jacobian of the change of variables $({\bf N}, {{\bf f}}) \to {\bf y}$, with:
\begin{equation}\label{eq:Volume}
  \begin{split}
    &\bar{\mathcal{J}}^{(\alpha)}(\hat{\bf x}) =\int \frac{du_1 du_2 e^{\frac{\hat{\xi}_0 u_1^2 + \hat{q}_0u_2^2 - 2\hat{z}u_1 u_2}{2(\hat{q}_0 \hat{\xi}_0 - \hat{z}^2)}} }{2\pi \sqrt{\hat{q}_0 \hat{\xi}_0 - \hat{z}^2}}\times\\
    & \log \Bigg[e^{-\hat{\phi}} \sqrt{\frac{\pi}{2(2\hat{q}_1 - \hat{q}_0)}} \Pi^{(u)}\left(\frac{\hat m-u_1}{\sqrt{2(2\hat{q}_1 - \hat{q}_0)}}\right)+ \sqrt{\frac{\pi}{2 (2\hat{\xi}_1 - \hat{\xi}_0)}} \Pi^{(\alpha)}\left(\frac{u_2-\hat{p}}{\sqrt{2(2\hat{\xi}_1 - \hat{\xi}_0)}}\right) \Bigg]
  \end{split}
\end{equation}
where
\begin{equation}
  \Pi^{(\alpha)}(x)=
  \begin{cases}
    e^{x^2} \text{Erfc}(x), & \text{if} \quad \alpha=u\\
    2 e^{x^2}, & \text{if} \quad \alpha=t.
  \end{cases}
\end{equation}
Only this term discriminates between the total ($\alpha=t$) and uninvadable ($\alpha=u$) complexities. The $\Sigma^{(\alpha)}_{\sigma,\gamma}(\phi)$
are obtained fixing $\phi$ and solving the 15 coupled equations $\nabla_{\bf y}\bar{\mathcal{A}}^{(\alpha)}({\bf y}; \phi)=0$. For $\gamma=0$, $\alpha=u$ one recovers \cite{rosGeneralizedLotkaVolterraEquations2023, rosQuenchedComplexityEquilibria2023}.\\

We remark that the integral representation \eqref{eq:Volume} of the Jacobian $\bar{\mathcal{J}}^{(\alpha)}(\hat{\bf x})$ is derived under the assumption that the matrix in the quadratic form at the exponent has  negative eigenvalues. By solving the saddle point equations in the ME phases, we find that when the diversity $\phi$ reaches from above a critical value $\phi_{\text{Match}}$, the linear combination $\hat z + \sqrt{\hat{q}_0 \hat{\xi}_0}$ approaches zero, implying that the quadratic form at the exponent of \eqref{eq:Volume} develops a divergent mode. At this point, the double integral converges to a single integral over an effective variable, that is an appropriate linear combination of $u_1$ and $u_2$. Exactly at this value of parameters, the complexities $\Sigma_{\sigma, \gamma}^{(\alpha)} (\phi)$ map into their annealed counterparts, that are obtained exchanging the order in which the logarithm and the average over the randomness are taken in \eqref{eq:CompUn} (equivalently, by taking $n=1$ in \eqref{eq:SPINT}):
\begin{equation}
  \Sigma^{(\alpha, A)}_{\sigma, \gamma} (\phi) = \lim_{S\to\infty} \frac{1}{S} \log \left \langle\mathcal{N}^{(\alpha)}_S(\phi) \right \rangle.
\end{equation}
In other words, at this point the large-$S$ scaling of the typical values of the random variables $\mathcal{N}^{(\alpha)}_S$, which is controlled by  $\Sigma_{\sigma, \gamma}^{(\alpha)}$, coincides with the large-$S$ scaling of the average values of the random variables $\mathcal{N}^{(\alpha)}_S$, controlled by  $\Sigma_{\sigma, \gamma}^{(\alpha, A)}$. In addition, we find that for $\alpha=u$, $\phi_{\text{Match}}$ and the typical properties of the equilibria at that diversity can be predicted with the cavity method \cite{opper1992phase, buninEcologicalCommunitiesLotkaVolterra2017} in its simplest form. This method \emph{assumes} the existence of a unique, uninvadable equilibrium, and derives its properties under this assumption. In principle, its limit of validity is restricted to the values of $\sigma, \gamma$ within the UFP phase. Our analysis shows that, when extended within the ME phase, it still correctly describes the properties of a specific family of equilibria, those of diversity
$\phi_{\text{Match}}$. This family is however not typical: they are not the most numerous at the given values of $\sigma, \gamma$, as indicated by the fact that $\Sigma_{\sigma, \gamma}^{(u)} (\phi)$ peaks at $\phi \neq \phi_{\text{Match}}$. \\
For $\phi < \phi_{\text{Match}}$, the solution of the quenched saddle point equations is numerically unaccessible, presumably due to the smallness (or vanishing) of $\hat{z}+\sqrt{\hat{q}_0 \hat{\xi}_0}$. In this regime, we expect that $\Sigma_{\sigma, \gamma}^{(\alpha)} (\phi)=\Sigma_{\sigma, \gamma}^{(\alpha, A)} (\phi)$ continues to hold. We solve a modified set of saddle point equations obtained enforcing $\hat z + \sqrt{\hat{q}_0 \hat{\xi}_0}=0$, and find that the resulting branch of complexity and the saddle point values of the order parameters are numerically indistinguishable from the values in the annealed counterpart, reinforcing our hypothesis. Notice that we consistently find $\phi_{\text{Match}} > \phi_{\rm May}$, and thus the transition line separating the robust and fragile ME phases lies within the region of parameters where this matching is expected.  We report the saddle-point equations in the Appendix, and focus now on the results.

\section{Results}

\begin{figure}
  \centering
  \includegraphics[width=0.6\textwidth]{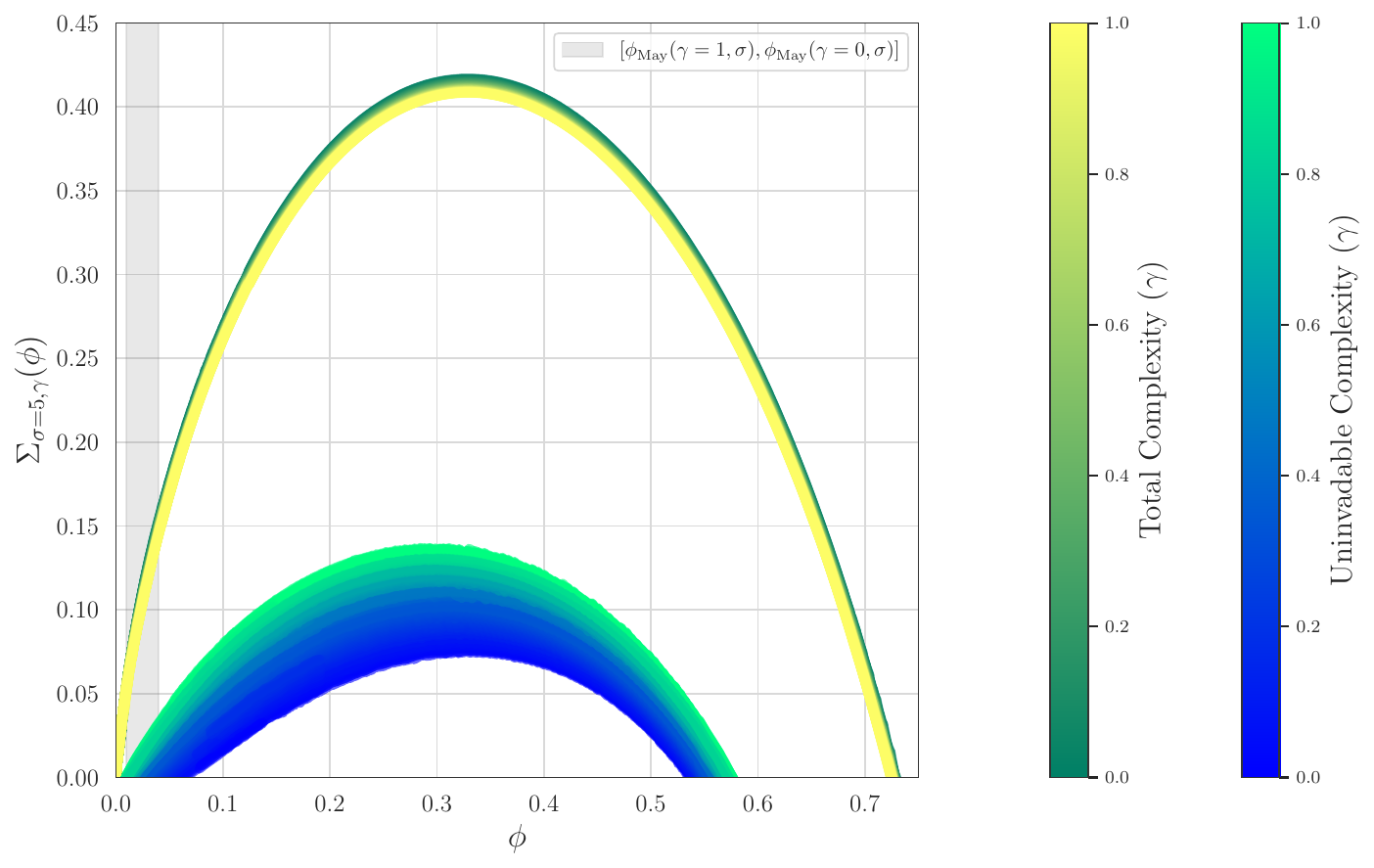}
  \caption{Total complexity $\Sigma^{(t)}_{\sigma, \gamma}(\phi)$ (upper curves) and uninvadable complexity $\Sigma^{(u)}_{\sigma, \gamma}(\phi)$ (lower curves) for $\sigma=5$ and various $\gamma$. The gray vertical band indicates the range of $\phi_{\rm May}=[{\sigma^2(1+\gamma)^2}]^{-1}$ for $\sigma=5$ and $\gamma \in [0,1]$.  }
  \label{fig:inv_vs_uninv}
\end{figure}

\begin{figure}
  \centering
  \includegraphics[width=.6\textwidth]{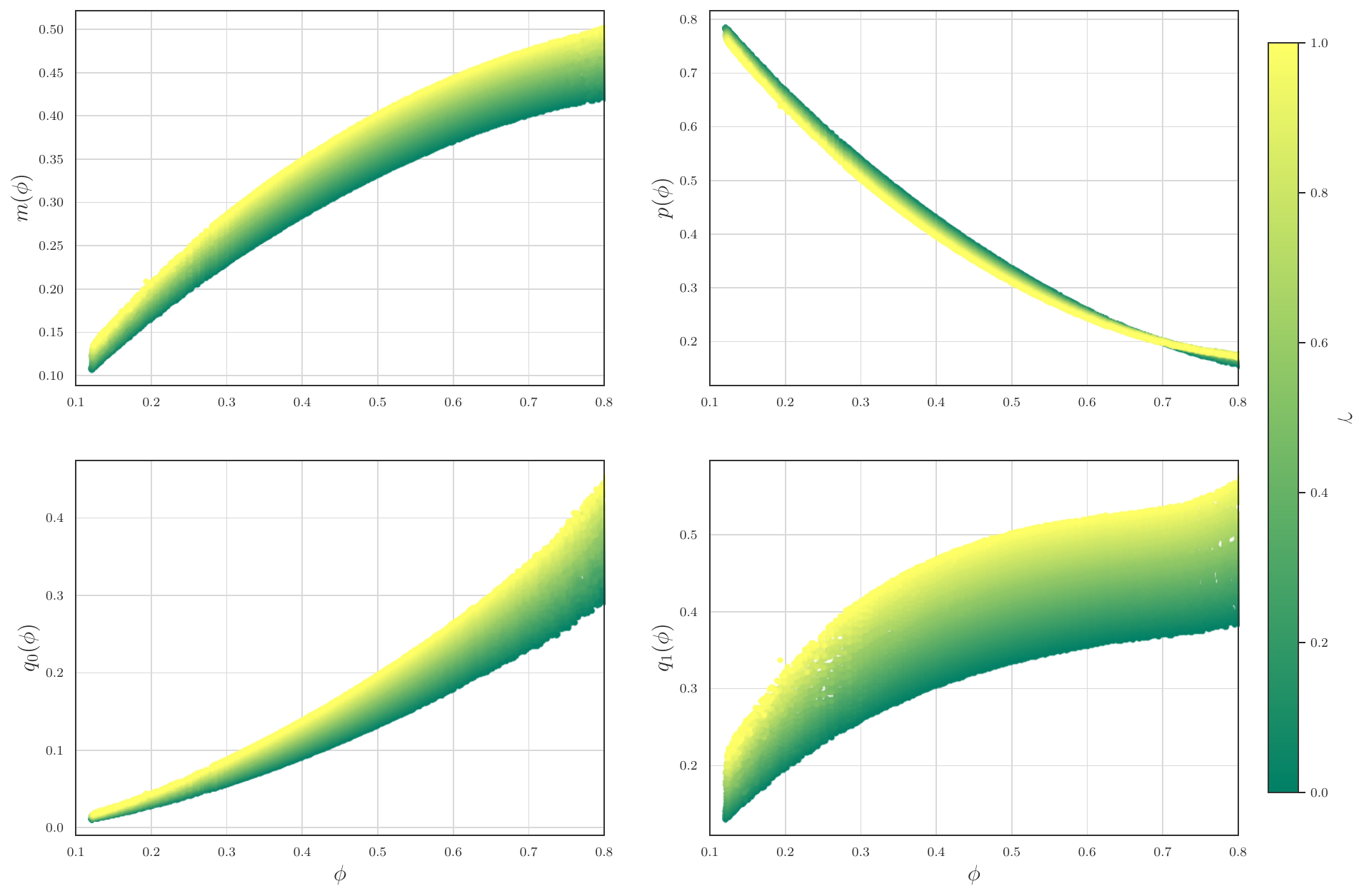}
  \caption{Order parameters of the typical equilibria at given diversity $\phi$, for $\sigma=2, \mu=1$ and various $\gamma$. The parameters are the average abundance $m$ (\emph{top left}), the average growth rate $p$ (\emph{top right}), the self similarity $q_1$ (\emph{bottom left}) and the mutual similarity $q_0$ (\emph{bottom right}). These values refer to invadable equilibria ($\alpha=t$), as also indicated by  $p>0$.}
  \label{fig:combined}
\end{figure}

Fig.~\ref{fig:inv_vs_uninv} shows $\Sigma^{(\alpha)}_{\sigma,\gamma}(\phi)$ for fixed variability $\sigma = 5$ and $\gamma \in [0,1]$. The complexity $\Sigma^{(u)}_{\sigma,\gamma}(\phi)$ is positive for all $\gamma$ over a finite interval of diversity, $\phi \in [\phi_{\rm min}^{(\alpha)}(\gamma), \phi_{\rm max}^{(\alpha)}(\gamma)]$: at this $\sigma$ the system is in a ME phase ($\sigma = 5> \sigma_c(\gamma)$, see Fig.~\ref{fig:gamma_crit}), with exponentially-many uninvadable fixed points of the dynamical equations. Equilibria with diversity lying outside this range exist with probability that is exponentially suppressed in $S$. As expected, $\Sigma^{(u)}_{\sigma,\gamma}(\phi)$ is much smaller than the total complexity $\Sigma^{(t)}_{\sigma,\gamma}(\phi)$, which counts both invadable and uninvadable equilibria: when $S$ is large, at any $\phi$ the total complexity is dominated by equilibria that are unstable to invasions, whose number exponentially exceeds that of the uninvadable ones. As $\gamma$ decreases, the distribution of diversities of the uninvadable equilibria becomes monotonically narrower, and the complexity at any fixed $\phi$ decreases. The opposite trend is observed for the total complexity.
The maximal diversity $\phi_{\rm max}^{(t)}(\gamma)$ remains always bounded away from one, in agreement with the statement that feasible solutions of~\eqref{eq:EqDef} (i.e., solutions with all entries strictly positive) occur with non-zero probability only when the interactions are further rescaled by a factor of order $\sqrt{\log S}$~\cite{bizeul2021positive}. For any $\gamma$, $\phi_{\rm max}^{(u)}(\gamma)< \phi_{\rm max}^{(t)}(\gamma)$: thus, there is a range of diversity $\phi \in [\phi_{\rm max}^{(u)}(\gamma), \phi_{\rm max}^{(t)}(\gamma)]$ where equilibria are exponentially numerous, but none of them are uninvadable when $S \gg 1$. Notice that the complexities $\Sigma^{(\alpha)}_{\sigma,\gamma}(\phi)$ turn out to be independent of  $\kappa, \mu$, that influence the typical properties $m, q_1, q_0$ of the equilibria, but not their number.  The values of $m, q_1$ increase with $\mu$ until divergence, that corresponds to the unbounded growth phase in \cite{buninEcologicalCommunitiesLotkaVolterra2017}. Fig.~\ref{fig:combined} shows the order parameters of the typical (invadable) equilibria at $\sigma=2, \mu=1$. At fixed $\phi$, the average abundance, self-similarity and  similarity between different equilibria all increase with $\gamma$. This generalizes the trend of the order parameters of the unique equilibrium in the UFP phase, and it can be understood from simple two-species arguments (the average abundance increases when interactions are more correlated). The average effective growth rate $p$ is instead non-monotonic in $\gamma$: at small $\phi$, equilibria have higher values of $p$ when the degree of reciprocity $\gamma$ is higher (invading species grow faster), while at large $\phi$ the trend is reversed. {This behavior is overall robust to changes in $\sigma$ in the ME phase.}
\paragraph*{Recovering the UFP phase.} As one approaches $\sigma \to \sigma_c(\gamma)$, the complexity $\Sigma^{(u)}_{\sigma, \gamma}(\phi)$ decreases and the range $[\phi_{\rm min}^{(u)}(\gamma), \phi_{\rm max}^{(u)}(\gamma)]$ shrinks, reducing to one single point at $\sigma=\sigma_c$: this point is exactly $\phi_{\rm May}$, and $\Sigma^{(u)}_{\sigma_c(\gamma), \gamma}(\phi_{\rm May})=0$: the transition to the UFP phase is recovered. However, $\Sigma^{(t)}_{\sigma, \gamma}(\phi)$ remains positive below the transition: in the UFP phase, there actually exists an exponentially large number of equilibria, some of which are even internally stable, but all of which are unstable to invasions. The dynamics is oblivious to these equilibria and converges to the unique uninvadable fixed point.

\paragraph*{Two distinct ME phases.} The fixed points counted by the complexities (whether invadable or not) are internally stable whenever \eqref{eq:Jacobian} has no eigenvalues with positive real part. In the ME phase, one has to characterize the statistics of the matrices $H_{ij}(\vec{N})$ conditional on $\vec{N}$  solving \eqref{eq:EqDef} and having a given $\phi$. The replicated Kac–Rice formalism is well suited for this analysis~\cite{rosQuenchedComplexityEquilibria2023, rosComplexEnergyLandscapes2019}. One finds that the Jacobian at fixed points is statistically equivalent to an elliptic Gaussian matrix of size $S \phi \times S \phi$, deformed by finite-rank perturbations and shifted by the identity matrix. Importantly, the finite-rank perturbations do not affect the eigenvalue density\footnote{We stress that the eigenvalue density does not account for possible isolated eigenvalues that provide subleading corrections to the spectrum.} for large $S$: eigenvalues are complex and uniformly distributed in the ellipse $\mathcal{S}_{\sigma, \gamma}=\left\{ z \in \mathbb{C}: \frac{(\Re z+1)^2}{\sigma^2 \phi (1+\gamma)^2}+ \frac{(\Im z)^2}{\sigma^2 \phi (1-\gamma)^2} \leq 1\right\}$, and therefore they all have negative real part whenever $\phi < \phi_{\rm May}$. By tracking the internal stability of the equilibria, we find that the ME region splits into two distinct phases, see Fig.~\ref{fig:gamma_crit}, which we call \emph{fragile} and \emph{robust}.
In the fragile phase, all uninvadable equilibria are internally unstable ($\phi^{(u)}_{\rm min} > \phi_{\rm May}$), and thus internally stable fixed points are invadable. In the robust phase, instead, $\phi^{(u)}_{\rm min} <\phi_{\rm May}$, so internally stable equilibria that are also resistant to invasions do exist. The two phases are separated by a critical curve $\gamma_{\text{FR}}(\sigma)$, defined by $\phi^{(u)}_{\rm min} =\phi_{\rm May}$. A numerical fit yields $\gamma_{\text{FR}}(\sigma \to \infty) \approx 0.549$. Thus, for $\gamma \in [0, 0.549]$ only the fragile ME phase exists, consistent with the $\gamma=0$ results~\cite{rosQuenchedComplexityEquilibria2023,rosGeneralizedLotkaVolterraEquations2023}. For $\gamma \in (0.549, 1)$, both phases coexist. At $\gamma=1$ the critical line meets the transition to the UFP phase, implying that no fragile ME phase exists in the conservative limit—consistent with the expected energy landscape, which possess many local minima (internally stable, uninvadable fixed points).

\section{Discussion.}

Assessing to what extent the chaotic dynamics of systems with non-reciprocal interactions can be understood in terms of the equilibria of the underlying (non-conservative) equations of motion is an open challenge. While a qualitative correspondence between equilibria properties and the dynamics can be identified~\cite{wainrib2013topological, stubenrauch2025fixed, yang2025relationship}, recent results indicate that establishing a quantitative connection is generally far from straightforward~\cite{fournier2025non}. Assuming nonetheless that out-of-equilibrium dynamics is influenced by fixed points, one can derive predictions for the rGLV dynamics from the phase diagram in Fig.~\ref{fig:gamma_crit}. In particular, one should expect two qualitatively different (chaotic) regimes; when $\gamma$ is small and the system is in the fragile ME phase, if internally stable fixed points attract the dynamics, they are eventually destabilized by the growth of species absent from the fixed point, as all internally stable equilibria are invadable. This seems consistent with ~\cite{arnoulxdepireyManySpeciesEcologicalFluctuations2024}, where (for $\gamma=0$) species abundances under weak immigration exhibit a separation of scales: they split into well-separated high- and low-abundance subsets, with dynamical turnover between them. The diversity of the high-abundance subset (extrapolated to vanishing immigration) lies below the stability threshold $\phi_{\rm May}$~\cite{arnoulxdepireyManySpeciesEcologicalFluctuations2024}, and in a region where we find $\Sigma^{(u)}_{\sigma, \gamma}$ to be high, though not maximal. This behavior can be interpreted as the system fluctuating between internally stable but invadable fixed points, with dynamics largely driven by migration/takeover of low-abundance species. From our complexity analysis, we expect this $\gamma=0$ phenomenology to extend across the fragile ME phase.
When $\gamma$ is larger, the system enters the robust ME phase, where internally stable and uninvadable fixed points exist in exponential number. Understanding which of these equilibria dominate the chaotic dynamics, if any, requires comparing their properties (computed here) with those of the configurations visited by chaotic trajectories at large times. Such a comparison was recently carried out in a family of models with random non-reciprocal interactions~\cite{fournier2025non}, showing that while the chaotic attractor may be largely contributed by fixed points, these are not the typical ones, even though correlations exist between properties of the attractor and those of the fixed points. Investigating these links in the gLVE dynamics, especially the qualitative changes in the dynamics across the two ME phases, and establishing quantitative connections with the properties of equilibria are natural perspectives of this work. Another natural direction is to generalize the complexity calculation to models with structured or stronger interactions~\cite{giral2024interplay, mallmin2024chaotic}.

\paragraph*{Acknowledgments}
VR acknowledges Thomas Roy for previous investigations on the subject of this paper and for valuable discussions, and acknowledges funding by the French government under the France 2030 program (PhOM - Graduate School of Physics) with
reference ANR-11-IDEX-0003. VR and TLS thank Juan Giral Martínez and Thibaut Arnoulx de Pirey for insightful discussions.

\bibliography{biblio}

\clearpage
\section*{Appendix}
\begin{appendix}
    
This appendix is structured as follows. In section \ref{sec:setup}, we recall useful definitions and results on the random Generalized Lotka Volterra (rGLV) equations. In section \ref{sec:LV} we recall the structure of the Kac-Rice calculation, that is behind the derivation of our results. In section \ref{sec:self-consistent}, we present the self-consistent equations obtained from this method. In section \ref{sec:results}, we present some additional results on the complexity calculation. Finally, section \ref{sec:appendix} contains further details on the derivation of the self-consistent equations.

\section{Model setup and definitions}\label{sec:setup}
\parhead{The model.} The \textit{Generalized Lotka Volterra} (GLV) equations read, in their standard rescaled form:
\begin{equation}
  \label{eq:LVmulti2}
  \frac{dN_i}{dt}
  = N_i\Bigl(\kappa_i - N_i -  \sum_{j=1}^S \alpha_{ij}\,N_j\Bigr),
  \quad i=1,\dots,S
\end{equation}
where $N_i$ is the abundance of species $i$, and where we have rescaled each population by its carrying capacity, which we assume to be uniform: $\kappa_i = \kappa$ for all $i=1, \dots , S$. This formulation allows for arbitrary networks of interactions between species, which can be competitive, predatory, or mutualistic. We assume that the elements \(\alpha_{ij}\) of the interaction matrix are drawn at random from a Gaussian distribution with mean \(\mu/S\) and variance \(\sigma^2/S\), so that
\begin{equation}\label{eq:StatInt}
  \alpha_{ij} = \frac{\mu}{S} + \frac{\sigma}{\sqrt{S}}\,a_{ij} \quad \text{with} \quad \bigl\langle a_{ij}a_{kl}\bigr\rangle
  =\delta_{ik}\,\delta_{jl} + \gamma\,\delta_{il}\,\delta_{jk}.
\end{equation}
Note that although it is customary to set $\alpha_{ii}=0$ in the GLV model, in what follows we allow Gaussian fluctuations of the diagonal elements as well, as specified in \eqref{eq:StatInt}; this does not affect any of the results obtained in the $S \to \infty$
limit considered below. In this limit, certain system-level trends are robust to changes in modeling choices, and the minimal Gaussian model for interaction statistics \eqref{eq:StatInt} appears appropriate, as suggested by arguments of Gaussian universality~\cite{barbierGenericAssemblyPatterns2018}. The reciprocity of the interactions is controlled by the parameter $\gamma \in [0,1]$.


\parhead{Definitions.}
A \textit{fixed point} or  \textit{equilibrium} of the GLV equations is a vector $\vec{N}^*=(N_1^*,\dots,N_S^*)$ which verifies
\begin{equation}
  N_i^* F_i(\vec{N}^*)=0  \text{ for all  }  i=1,\dots, S, \quad \quad F_i(\vec{N}):=  \kappa - \frac{\mu}{S}\sum_{i} N_i - N_i  - \frac{\sigma}{\sqrt{S} }\sum_{j} a_{ij}N_j,
\end{equation}
where $\vec{F}(\vec{N})$ is the vector of \textit{forces} or \textit{effective growth rates} associated to each species.
In a fixed point configuration, for all $i=1,\dots, S$ either $N_i^* =0$ (the species is absent in the equilibrium) or $F_i(N_i^*)=0$. We define the index set $I(\vec{N}^*) = (i_1, \dots, i_s)$ collecting the labels of the species coexisting in the equilibrium. The equilibrium \textit{diversity} and \textit{average abundance} are given by
\begin{equation}
  \phi_S(\vec{N}^*) = \frac{|I(\vec{N}^*)|}{S}, \quad \quad  \quad   m_S(\vec{N}^*) = \frac{1}{S}\sum_{i=1}^S N_i.
\end{equation}
Henceforth, we denote simply with $\phi, m$ the $S \to \infty$ limit of these quantities. The Kac-Rice calculation allows to determine the distribution of $\phi, m$ over the multiple fixed points of the GLV equations.

Given the factorized structure of the fixed point condition $N_i^*F_i(\vec{N}^*) = 0$, we define two types of stability of each equilibrium. \textit{Uninvadibility} is the  stability with respect to the injections of the species that are absent in the fixed point, and it is guaranteed whenever $ F_i(\vec{N}^*) < 0$ for all $ i \notin I(\vec{N}^*)$. The  \textit{internal stability} of a fixed point is instead the stability with respect to small fluctuations in the abundance of the existing species. Mathematically, this is guaranteed by the fact that the eigenvalues of the stability matrix
\begin{equation}
  H_{ij} = \left( \frac{\partial F_i(\vec{N}^*)}{\delta N_j} \right)_{\substack{i,j \in I(\vec{N}^*)}}
\end{equation}
all have a real part that is negative. For the GLV model with Gaussian interactions, stable equilibria are those that verify $\phi \leq \phi_{\text{May}}= [\sigma(1+\gamma)]^{-2}$, as it follows from a random matrix theory argument recalled in the main text \cite{rosGeneralizedLotkaVolterraEquations2023}.

\parhead{Multiple equilibria and complexity.} We recall here the phase diagram obtained in \cite{buninEcologicalCommunitiesLotkaVolterra2017} using the so called cavity method \cite{mezardSpinGlassTheory1986}, in $(\mu,\sigma)$ space. For general $\gamma$, the GLV model exhibits three phases: (i) A \textit{Unique Fixed Point} (UFP) phase, (ii) an \textit{Unbounded Growth} phase and  (iii) a \textit{Multiple Equilibria} (ME) phase. The results of \cite{rosGeneralizedLotkaVolterraEquations2023, rosQuenchedComplexityEquilibria2023} for $\gamma=0$ show that both the Multiple Equilibria phase and the unbounded growth phase are characterized by the existence of an exponentially large (in $S$) number of fixed points of the GLV equations. In the unbounded growth phase, a fraction of these fixed points is characterized by a divergent value of their average abundance. The line separating UFP and ME phases lies at $\sigma_c(\gamma) = {\sqrt{2}}/(1+\gamma)$, as first obtained in \cite{rieger1989solvable}.

We denote with $\mathcal{N}^{(u)}_{S}(\phi)$ and $\mathcal{N}^{(t)}_{S}(\phi)$ the number of uninvadable equilibria and the total number of equilibria (both invadable and uninvadable) at fixed diversity $\phi$. We use the notation $\mathcal{N}^{(\alpha)}_{S}(\phi)$ with $\alpha\in \left\{ u, t \right\}$. Our object of interest is the \textit{topological complexity} $\Sigma_S(\phi) = S^{-1}\log \mathcal{N}_{S}(\phi)$. More precisely, we aim at the deterministic functions:
\begin{align}
  \Sigma^{(\alpha)}_{\sigma,\gamma} (\phi) := \lim_{S\to\infty} \frac{1}{S}\langle \log \mathcal{N}^{(\alpha)}_S(\phi) \rangle, \quad \quad
  \Sigma^{(\alpha, A)}_{\sigma,\gamma} (\phi) := \lim_{S\to\infty} \frac{1}{S}\log \langle \mathcal{N}^{(\alpha)}_S(\phi) \rangle,
\end{align}
where $\Sigma^{(t)}_{\sigma,\gamma} (\phi) $ is the \textit{quenched} total complexity, $\Sigma^{(t,A)}_{\sigma,\gamma} (\phi)$ the \textit{annealed} total complexity, and  $\Sigma^{(u)}_{\sigma,\gamma} (\phi) $ and $\Sigma^{(u,A)}_{\sigma,\gamma} (\phi) $ are the uninvadable counterparts. The distinction between quenched and annealed averages is central in disordered systems, where the quenched complexity is the relevant quantity to consider, as it describes the correct asymptotic behavior of the random variable $\Sigma_S(\phi)$, which is self-averaging and whose distribution concentrates around its mean when $S \to \infty$. The annealed complexity is much simpler to calculate, though, and often considered in the mathematical literature as it can be derived within a rigorous formalism; it furnishes an upper bound to the quenched complexity, $\Sigma^{(\alpha)}_{\sigma,\gamma} (\phi) \leq \Sigma^{(A,\alpha)}_{\sigma,\gamma} (\phi)$ with $\alpha\in \left\{ u, t \right\}$. Below, we present the derivation of the formulas of both quenched and annealed complexities.

\section{The replicated Kac-Rice calculation: the structure}\label{sec:LV}
In this Section, we discuss the step leading to the integral representation (10) in the main text. The structure of the calculation is very similar to that presented in Ref.~\cite{rosQuenchedComplexityEquilibria2023}, and we report it here for completeness. It relies on the replicated Kac-Rice method first introduced in~\cite{rosComplexEnergyLandscapes2019}. For more details on this method, see \cite{rosHighdimensionalRandomLandscapes2025b}.

\subsection{The replicated Kac-Rice formula}
We begin by deriving the expression of the density $\rho_{S,n}^{(\alpha)}$ appearing in Eq.~(8) in the main text. We introduce
\begin{equation}
  \Theta^{(\alpha)}_{I}(\vec{N}, \vec{f}) =\prod_{i\in I} \theta(N_i) \delta(f_i) \prod_{i \notin I} \delta(N_i) \chi^{(\alpha)}(f_i), \quad \quad   \chi^{(\alpha)}(f)=
  \begin{cases}
    \theta(-f) \quad \text{if} \quad \alpha=u\\
    1 \quad \text{if} \quad \alpha=t,
  \end{cases}
\end{equation}
where $\theta (x) = 1 $ for $x>0$ and $0$ otherwise. For a fixed realization of the random GLV equations, the  number of fixed points  $\mathcal{N}^{(\alpha)}_{S}(\phi)$ admits the following integral representation:
\begin{equation}\label{eq:IntegralN}
  \mathcal{N}^{(\alpha)}_{S}(\phi) = \sum_{I : |I| = S\phi} \int_{\mathbb{R}^{S} \times \mathbb{R}^{S}}\dd \vec{N} \dd \vec{f} \; \Theta^{(\alpha)}_{I}(\vec{N}, \vec{f}) \,  \left|\det \left(\frac{\partial F_i}{\partial N_j}\right)_{ij \in I }\right| \delta\left(\vec{F}(\vec{N}) - \vec{f}\right).
\end{equation}
The last term in the integral enforces that the vector of effective growth rates takes values $\vec{f}$. For any possible choice of the index set $I$ compatible with the diversity $\phi$, $\Theta^{(\alpha)}_{I}(\vec{N}, \vec{f})$ enforces that the configuration $\vec{N}$ is an equilibrium, and that it is uninvadable when $\alpha=u$. The term containing the determinant is a Jacobian factor accounting for the multiplicity of possible solutions of the fixed point constraint. Notice that the only difference between the total number of equilibria and the number of uninvadable ones is in the factors $\theta(-f_i)$, that are absent for $\alpha=t$.

The replica trick requires to compute higher moments of the random variable \eqref{eq:IntegralN}. From \eqref{eq:IntegralN} we get:
\begin{align}\label{eq:CleanKR}
  &\quad \quadre{\mathcal{N}^{(\alpha)}_{S}(\phi)}^n \notag \\
  & =\sum_{\substack{(I_1, \cdots, I_n)\\  |I_a| = S\phi}}  \int_{\mathbb{R}^{Sn} \times \mathbb{R}^{Sn}} \prod_{i=1}^S \prod_{a=1}^n d {N}^a_i  d {f}^a_i \; \prod_{b=1}^n\Theta^{(\alpha)}_{I_b}(\vec{N}^b, \vec{f}^b)  \;  \left|\det \left(\frac{\partial F_i^b}{\partial N^b_j}\right)_{ij \in I_b }\right|\;  \delta\left(\vec{F}^b- \vec{f}^b\right),
\end{align}
where we set $I_a \equiv I(\vec{N}^a)$ and $\vec{F}^a \equiv F(\vec{N}^a)$.
The replicated Kac-Rice formula is obtained averaging \eqref{eq:CleanKR} over the random couplings $\alpha_{ij}$. Denoting with $\mathbf{N} = (\vec{N}^1,...,\vec{N}^n)$ the concatenation of the configurations and with $\mathbf{F}(\mathbf{N}) = (\vec{F}^1,\dots,\vec{F}^n)$ the concatenation of effective growth rates, we can write:
\begin{equation}
  \mathcal{W}_{\mathbf{N}}^{(n)} (\mathbf{f}):= \left\langle \prod_{b=1}^n \left|\det \left(\frac{\partial F_i^b}{\partial N^b_j}\right)_{ij \in I_b }\right|\;  \delta\left(\vec{F}^b - \vec{f}^b\right) \right\rangle=\mathcal{D}_{\mathbf{N}}^{(n)} (\mathbf{f}) \times \mathcal{P}_{\mathbf{N}}^{(n)}({\bf f}),
\end{equation}
where
\begin{equation}
  \mathcal{D}_{\mathbf{N}}^{(n)} (\mathbf{f}) = \left\langle \prod_{a=1}^n \left| \det\left(\frac{\delta F_i^{a}}{\delta N_j^a} \right)_{i,j \in I_a}\right| \Big | \mathbf{F}(\mathbf{N}) = \mathbf{f}\right\rangle
  \label{eq:cond_exp}
\end{equation}
is the conditional expectation of the product of determinants given that $\mathbf{F}(\mathbf{N}) = \mathbf{f}$, while  $\mathcal{P}_{\mathbf{N}}^{(n)}({\bf f}) = \left\langle  \delta\left(  \mathbf{F}(\mathbf{N}) - \mathbf{f}\right) \right \rangle$
is the probability that $\mathbf{F}(\mathbf{N}) = \mathbf{f}$. We then recognize that
\begin{align}\label{eq:KRApp}
  \left \langle \quadre{\mathcal{N}^{(\alpha)}_{S}(\phi)}^n  \right \rangle &=\int_{\mathbb{R}^{Sn}} \prod_{a=1}^n d \vec{N}^a   \; \rho_{S,n}^{(\alpha)}\tonde{\grafe{\vec{N}^a}_{a=1}^n ; \phi}\\
  \rho_{S,n}^{(\alpha)}=& \sum_{\substack{(I_1, \cdots, I_n)\\  |I_a| = S\phi}}  \int_{\mathbb{R}^{Sn}}  \prod_{a=1}^n d \vec{f}^a   \; \Theta^{(\alpha)}_{I_a}(\vec{N}^a, \vec{f}^a) \mathcal{W}_{\mathbf{N}}^{(n)} (\mathbf{f}),
\end{align}
as reported in the main text.
Eq.~\eqref{eq:KRApp} is the \textit{replicated Kac-Rice formula}.
Let's stress that the general $n$ case is relevant for the quenched computation, while $n=1$ gives us the annealed complexity.

To proceed, we need to determine \eqref{eq:KRApp} to leading exponential order in $S$ in order to perform a saddle point calculation. In $\mathcal{W}_{\mathbf{N}}^{(n)} (\mathbf{f})$, the objects containing the randomness are the effective growth rates $\vec{F}^a$: these are vectors with Gaussian statistics, owning to the Gaussianity of the couplings $\alpha_{ij}$; moreover, the statistics of the vectors components is isotropic: their first and second moments depend on the components $N^a_i$ only through rotationally invariant combinations, that are, scalar products. In other words,  $\mathcal{W}_{\mathbf{N}}^{(n)} (\mathbf{f})$ depends on the $\vec{N}^a$ and $\vec{f}^a$ only through the functions:
\begin{align}\label{eq:OP}
  &q(\vec{N}^a, \vec{N}^b) = \frac{\vec{N}^a \cdot \vec{N}^b}{S}, \quad  \xi(\vec{f}^a, \vec{f}^b) = \frac{\vec{f}^a \cdot \vec{f}^b}{S}, \\
  &  z(\vec{N}^a, \vec{f}^b) = \frac{\vec{N}^a \cdot \vec{f}^b }{S}, \quad   m(\vec{N}^a) = \frac{\vec{N}^a \cdot \vec{1}}{S}, \quad p(\vec{f}^a) = \frac{\vec{f}^a \cdot \vec{1}}{N}
\end{align}
for all $a \leq b =1, \cdots, b$. In here, we denoted  $\vec{1}=(1,1,\cdots, 1)^T$. We can therefore conveniently perform a change of variable and rewrite the integration over the $(Sn)^2$ parameters $N_i^a, f^a_i$ as an integration over the variables \eqref{eq:OP}. Denoting with $q_{ab}$ the value taken by the function $q (\vec{N}^a, \vec{N}^b)$ and similarly for the others, and defining with  $\mathbf{x} = \left\{q_{ab},\xi_{ab},z_{ab},m_a,p_a\right\}$ the collection of such values for all $a \leq b$ and with  $\hat{\mathbf{x}} = \left\{\hat{q}_{ab},\hat{\xi}_{ab},\hat{z}_{ab},\hat{m}_a,\hat{p}_a\right\}$ a set of auxiliary variables, we obtain the integral representation
\begin{equation}  \label{eq:N^n}
  \left \langle \quadre{\mathcal{N}^{(\alpha)}_{S}(\phi)}^n  \right \rangle = \int \frac{ \dd \mathbf{x} \dd \hat{\mathbf{x}}}{(2\pi)^{2n^2 + 3n}} e^{S g_n(\mathbf{x},\hat{\mathbf{x}},\phi)}\mathcal{V}^{(\alpha)}_n(\hat{\mathbf{x}}) \mathpzc{d}_n(\mathbf{x},\phi)\mathcal{P}_n(\mathbf{x})
\end{equation}
where $\mathpzc{d}_n$ and $\mathcal{P}_n$ denote the terms $\mathcal{D}_{\mathbf{N}}^{(n)}$ and $\mathcal{P}_{\mathbf{N}}^{(n)}$ now expressed as a function of the variables ${\bf x}$, and
\begin{equation}
  \begin{split}
    &g_n(\mathbf{x}, \hat{\mathbf{x}}, \phi) = \sum_{a=1}^n \left( \hat{m}_a m_a + \hat{p}_a p_a + \hat{\phi}_a \phi \right) + \sum_{ a \leq b=1}^n \left( \hat{q}_{ab} q_{ab} + \hat{\xi}_{ab} \xi_{ab} \right) + \sum_{a \neq b} \hat{z}_{ab} z_{ab}.
  \end{split}
\end{equation}
The `volume term' $\mathcal{V}^{(\alpha)}_n(\mathbf{x},\hat{\mathbf{x}})$ contains the integration over the variables $N^a_i, f^a_i$ and it reads explicitly:
\begin{equation}
  \begin{split}\label{eq:VolumeKR}
    &\mathcal{V}^{(\alpha)}_n( \hat{\mathbf{x}}) = S^{4n(n+1)+n} \sum_{\tau_i^1 = 0,1} \cdots \sum_{\tau_i^n = 0,1} \int \prod_{a=1}^n d\vec{N}^a \, d\vec{f}^a \, e^{-\hat{m}_a \, \vec{N}^a \cdot \vec{1} - \hat{p}_a \, \vec{f}^a \cdot \vec{1} - \hat{\phi}_a \, \vec{\tau}^a \cdot \vec{\tau}^a} \times   \\
    & \times \prod_{\substack{a,b=1\\ a \ne b}}^n e^{-\hat{z}_{ab} \, \vec{N}^a \cdot \vec{f}^b} \prod_{a\leq b=1}^n e^{-\hat{q}_{ab} \, \vec{N}^a \cdot \vec{N}^b - \hat{\xi}_{ab} \, \vec{f}^a \cdot \vec{f}^b} \prod_{a=1}^n \left( \prod_{i : \tau_i^a = 1} \theta(N_i^a) \delta(f_i^a) \prod_{i : \tau_i^a = 0} \delta(N_i^a) \chi^{(\alpha)}(f_i^a) \right),
  \end{split}
\end{equation}
where we have introduced the $S$-dimensional vectors $\vec{\tau}^a$ that have components $\tau^a_i=1$ if $i \in I_a$ (meaning, if the species labeled by $i$ is present in the given configuration), and  $\tau^a_i=0$ otherwise. These expressions are obtained enforcing that the functions \eqref{eq:OP} take values ${\bf x}$ through delta functions, which are then expressed as $ \delta(x) =\frac{1}{2\pi} \int \dd{\hat{x}} e^{i x \hat{x}}$ with the help of the conjugate parameters $\hat{\bf x}$.  The constraint $\vec{\tau}^a \cdot \vec{\tau}^a = S\phi$ is enforced by the conjugate parameter $\hat{\phi}_a$.
The introduction of the  \textit{order parameters} $\mathbf{x}$ allows for a drastic reduction in dimensionality, as is standard in statistical physics calculations for high-dimensional fully connected systems.

\subsection{The replica symmetric assumption}

We proceed making a  \textit{replica symmetric (RS)} assumption, that corresponds to:
\begin{align}
  &q_{ab} = \delta_{ab} q_1 + (1-\delta_{ab}) q_0, \quad  \hat{q}_{ab} = \delta_{ab} \hat{q}_1 + (1-\delta_{ab}) \hat{q}_0 \\
  &\xi_{ab} = \delta_{ab} \xi_1 + (1-\delta_{ab}) \xi_{0} , \quad \hat{\xi}_{ab} = \delta_{ab} \hat{\xi}_1 + (1-\delta_{ab}) \hat{\xi}_{0} \\
  &z_{ab} =  (1-\delta_{ab}) z , \quad \hat{z}_{ab} =  (1-\delta_{ab}) \hat{z} \\
  &m_a = m , \quad \hat{m}_a = \hat{m} \\
  &p_a = p , \quad \hat{p}_a = \hat{p} \text{, }
\end{align}
where $z_{ab} = 0$ for $a=b$ because of the equilibrium condition. For a discussion on the relevance of this assumption, see~\cite{rosQuenchedComplexityEquilibria2023}. We now give, under this assumption, the value of each of the terms appearing in \eqref{eq:N^n}, to leading exponential order in $S$. We remark that the derivation of the asymptotics of the terms $\mathcal{P}_n({\bf x})$ and $\mathpzc{d}_n(\mathbf{x},\phi)$ is discussed in Ref.~\cite{rosQuenchedComplexityEquilibria2023}, and we refer the reader to that reference for further details. As in that work, we find:
\begin{equation}\label{eq:Asy1}
  \begin{aligned}
    p_n(\mathbf{x}) := \lim_{S \to \infty} \frac{\log \mathcal{P}_n({\bf x})}{S} =& -\frac{1}{2\sigma^2 (1 + \gamma)} \frac{n\, U_n(\mathbf{x})}{(q_1 - q_0)^2 \left[ q_1 + (n - 1) q_0 \right]^2}  \\
    &- \frac{n}{2} \log(2\pi\sigma^2) - \frac{n - 1}{2} \log(q_1 - q_0)\\
    &- \frac{1}{2} \log\left[q_1 + (n - 1) q_0\right]
  \end{aligned}
\end{equation}
with
\begin{equation}
  \begin{aligned}
    U_n(\mathbf{x}) &= (\kappa - \mu m)^2 (q_1 - q_0)^2 \left\{ (1 + \gamma)\left[q_1 + (n - 1) q_0\right] - \gamma n m^2 \right\}\\
    &\quad + (1 + \gamma)\xi_1 (q_1 - q_0) \left[q_1 + (n - 2) q_0\right] \left[q_1 + (n - 1) q_0\right]\\
    &\quad - 2(\kappa - \mu m)(q_1 - q_0)^2 \left\{ m \left[q_1 + (n - 1)(q_0 - \gamma z)\right] + (1 + \gamma)p \left[q_1 + (n - 1) q_0\right] \right\} \\
    &\quad  - \gamma (n - 1) z^2 \left[q_1^2 + (n - 1) q_0^2\right]  \\
    &\quad + (q_1 - q_0)^2 \left[q_1 + (n - 1) q_0\right]^2 - (n - 1)(1 + \gamma)\xi_0 q_0 (q_1 - q_0) \left[q_1 + (n - 1) q_0\right]  \\
    &\quad - 2(n - 1) q_0 z (q_1 - q_0) \left[q_1 + (n - 1) q_0\right].
  \end{aligned}
\end{equation}
Similarly, to leading order in $S$:
\begin{align}\label{eq:Asy2}
  &\quad n\,\mathpzc{d}(\phi):= \lim_{S \to \infty} \frac{\mathpzc{d}_n(\mathbf{x},\phi)}{S} \\
  &=n
  \begin{cases}
    \displaystyle
    \frac{1}{4 \gamma \sigma^2} \left( 1 - \sqrt{1 - 4 \gamma \sigma^2 \phi} \right) + \phi \log \left( 1 + \sqrt{1 - 4 \gamma \sigma^2 \phi} \right) - \phi \left( \frac{1}{2} + \log 2 \right) &  a_{\sigma, \gamma}(\phi) < 1 \\[1em]
    \displaystyle
    \frac{1}{2 \sigma^2} \frac{1}{1 + \gamma} - \frac{\phi}{2} + \frac{\phi}{2} \log(\sigma^2 \phi) &  a_{\sigma, \gamma}(\phi) > 1
  \end{cases}
\end{align}
with $a_{\sigma, \gamma}(\phi)= \sigma \sqrt{\phi}(1 + \gamma)$. The asymptotics of the volume term $\mathcal{V}^{(\alpha)}_n(\hat{\mathbf{x}})$ is obtained in~\cite{rosQuenchedComplexityEquilibria2023} for the case $\alpha=u$. A generalization of that calculation, whose details can be found in Sec.~\ref{sec:AppVolume} of this appendix, gives:
\begin{align}\label{eq:AsyVOl}
  &\quad \mathcal{J}_n^{(\alpha)}(\hat{\mathbf{x}}):=\lim_{S \to \infty} \frac{\log \mathcal{V}^{(\alpha)}_n(\hat{\mathbf{x}}) }{S}\\
  &=\log \left[\int \dd{u_1} \dd{u_2} \mathcal{G}_{\mathbf{\hat{x}}}(u_1,u_2) \left(e^{-\hat{\phi}}g^{(u)}(\hat{m} - u_1 ; 2\hat{q_1}-\hat{q}_0) +  g^{(\alpha)}(u_2 - \hat{p} ; 2\hat{\xi}_1 - \hat{\xi}_0)\right)^n  \right]
\end{align}
where
\begin{equation}
  \begin{split}
    \mathcal{G}_{\hat{\mathbf{x}}}(u_1,u_2) &= \frac{1}{2\pi \sqrt{\hat{q}_0 \hat{\xi}_0 -\hat{z}^2}}\exp\left(\frac{\hat{\xi}_0 u_1^2 + \hat{q}_0u_2^2 - 2\hat{z}u_1 u_2}{2(\hat{q}_0 \hat{\xi}_0 - \hat{z}^2)}\right),\\
    g^{(\alpha)}(x,y)&=
    \sqrt{\frac{\pi}{2y}} e^{\frac{x^2}{2y}} \tonde{2 \delta_{\alpha, t} + \delta_{\alpha, u} \text{Erfc}\left( \frac{x}{\sqrt{2y}} \right)}.
  \end{split}
\end{equation}
This expression is obtained assuming that the matrix in the quadratic form at the exponent of the Gaussian kernel is positive definite, that is, assuming that $\hat{q}_0 \hat{\xi}_0 - \hat{z}^2>0$. When $\hat{q}_0 \hat{\xi}_0 - \hat{z}^2\to 0$, the asymptotics of the volume reduces to:
\begin{align}\label{eq:VolumeDelta0}
  &\lim_{\hat q_0 \hat \xi_0- \hat z^2\to0}  \mathcal{J}_n^{(\alpha)}(\hat{\mathbf{x}}) \notag \\
  &= \log \left[\int\dd{u} e^{-\frac{ u^2}{2}} \left(e^{-\hat{\phi}} g^{(u)}(\hat{m} - u\sqrt{-\hat{q}_0} ; 2\hat{q_1}-\hat{q}_0) +  {g}^{(\alpha)}( u\sqrt{-\hat{\xi}_0} - \hat{p} ; 2\hat{\xi}_1 - \hat{\xi}_0)\right)^n  \right],
\end{align}
as we show in Sec.~\ref{sec:AppVolume}. Combining everything, within the Replica Symmetric assumption we get
\begin{equation}
  \begin{split}
    \left \langle \quadre{\mathcal{N}^{(\alpha)}_{S}(\phi)}^n  \right \rangle &= \int  \dd \mathbf{x} \dd \hat{\mathbf{x}} e^{S \mathcal{A}_n^{(\alpha)}(\mathbf{x},\hat{\mathbf{x}},\phi) + o(S)},\\
    \mathcal{A}_n^{(\alpha)}(\mathbf{x}, \hat{\mathbf{x}}, \phi) = p_n(\mathbf{x}) + n\, \mathpzc{d}(\phi) + &n \left( \hat{q}_1 q_1 + \hat{\xi}_1 \xi_1 + \hat{m} m + \hat{p} p + \hat{\phi} \phi \right)\\
    &+ \frac{n(n-1)}{2} \left( \hat{q}_0 q_0 + \hat{\xi}_0 \xi_0 + 2 \hat{z} z \right) + \mathcal{J}_n^{(\alpha)}(\hat{\mathbf{x}}).
  \end{split}
\end{equation}

\subsection{The replica trick and the saddle point}
The complexities $\Sigma^{(\alpha)}_{\sigma,\gamma} ,\Sigma^{(\alpha, A)}_{\sigma,\gamma}$ are obtained via a saddle point argument. We distinguish between the quenched and annealed calculations.\\

\parhead{Quenched Case.} For the quenched case, we need to analytically continue $\mathcal{A}_n^{(\alpha)}$ to take the limit $n\to 0$.  Expanding to linear order in $n$  we find $\mathcal{A}_n^{(\alpha)}= n \bar{\mathcal{A}}^{(\alpha)}(\mathbf{x}, \hat{\mathbf{x}}, \phi) + O(n^2)$ with
\begin{equation}  \label{eq:barA}
  \begin{aligned}
    \bar{\mathcal{A}}^{(\alpha)}(\mathbf{x}, \hat{\mathbf{x}}, \phi) &= \bar{p}(\mathbf{x}) + \mathpzc{d}(\phi) + \hat{q}_1 q_1 + \hat{\xi}_1 \xi_1 + \hat{m} m + \hat{p} p + \hat{\phi} \phi\\
    &\quad- \frac{1}{2} \left( \hat{q}_0 q_0 + \hat{\xi}_0 \xi_0 \right) - \hat{z} z + \bar{\mathcal{J}}^{(\alpha)}(\hat{\mathbf{x}}),
  \end{aligned}
\end{equation}
where
\begin{equation}
  \begin{aligned}
    \bar{p}(\mathbf{x}) &= \frac{(\kappa - \mu m)\, m (q_1 - q_0 + z \gamma)}{\sigma^2 (1 + \gamma)(q_1 - q_0)^2}
    + \frac{(\kappa - \mu m)\, p}{\sigma^2 (q_1 - q_0)}
    - \frac{\gamma}{2\sigma^2 (1 + \gamma)} \frac{z^2 (q_1 + q_0)}{(q_1 - q_0)^3}  \\
    &\quad - \frac{\xi_1}{2\sigma^2 (q_1 - q_0)}
    - \frac{q_0 (\xi_0 - \xi_1)}{2\sigma^2 (q_1 - q_0)^2}
    - \frac{1}{2\sigma^2 (1 + \gamma)} \left[ 1 + \frac{2 q_0 z}{(q_1 - q_0)^2} \right]  \\
    &\quad - \frac{1}{2\sigma^2} \frac{(\kappa - \mu m)^2}{q_1 - q_0}
    - \frac{\log\left[2\pi\sigma^2(q_1 - q_0)\right]}{2}
    - \frac{q_0}{2[q_1 - q_0]},
  \end{aligned}
\end{equation}

and
\begin{equation}\label{eq:InQ}
  \begin{aligned}  \bar{\mathcal{J}}^{(\alpha)}(\hat{\mathbf{x}}) = \int \dd{u_1} \dd{u_2} \mathcal{G}_{\mathbf{\hat{x}}}(u_1,u_2) \log \Bigg[e^{-\hat{\phi}}g(\hat{m} - u_1 ; 2\hat{q_1}-\hat{q}_0) +  g^{(\alpha)}(u_2 - \hat{p} ; 2\hat{\xi}_1 - \hat{\xi}_0) \Bigg].
  \end{aligned}
\end{equation}
The saddle point equations for the quenched case read
\begin{equation}
  \frac{\partial \bar{\mathcal{A}}^{(\alpha)}(\mathbf{x},\hat{\mathbf{x}},\phi)}{\partial \mathbf{x}}\Bigg|_{\mathbf{x}^*,\hat{\mathbf{x}}^*} = 0 , \quad \quad \frac{\partial \bar{\mathcal{A}}^{(\alpha)}(\mathbf{x},\hat{\mathbf{x}},\phi)}{\partial \hat{\mathbf{x}}}\Bigg|_{\mathbf{x}^*,\hat{\mathbf{x}}^*} = 0
  \label{eq:quenched_sad}
\end{equation}
and we get directly the quenched complexity $\Sigma_{\sigma,\gamma}^{(\alpha)}(\phi) = \bar{\mathcal{A}}^{(\alpha)}(\mathbf{x}^*,\hat{\mathbf{x}}^*,\phi)$. For ease of notation, we are not indicating explicitly the dependence of the saddle point solutions ${\bf x}, \hat{\bf x}$ on the choice of $\alpha$.

\parhead{Annealed Case.} By choosing $n = 1$, we obtain instead:
\begin{equation}\label{eq:Aann}
  \mathcal{A}_1^{(\alpha)}(\mathbf{x}, \hat{\mathbf{x}}, \phi) = p_1(\mathbf{x}) + \mathpzc{d}(\phi) +  \hat{q}_1 q_1 + \hat{\xi}_1 \xi_1 + \hat{m} m + \hat{p} p + \hat{\phi} \phi + \mathcal{J}^{(\alpha)}_1(\hat{\mathbf{x}}),
\end{equation}
with
\begin{equation}
  \begin{aligned}
    p_1(\mathbf{x}) &= -\frac{1}{2 \sigma^2 q_1^2} \left[ (\kappa - \mu m)^2 \left( q_1 - \frac{\gamma m^2}{1 + \gamma} \right) - 2(\kappa - \mu m) q_1 \left( p + \frac{m}{1 + \gamma} \right) + \xi_1 q_1 \right] \\
    &\quad - \frac{1}{2} \log(2 \pi \sigma^2 q_1) - \frac{1}{2 \sigma^2 (1 + \gamma)},
  \end{aligned}
\end{equation}
and (see Sec.~\ref{sec:AppVolume} of this appendix for further details):
\begin{equation}\label{eq:ValA}
  \mathcal{J}^{(\alpha)}_1(\hat{\mathbf{x}}) = \log \left[ \frac{e^{-\hat{\phi}}}{2} \sqrt{\frac{\pi}{\hat{q}_1}} e^{\frac{\hat{m}^2}{4 \hat{q}_1}} \operatorname{Erfc} \left( \frac{\hat{m}}{2 \sqrt{\hat{q}_1}}\right)+\frac{1}{2} \sqrt{\frac{\pi}{\hat{\xi}_1}} e^{\frac{\hat{p}^2}{4 \hat{\xi}_1}}\tonde{2\delta_{\alpha, t} + \delta_{\alpha, u} \operatorname{Erfc} \left(-\frac{\hat{p}}{2 \sqrt{\hat{\xi}_1}}\right)}
  \right].
\end{equation}
As expected we obtain a functional that does not depend on $q_0,\hat{q}_0,\xi_0,\hat{\xi}_0,z,\hat{z}$ which exist only when more than one replica are present. The saddle point equations for the annealed complexity read
\begin{equation}
  \frac{\partial \mathcal{A}^{(\alpha)}_1(\mathbf{x},\hat{\mathbf{x}},\phi)}{\partial \mathbf{x}}\Bigg|_{\mathbf{x}^*,\hat{\mathbf{x}}^*} = 0 , \quad \frac{\partial \mathcal{A}^{(\alpha)}_1(\mathbf{x},\hat{\mathbf{x}},\phi)}{\partial \hat{\mathbf{x}}}\Bigg|_{\mathbf{x}^*,\hat{\mathbf{x}}^*} = 0,
  \label{eq:annealed_sad}
\end{equation}
and the annealed complexity is obtained as $\Sigma_{\sigma,\gamma}^{(\alpha, A)}(\phi) = \mathcal{A}^{(\alpha)}_1(\mathbf{x}^*,\hat{\mathbf{x}}^*,\phi)$. Again, for ease of notation we are not indicating explicitly the dependence of the saddle point solutions ${\bf x}, \hat{\bf x}$ on the choice of $\alpha$.

\section{Self-consistent equations and their numerical solution}\label{sec:self-consistent}
The saddle point equations obtained from  \eqref{eq:quenched_sad} and \eqref{eq:annealed_sad} correspond to a set of coupled equations for the conjugate parameters $\hat{\bf x}$ and the order parameters ${\bf x}$. In particular, by differentiating \eqref{eq:barA} or \eqref{eq:Aann} with respect to the order parameters $\mathbf{x}$ we get a representation of the conjugate parameters $\hat{\mathbf{x}}$ as a function of ${\bf x}$, and conversely we get a representation of $\mathbf{x}$ by differentiating with respect to $\hat{\mathbf{x}}$. Therefore, plugging the second set of equations back into the first set of equations, one gets a set of \emph{self-consistent} relations for the parameters $\hat{\bf x}$. Once the solutions to these self-consistent equations is found, the value of the order parameters ${\bf x}$ can be obtained straightforwardly using the parametrization given by the second set of equations.\\
We report the self-consistent equations written in the simplest form, that is also convenient for their numerical solution, in Sec.~\ref{sec:SCE_q} for the quenched calculation and in Sec.~\ref{supp:AnnShort} for the annealed one. In Sec.~\ref{sec:delta_zero} we discuss the mapping of the quenched set of equations in to the annealed one, achieved at a critical value of parameters. The equations presented here follow from several manipulations of the original equations that are obtained from the variational problems \eqref{eq:quenched_sad} and \eqref{eq:annealed_sad}. For the sake of completeness, we provide the original equations and discuss their manipulations in Sec.~\ref{supp:OriginalSPE}.

\subsection{Quenched self-consistent equations and the procedure to solve them}\label{sec:SCE_q}
The self-consistent equations for the conjugate parameters are more conveniently expressed in terms of the rescaled variables:
\begin{equation}\label{eq:ROP}
  \begin{aligned}
    x_1 &= \frac{\hat{m}}{\sqrt{2\hat{q}_1 - \hat{q}_0}}, \quad
    x_2 = \frac{\hat{p}}{\sqrt{2\hat{\xi}_1 - \hat{\xi}_0}}, \quad
    y = \sqrt{2\hat{\xi}_1 - \hat{\xi}_0}, \quad
    r = \sqrt{\frac{2\hat{q}_1 - \hat{q}_0}{2\hat{\xi}_1 - \hat{\xi}_0}}, \\
    \beta_1 &= \frac{\hat{q}_0}{y^2}, \quad
    \beta_2 = \frac{\hat{\xi}_0}{y^2}, \quad
    \beta_3 = \frac{\hat{z}}{y^2}.
  \end{aligned}
\end{equation}
We define the function
\begin{equation}\label{eq:DefK}
  K(x) = e^{x^2/2}\text{Erfc}\left(\frac{x}{\sqrt{2}}\right),
\end{equation}
and the Gaussian kernel:
\begin{equation}
  \mathpzc{g}_{\mathbf{\hat{x}}}(u_1,u_2) = y r\,\mathcal{G}_{\mathbf{\hat{x}}}(r u_1,u_2) = \frac{r}{2\pi\sqrt{\beta_1 \beta_2 - \beta_3^2}}  \exp\left(-\frac{-u_1^2 \beta_2r^2 + 2 \beta_3  r u_1 u_2 - u_2^2 \beta_1}{2 (\beta_1 \beta_2 - \beta_3^2)}\right).
\end{equation}
From the second set of equations obtained from \eqref{eq:quenched_sad}, one sees that the order parameters, properly rescaled by factors of $y$, admit the following integral representations in terms of the quantities \eqref{eq:ROP} and of $\hat \phi$:
\begin{align}
  my &= \int \dd{u_1} \dd{u_2} \, \mathpzc{g}_{\mathbf{\hat{x}}}(u_1,u_2)\frac{1}{r} \frac{ \sqrt{\frac{2}{\pi}}  - (x_1 - u_1) K(x_1- u_1)}{K(x_1- u_1) + e^{\hat\phi}r [\delta_{\alpha, t} 2 e^{\frac{(u_2-x_2)^2}{2}}+ \delta_{\alpha, u}K(u_2-x_2)]} \label{eq:mInt}\\
  py &= \int \dd{u_1} \dd{u_2} \, \mathpzc{g}_{\mathbf{\hat{x}}}(u_1,u_2) \, re^{\hat\phi} \frac{-\delta_{\alpha, u}\sqrt{\frac{2}{\pi}} - (x_2 - u_2) [\delta_{\alpha, t} 2 e^{\frac{(u_2-x_2)^2}{2}}+ \delta_{\alpha, u}K(u_2-x_2)]}{K(x_1- u_1) + e^{\hat\phi}r [\delta_{\alpha, t} 2 e^{\frac{(u_2-x_2)^2}{2}}+ \delta_{\alpha, u}K(u_2-x_2)]} \\
  q_1 y^2 &= \int \dd{u_1} \dd{u_2} \, \mathpzc{g}_{\mathbf{\hat{x}}}(u_1,u_2) \frac{1}{r^2} \frac{-\sqrt{\frac{2}{\pi}} (x_1 - u_1) + \left(1 + (x_1 - u_1)^2\right) K(x_1- u_1) }{K(x_1- u_1) + e^{\hat\phi}r [\delta_{\alpha, t} 2 e^{\frac{(u_2-x_2)^2}{2}}+ \delta_{\alpha, u}K(u_2-x_2)]} \\
  \xi_1 y^2 &= \int \dd{u_1} \dd{u_2} \, \mathpzc{g}_{\mathbf{\hat{x}}}(u_1,u_2)r e^{\hat\phi} \notag\\
  &\quad \times \frac{\delta_{\alpha, u}\sqrt{\frac{2}{\pi}} (x_2 - u_2) + (1 + (x_2 - u_2)^2) [\delta_{\alpha, t} 2 e^{\frac{(u_2-x_2)^2}{2}}+ \delta_{\alpha, u}K(u_2-x_2)]}{K(x_1- u_1) + e^{\hat\phi}r [\delta_{\alpha, t} 2 e^{\frac{(u_2-x_2)^2}{2}}+ \delta_{\alpha, u}K(u_2-x_2)]} \\
  q_0 y^2 &= \int \dd{u_1} \dd{u_2} \, \mathpzc{g}_{\mathbf{\hat{x}}}(u_1,u_2) \frac{1}{r^2} \left[ \frac{\sqrt{\frac{2}{\pi}} - (x_1 - u_1) K(x_1- u_1) }{K(x_1- u_1) + e^{\hat\phi}r [\delta_{\alpha, t} 2 e^{\frac{(u_2-x_2)^2}{2}}+ \delta_{\alpha, u}K(u_2-x_2)]} \right]^2\\
  \xi_0 y^2 &= \int \dd{u_1} \dd{u_2} \, \mathpzc{g}_{\mathbf{\hat{x}}}(u_1,u_2) r^2e^{2\hat\phi} \left[ \frac{-\delta_{\alpha, u}\sqrt{\frac{2}{\pi}} - (x_2 - u_2) [\delta_{\alpha, t} 2 e^{\frac{(u_2-x_2)^2}{2}}+ \delta_{\alpha, u}K(u_2-x_2)] }{K(x_1- u_1) + e^{\hat\phi}r [\delta_{\alpha, t} 2 e^{\frac{(u_2-x_2)^2}{2}}+ \delta_{\alpha, u}K(u_2-x_2)]}\right]^2 \\
  z y^2 &= \int \dd{u_1} \dd{u_2} \, \mathpzc{g}_{\mathbf{\hat{x}}}(u_1,u_2) e^{\hat\phi}\frac{\sqrt{\frac{2}{\pi}} - (x_1 - u_1) K(x_1- u_1) }{K(x_1- u_1) + e^{\hat\phi}r [\delta_{\alpha, t} 2 e^{\frac{(u_2-x_2)^2}{2}}+ \delta_{\alpha, u}K(u_2-x_2)]} \notag\\
  &\hspace{1in} \times \frac{-\delta_{\alpha, u}\sqrt{\frac{2}{\pi}} - (x_2 - u_2) [\delta_{\alpha, t} 2 e^{\frac{(u_2-x_2)^2}{2}}+ \delta_{\alpha, u}K(u_2-x_2)] }{K(x_1- u_1) + e^{\hat\phi}r [\delta_{\alpha, t} 2 e^{\frac{(u_2-x_2)^2}{2}}+ \delta_{\alpha, u}K(u_2-x_2)]}\label{eq:zInt} \\
  \phi &= \int \dd{u_1} \dd{u_2} \, \mathpzc{g}_{\mathbf{\hat{x}}}(u_1,u_2) \frac{K(x_1- u_1)}{K(x_1- u_1) + e^{\hat\phi}r [\delta_{\alpha, t} 2 e^{\frac{(u_2-x_2)^2}{2}}+ \delta_{\alpha, u}K(u_2-x_2)]}\label{eq:PhiInt},
\end{align}
where we remark that the integrands depend on the choice of $\alpha$, i.e., on whether we are considering the total or uninvadable complexity. These integral representations express how the order parameters ${\bf x}$ can be obtained in terms of the conjugate parameters $\hat{\bf x}$; we see that the right-hand sides depend on all the rescaled conjugate parameters except $y$.

Manipulating the first set of equations obtained from \eqref{eq:quenched_sad}, one finds the following six self-consistent relations coupling the parameters $x_1, x_2, \beta_1, \beta_2, \beta_3$ and $r$:
\begin{align}
  &\sigma^2  [ q_0 y^2]=-\beta_2 \label{eq:VRtop} \\
  &\sigma^2  [ q_1 y^2] =1-\beta_2\label{eq:map1}\\
  &\sigma^2\, [py]= \frac{1+\gamma}{\gamma}\frac{r^2 x_1^2 (1-2\beta_2)}{x_2^3}
  + \frac{r x_1(\beta_2 -1 + \beta_3 + \beta_3\gamma)}{x_2^2\gamma}
  - \sigma^2 x_2,\\
  &\sigma^2\, [my]=\frac{1+\gamma}{\gamma}\frac{\beta_2}{x_2}\!\left( \frac{2 r x_1}{x_2} - \frac{1}{1+\gamma} \right)
  - \frac{1+\gamma}{\gamma}\frac{r x_1}{x_2^2}
  + \frac{1+\gamma}{\gamma}\frac{1}{x_2}\tonde{\frac{1}{1+\gamma}
  - \beta_3}\label{eq:map2}\\
  &\sigma^2\,[\xi_0 y^2]=-\frac{1+\gamma}{\gamma}\frac{2 r^2 x_1^2 (1+\beta_2)}{x_2^2}
  + \frac{2 r x_1 (1+\beta_2+\beta_3+\beta_3\gamma)}{x_2 \gamma} \\
  &\quad -  \beta_1+ 2 r^2 \beta_2
  - \frac{2\beta_3}{\gamma}
  + x_2^2 \sigma^2
  - \beta_2 \sigma^2,\\
  &\sigma^2\,[\xi_1 y^2]={\beta_2(2 r^2 - \sigma^2)
    + \sigma^2 - r^2- \beta_1
  + \sigma^2 x_2^2}\\
  &\quad - \frac{1+\gamma}{\gamma}\tonde{\frac{2 r^2 x_1^2 \beta_2}{x_2^2}
  -\frac{2 r x_1 \beta_3 }{x_2}}
  + \frac{2 r x_1 \beta_2}{x_2 \gamma}
  - \frac{2\beta_3}{\gamma}\label{eq:VRbottom},
\end{align}
where the notation $[\cdot]$ indicates that the quantity in brackets is expressed as the corresponding integral representation~\eqref{eq:mInt}-\eqref{eq:zInt}.
These equations are here written in  a form that assumes $\gamma \neq 0$ (for the case $\gamma=0$, see \cite{rosQuenchedComplexityEquilibria2023}) and $y>0$. Their derivation is explained in detailed in Sec.~\ref{supp:OriginalSPE}. These six self-consistent equations do not depend on the parameter $y$, and can be solved for each value of the parameter $\hat \phi$. Once these equations are solved, the parameter $y$ can be obtained using the identity:
\begin{align}
  \kappa y =-x_2+ \mu [m y].
\end{align}
Therefore, the only conjugate parameter that depends explicitly on the constants $\kappa$ and $\mu$ is $y$. Once the conjugate parameters are determined (for fixed $\hat \phi$), the order parameters can be obtained using Eqs.~\eqref{eq:mInt}-\eqref{eq:zInt}. Finally, one can use Eq.~\eqref{eq:PhiInt} to determine the value of $\phi$ corresponding to the considered $\hat{\phi}$: following this procedure, one finally obtains the typical values of the order parameters as a function of $\phi$. The quenched complexity admits the following representation in terms of the conjugate parameters, as we also derive in Sec.~\ref{supp:COmpXhat}:
\begin{equation}\label{eq:QuenchedSigma}
  \begin{aligned}
    \Sigma^{(\alpha)}_{\sigma,\gamma}(\phi) &= \frac{1}{2}\left(1 - r^2 [q_0y^2] - [\xi_1y^2] - \frac{1}{\sigma^2(\gamma + 1)}   \right)\\
    &\quad -\sigma^2 \frac{\gamma}{2(\gamma + 1)}[zy^2]^2 + \mathpzc{d}(\phi)  + \hat{\phi}(\phi - 1) - \log (2 r)\\
    &+\int \dd{u_1} \dd{u_2}\mathcal{G}_{\mathbf{\hat{x}}}(u_1,u_2) \log\left(K(x_1 - u_1) + e^{\hat\phi}r  [\delta_{\alpha, t} 2 e^{\frac{(u_2-x_2)^2}{2}}+ \delta_{\alpha, u}K(u_2-x_2)]\right).
  \end{aligned}
\end{equation}

\subsection{Annealed self-consistent equations and the procedure to solve them}\label{supp:AnnShort}
In the annealed calculation, the self-consistent equations for the conjugate parameters can be also expressed in terms of some rescaled variables, in this case given by:
\begin{equation}\label{eq:NewParametersA}
  \begin{split}
    \mathtt{x}_1=\frac{\hat m}{\sqrt{2 \hat q_1}}, \quad
    \mathtt{x}_2=\frac{\hat p}{\sqrt{2 \hat \xi_1}}, \quad
    \mathtt{y}={\sqrt{2 \hat \xi_1}}, \quad
    \mathtt{r} =\sqrt{\frac{\hat q_1}{ \hat \xi_1}}.
  \end{split}
\end{equation}
Notice that these variables do not coincide with the quenched ones, and we indicate this exploiting a different font.
Similarly to the quenched case, the second set of equations obtained from \eqref{eq:annealed_sad} give a parametrization of the order parameters in terms of the rescaled conjugate ones, that in the annealed case reads:
\begin{align}
  m\mathtt{y} &= \frac{1}{\mathtt{r}} \frac{\sqrt{\frac{2}{\pi}} -\mathtt{x}_1 K(\mathtt{x}_1)}{K(\mathtt{x}_1)+ e^{\hat{\phi}}\mathtt{r} [\delta_{\alpha, t} 2e^{\frac{\mathtt{x}_2^2}{2}} + \delta_{\alpha, u} K(-\mathtt{x}_2)] } \label{eq:my_ann}\\
  p\mathtt{y} &= -\mathtt{r} e^{\hat{\phi}} \frac{\delta_{\alpha, u} \sqrt{\frac{2}{\pi}}+\mathtt{x}_2 [\delta_{\alpha, t} 2e^{\frac{\mathtt{x}_2^2}{2}} + \delta_{\alpha, u} K(-\mathtt{x}_2)]  }{K(\mathtt{x}_1)+ e^{\hat{\phi}}\mathtt{r} [\delta_{\alpha, t} 2e^{\frac{\mathtt{x}_2^2}{2}} + \delta_{\alpha, u} K(-\mathtt{x}_2)] }\\
  q_1\mathtt{y}^2 &= \frac{1}{\mathtt{r}^2} \frac{-\sqrt{\frac{2}{\pi}}\mathtt{x}_1 + (1 +\mathtt{x}_1^2) K(\mathtt{x}_1)}{K(\mathtt{x}_1)+ e^{\hat{\phi}}\mathtt{r} [\delta_{\alpha, t} 2e^{\frac{\mathtt{x}_2^2}{2}} + \delta_{\alpha, u} K(-\mathtt{x}_2)] } \\
  \xi_1 \mathtt{y}^2 &= e^{\hat{\phi}}\mathtt{r} \frac{\delta_{\alpha, u}\mathtt{x}_2 \sqrt{\frac{2}{\pi}} + (1 +\mathtt{x}_2^2) [\delta_{\alpha, t} 2e^{\frac{\mathtt{x}_2^2}{2}} + \delta_{\alpha, u} K(-\mathtt{x}_2)]}{K(\mathtt{x}_1)+ e^{\hat{\phi}}\mathtt{r} [\delta_{\alpha, t} 2e^{\frac{\mathtt{x}_2^2}{2}} + \delta_{\alpha, u} K(-\mathtt{x}_2)]} \label{eq:xi1y2_ann}\\
  \phi &=  \frac{K(\mathtt{x}_1)}{K(\mathtt{x}_1)+ e^{\hat{\phi}}\mathtt{r} [\delta_{\alpha, t} 2e^{\frac{\mathtt{x}_2^2}{2}} + \delta_{\alpha, u} K(-\mathtt{x}_2)]}\label{eq:phi_ann2}.
\end{align}
The right-hand sides again depend on all the rescaled conjugate parameters except $\mathtt{y}$. Manipulating the remaining saddle point equations, for $\gamma \neq 0$ one finds the following three self-consistent equations coupling $\mathtt{x}_1,\mathtt{x}_2$ and $\mathtt{r}$:
\begin{align}
  &\sigma^2 [q_1\mathtt{y}^2]=1\label{eq:q1_ann} \\
  &\sigma^2 [m\mathtt{y}^2]=\frac{1}{\gamma\mathtt{x}_2}-\frac{1+\gamma}{\gamma} \frac{\mathtt{r}\mathtt{x}_1}{\mathtt{x}_2^2}\label{eq:m_ann}\\
  &\sigma^2 [\xi_1\mathtt{y}^2]=\sigma^2-\mathtt{r}^2+ \sigma^2\mathtt{x}_2^2.\label{eq:tau2_ann}
\end{align}
where the notation $[\cdot]$ now indicates that the quantity in brackets is expressed as the corresponding representation \eqref{eq:my_ann}-\eqref{eq:xi1y2_ann}. One can check that the additional relation $\sigma^2 [p\mathtt{y}]=-\frac{\mathtt{x}_1\mathtt{r}}{\gamma \mathtt{x}_2^2}-\sigma^2\mathtt{x}_2+\frac{1+\gamma}{\gamma}\frac{\mathtt{x_1^2}\mathtt{r}^2}{ \mathtt{x}_2^3}$ holds true. As for the quenched case, the equation in this form are derived under the assumption that $\gamma \neq 0$ (for $\gamma=0$ see \cite{rosQuenchedComplexityEquilibria2023}) $\mathtt{y}>0$. They can be solved for fixed value of $\hat{\phi}$, and the remaining parameter $\mathtt{y}$ is obtained from the resulting solutions through the identity
\begin{equation}
  \mathtt{x}_2 = -\kappa\mathtt{y} + \mu [m\mathtt{y}].
\end{equation}
Exactly as in the quenched calculation, $\mathtt{y}$ is the only conjugate parameter that depends on $\kappa, \mu$.  Once the conjugate parameters are solved for at fixed $\hat \phi$, the order parameters are retrieved from equations \eqref{eq:my_ann}-\eqref{eq:xi1y2_ann}, and the corresponding value of diversity $\phi$ from \eqref{eq:phi_ann2}. Finally, the annealed complexity can be expressed as a function of the rescaled conjugate parameters as:
\begin{equation}\label{eq:AnnealedSigma}
  \begin{aligned}
    \Sigma^{(\alpha, A)}_{\sigma,\gamma} (\phi) =&\frac{1}{2}\left(1 - [\xi_1\mathtt{y}^2] - \frac{ \gamma}{\gamma +1}\sigma^2[m\mathtt{y}^2]\mathtt{x}_2^2 - \frac{1}{ \sigma^2 (1 + \gamma)}\right)+ \hat{\phi}(\phi - 1) + \mathpzc{d}(\phi)-\log(2\mathtt{r}) \\
    &+ \log\left(K(\mathtt{x}_1)+ e^{\hat{\phi}}\mathtt{r} [\delta_{\alpha, t} 2e^{\frac{\mathtt{x}_2^2}{2}} + \delta_{\alpha, u} K(-\mathtt{x}_2)] \right).
  \end{aligned}
\end{equation}
For an account on the derivation of these formulas, we refer the reader to Sec.~\ref{supp:COmpXhat}.

\begin{figure}
  \centering
  \begin{subfigure}[c]{0.48\textwidth}
    \centering
    \includegraphics[width=\textwidth]{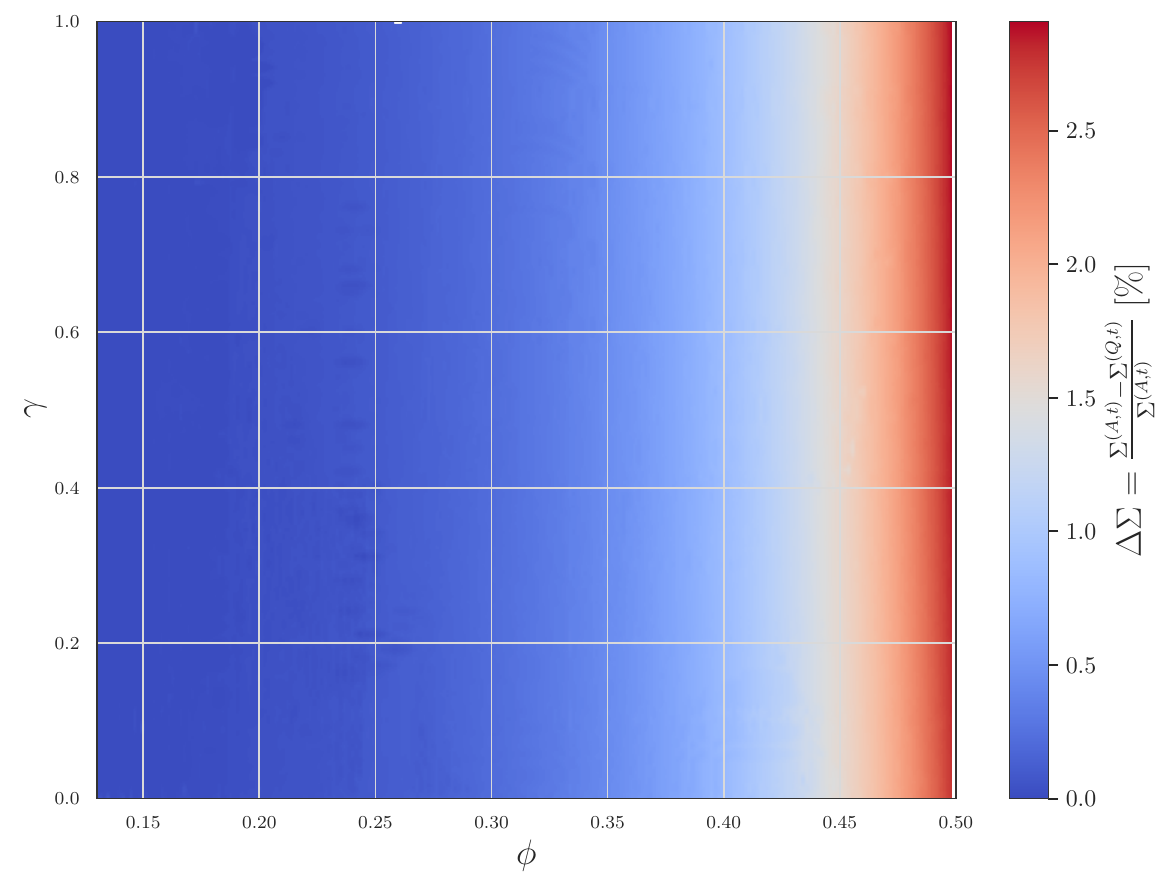}
  \end{subfigure}
  \hfill
  \begin{subfigure}[c]{0.48\textwidth}
    \centering
    \includegraphics[width=\textwidth]{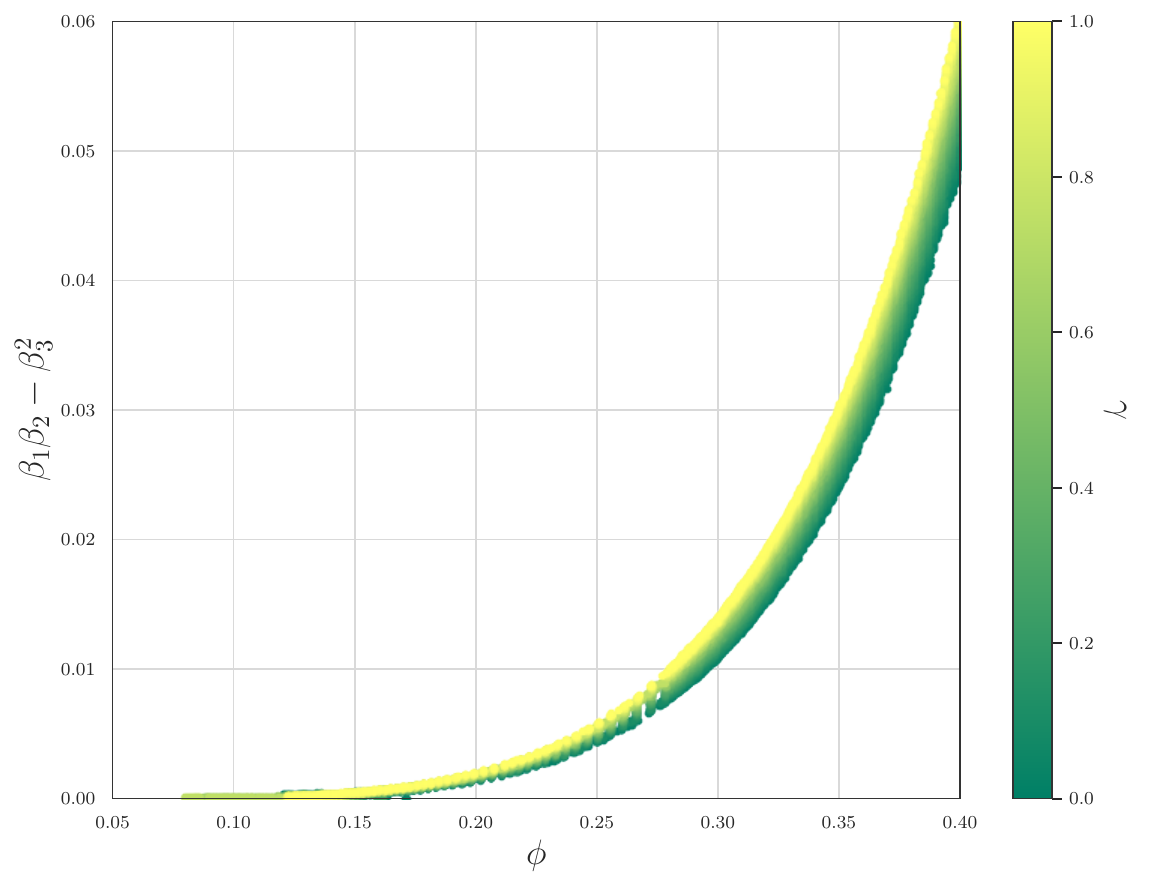}
  \end{subfigure}
  \caption{All quantities in is plot are with respect to total complexity. \emph{Left.} Relative difference between annealed and quenched total complexities for different $\gamma$ and $\phi$ at $\sigma=5$, showing convergence for small $\phi$. \emph{Right.} Convergence of $\beta_1\beta_2 - \beta_3^2$ to zero as $\phi \to \phi^+_{\text{Match}}(\gamma,\sigma)$, indicating divergence of the Gaussian determinant.}\label{fig_combined1}
\end{figure}

\subsection{The quenched-annealed matching point}\label{sec:delta_zero}
The quenched saddle point equations reported in Sec.~\ref{sec:SCE_q} are derived under the assumption that $\beta_1\beta_2 - \beta_3^2>0$.
Solving them numerically, we remark that when $\phi$ approaches (from above) a critical value $\phi_{\text{Match}}(\gamma,\sigma)$, parametrized in terms of the conjugate parameter $ \hat{\phi}_{\text{Match}}(\gamma,\sigma)$, this bounds saturates and the argument of the Gaussian kernel $\mathcal{G}_{\hat{\mathbf{x}}}$ diverges. This happens for both the uninvadable and total complexities. Mathematically, the solutions of the six equations \eqref{eq:VRtop}-\eqref{eq:VRbottom} satisfy:
\begin{equation}\label{eq:Delta0}
  \Delta(\beta_1, \beta_2, \beta_3) := \beta_1\beta_2 - \beta_3^2 \xlongrightarrow[\hat{\phi} \to \hat{\phi}_{\text{Match}}^-]{} 0.
\end{equation}
Fig.~\ref{fig_combined1} (\emph{right}) shows the behavior of $\Delta(\beta_1, \beta_2, \beta_3)$ for $\sigma=5$ and various $\gamma$, for the case $\alpha=u$.
The limit \eqref{eq:Delta0} is attained while $\beta_1, \beta_2$ and $\beta_3$ remain different from zero, and it corresponds to the boundary of stability of the quenched saddle point equations of Sec.~\ref{sec:SCE_q}.
We now discuss how to treat the equations for $\hat{\phi}\geq \hat{\phi}_{\text{Match}}$, corresponding to $\phi \leq {\phi}_{\text{Match}}$.

\subsubsection{The matching point: convergence to the annealed and cavity methods}

When the limit \eqref{eq:Delta0} is attained, the integral representations Eqs.~\eqref{eq:mInt}-\eqref{eq:PhiInt} take a simplified form, that is consistent with the limiting value of the volume term given in \eqref{eq:VolumeDelta0}. In particular, setting
\begin{equation}
  \mathpzc{g}_{\hat{\bf x}}(u):= \frac{re^{-\frac{1}{2}\frac{r^2 u^2}{-\beta_1}}}{\sqrt{2\pi (-\beta_1)}}, \quad \quad \lambda(r, \beta_1, \beta_2):=r\sqrt{\frac{\beta_2}{\beta_1}},
\end{equation}
one has:
\begin{align}
  my &\to \int  \dd{u}\,    \mathpzc{g}_{\hat{\bf x}}(u)\, \frac{1}{r} \frac{ \sqrt{\frac{2}{\pi}}  - (x_1 - u) K(x_1- u)}{K(x_1- u) + e^{\hat\phi}r [\delta_{\alpha, t} 2 e^{\frac{1}{2}\tonde{\lambda u-x_2}^2}+ \delta_{\alpha, u}K\tonde{\lambda u-x_2} ]} \label{eq:mIntD0}\\
  py &\to \int \dd{u}\,    \mathpzc{g}_{\hat{\bf x}}(u)\,  re^{\hat\phi} \frac{-\delta_{\alpha, u}\sqrt{\frac{2}{\pi}} +\tonde{\lambda u-x_2} [\delta_{\alpha, t} 2 e^{\frac{1}{2}\tonde{\lambda u-x_2}^2}+ \delta_{\alpha, u}K\tonde{\lambda u-x_2} ]}{K(x_1- u) + e^{\hat\phi}r [\delta_{\alpha, t} 2 e^{\frac{1}{2}\tonde{\lambda u-x_2}^2}+ \delta_{\alpha, u}K\tonde{\lambda u-x_2} ]} \\
  q_1 y^2 &\to \int \dd{u}\,    \mathpzc{g}_{\hat{\bf x}}(u)\,  \frac{1}{r^2} \frac{-\sqrt{\frac{2}{\pi}} (x_1 - u) + \left(1 + (x_1 - u)^2\right) K(x_1- u) }{K(x_1- u) + e^{\hat\phi}r [\delta_{\alpha, t} 2 e^{\frac{1}{2}\tonde{\lambda u-x_2}^2}+ \delta_{\alpha, u}K\tonde{r\lambda u-x_2} ]} \\
  \xi_1 y^2 &\to \int \dd{u}\,    \mathpzc{g}_{\hat{\bf x}}(u)\,  r e^{\hat\phi} \notag\\
  &\quad \times\frac{-\delta_{\alpha, u}\sqrt{\frac{2}{\pi}} \tonde{\lambda u-x_2} + [1 + \tonde{\lambda u-x_2}^2] [\delta_{\alpha, t} 2 e^{\frac{1}{2}\tonde{\lambda u-x_2}^2}+ \delta_{\alpha, u}K\tonde{\lambda u-x_2} ]}{K(x_1- u) + e^{\hat\phi}r [\delta_{\alpha, t} 2 e^{\frac{1}{2}\tonde{\lambda u-x_2}^2}+ \delta_{\alpha, u}K\tonde{\lambda u-x_2}]} \label{eq:xi1IntD0}\\
  q_0 y^2 &\to \int \dd{u}\,    \mathpzc{g}_{\hat{\bf x}}(u)\,  \frac{1}{r^2} \left[ \frac{\sqrt{\frac{2}{\pi}} - (x_1 - u) K(x_1- u) }{K(x_1- u) + e^{\hat\phi}r [\delta_{\alpha, t} 2 e^{\frac{1}{2}\tonde{\lambda u-x_2}^2}+ \delta_{\alpha, u}K\tonde{\lambda u-x_2}]} \right]^2\\
  \xi_0 y^2 &\to \int \dd{u}\,    \mathpzc{g}_{\hat{\bf x}}(u)\,  r^2e^{2\hat\phi} \left[ \frac{-\delta_{\alpha, u}\sqrt{\frac{2}{\pi}} + (\lambda u -x_2) [\delta_{\alpha, t} 2 e^{\frac{1}{2}\tonde{\lambda u-x_2}^2}+ \delta_{\alpha, u}K\tonde{\lambda u-x_2}] }{K(x_1- u) + e^{\hat\phi}r [\delta_{\alpha, t} 2 e^{\frac{1}{2}\tonde{\lambda u-x_2}^2}+ \delta_{\alpha, u}K\tonde{\lambda u-x_2}]}\right]^2 \\
  z y^2 &\to \int \dd{u}\,    \mathpzc{g}_{\hat{\bf x}}(u)\,  r e^{\hat\phi}\frac{\sqrt{\frac{2}{\pi}} - (x_1 - u) K(x_1- u) }{K(x_1- u) + e^{\hat\phi}r [\delta_{\alpha, t} 2 e^{\frac{1}{2}\tonde{\lambda u-x_2}^2}+ \delta_{\alpha, u}K\tonde{\lambda u-x_2}]}\notag \\
  &\hspace{1in} \times\frac{-\delta_{\alpha, u}\sqrt{\frac{2}{\pi}} + (\lambda u  x_2 )  [\delta_{\alpha, t} 2 e^{\frac{1}{2}\tonde{\lambda u-x_2}^2}+ \delta_{\alpha, u}K\tonde{\lambda u-x_2}]  }{K(x_1- u) + e^{\hat\phi}r [\delta_{\alpha, t} 2 e^{\frac{1}{2}\tonde{\lambda u-x_2}^2}+ \delta_{\alpha, u}K\tonde{\lambda u-x_2}]}\label{eq:zIntD0} \\
  \phi &\to \int \dd{u}\,    \mathpzc{g}_{\hat{\bf x}}(u)\, \frac{K(x_1- u)}{K(x_1- u) + e^{\hat\phi}r [K(x_1- u) + e^{\hat\phi}r [\delta_{\alpha, t} 2 e^{\frac{1}{2}\tonde{\lambda u-x_2}^2}+ \delta_{\alpha, u}K\tonde{\lambda u-x_2}]}\label{eq:PhiIntD0}.
  \end{align}
  We now argue that in this limit, the quenched calculation maps into the annealed one, as shown in
  \cite{rosQuenchedComplexityEquilibria2023} for the special case $\alpha=u$ and $\gamma=0$. Moreover, when $\alpha=u$ this matching point is described by the cavity equations \cite{opper1992phase, galla2018dynamically, buninEcologicalCommunitiesLotkaVolterra2017}.  \\

\begin{figure}
    \begin{subfigure}[t]{0.48\textwidth}
      \includegraphics[width = 0.9\textwidth]    {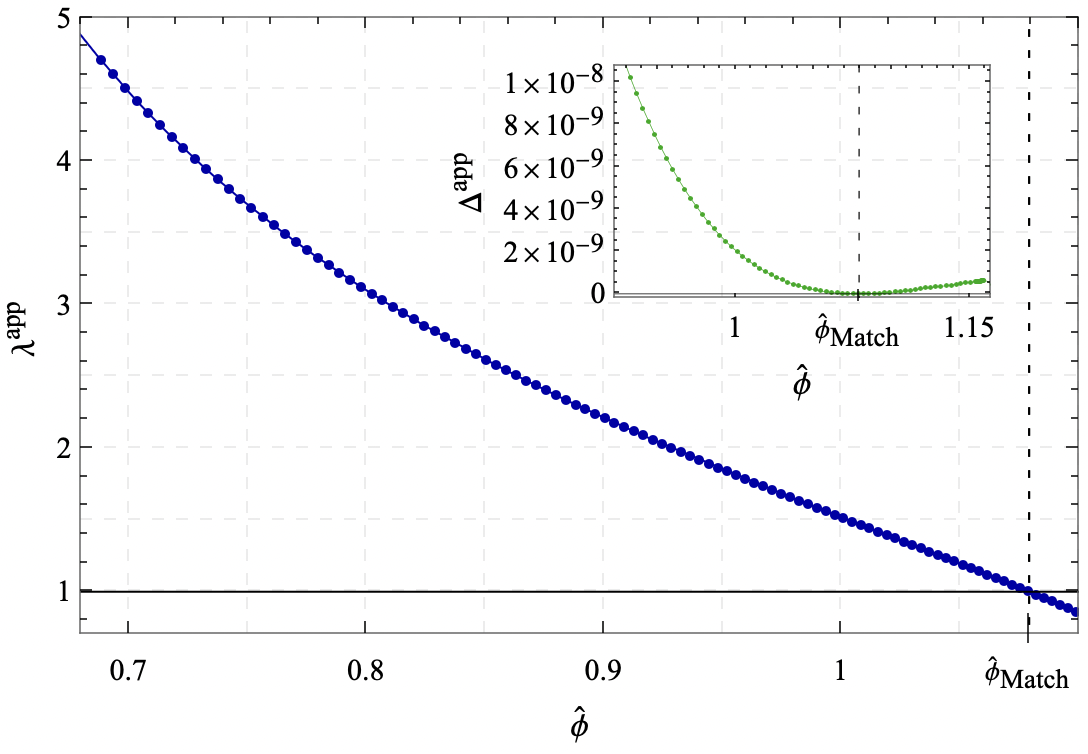}
    \end{subfigure}
    \begin{subfigure}[t]{0.48\textwidth}
      \includegraphics[width = 0.9\textwidth]{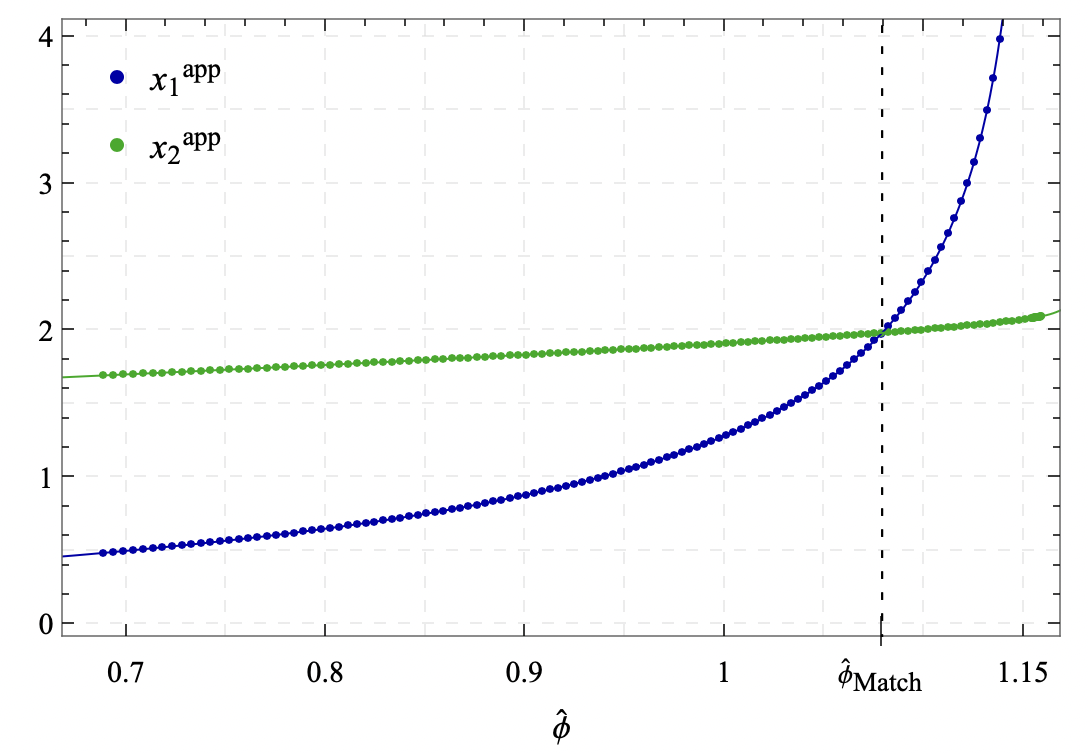}
    \end{subfigure}
    \caption{Behavior of $\lambda^{\text{app}}=r^{\text{app}}\sqrt{{\beta_2^{\text{app}}}/{\beta_1^{\text{app}}}}$ (\emph{left}) and $x_1^{\text{app}}, x_2^{\text{app}}$ (\emph{right}) as a function of $\hat{\phi}$, for $\sigma=4.3, \gamma=1/2$ and $\alpha=u$. Here $x_1^{\text{app}}, x_2^{\text{app}}, r^{\text{app}}, \beta_1^{\text{app}}, \beta_2^{\text{app}} $ solve the approximate equations obtained plugging \eqref{eq:mIntD0}-\eqref{eq:PhiIntD0} into \eqref{eq:VRtop}-\eqref{eq:VRbottom}, and coincide with the solutions of the exact equations only at $\hat{\phi}= \hat{\phi}_{\text{Match}}= 1.079$, where $\Delta^{\text{app}}=\beta_1^{\text{app}} \beta_2^{\text{app}}- (\beta_3^{\text{app}})^2=0$ (\emph{inset}). At this point, $\lambda^{\text{app}}=1$ and $x_1^{\text{app}}= x_2^{\text{app}}$. }
    \label{fig:LimitsU}
  \end{figure}

  \emph{Complexity of uninvadable equilibria ($\alpha=u$).}
  In this case, numerically, one finds that when \eqref{eq:Delta0} is attained, the remaining order parameters satisfy also the following limiting conditions:
  \begin{align}\label{eq:LimitAnnealed}
    \delta(x_1, x_2)&:= x_1-x_2 \xlongrightarrow[\hat{\phi} \to \hat{\phi}_{\text{Match}}]{} 0, \\
    \lambda(r, \beta_1, \beta_2)&=r \sqrt{\frac{\beta_2}{\beta_1}}\xlongrightarrow[\hat{\phi} \to \hat{\phi}_{\text{Match}}]{} 1 \\
    a(r, \hat \phi)&:=r e^{\hat \phi}\xlongrightarrow[\hat{\phi} \to \hat{\phi}_{\text{Match}}]{} 1.
  \end{align}
  When these limiting values are attained, the integral representations \eqref{eq:mIntD0}-\eqref{eq:PhiIntD0} map into the representation \eqref{eq:my_ann}-\eqref{eq:phi_ann2} obtained within the annealed calculation, provided that one identifies:
  \begin{equation}
    \mathtt{y}=\frac{y}{\sqrt{1-\beta_2}}, \quad \quad \mathtt{x}_2=\frac{x_2}{\sqrt{1-\beta_2}}, \quad \quad \mathtt{x}_1=\frac{x_1}{\sqrt{1-\beta_2}}, \quad \quad \mathtt{r}=r
  \end{equation}
  and exploits the fact that $\mathtt{x}_1=\mathtt{x}_2$ and $ \mathtt{r} e^{\hat \phi}=1$. Moreover, the equations \eqref{eq:map1}, \eqref{eq:map2} and \eqref{eq:VRbottom} map into \eqref{eq:q1_ann}-\eqref{eq:tau2_ann}.
  Therefore, similarly to the $\gamma=0$ case \cite{rosQuenchedComplexityEquilibria2023}, also for general $\gamma$ when \eqref{eq:Delta0} is attained, the quenched calculation of the complexity of uninvadable equilibria maps into the annealed one. Notice that solving the quenched self-consistent equations becomes numerically difficult when $\hat{\phi}$ approaches $ \hat{\phi}_{\text{Match}}(\gamma,\sigma)$ and $\Delta$ gets small, due to the divergence of the Gaussian kernel; therefore, to check that \eqref{eq:LimitAnnealed} are satisfied at $\hat{\phi}_{\text{Match}}$, for the values of $\hat{\phi}$ for which $\Delta \leq 10^{-4}$ we solve the approximate set of equations obtained placing the representations \eqref{eq:mIntD0}-\eqref{eq:PhiIntD0} into \eqref{eq:VRtop}-\eqref{eq:VRbottom}. We denote with $x_1^{\rm app}, x_2^{\rm app}, \beta_1^{\rm app},\beta_2^{\rm app},\beta_3^{\rm app}, r^{\rm app}$ the solutions of this approximate set of equations. These solutions are exact, i.e., they coincide with the solutions of \eqref{eq:VRtop}-\eqref{eq:VRbottom} without approximations only for the value of $\hat{\phi}$ for which \eqref{eq:Delta0} is satisfied: this allows to identify $\hat{\phi}_{\text{Match}}$ and to show that \eqref{eq:LimitAnnealed} are indeed satisfied, as Fig.~\ref{fig:LimitsU} illustrates. \\

  We now show that the typical properties  (i.e., the values of $m$ and $q_1$) of the equilibria having diversity $\phi=\phi_{\text{Match}}$ are described by the cavity method \cite{mezard1987sk}. The core idea of the method is to study the effect of adding one extra species to an interacting ecosystem, and to link the properties of the enlarged ($S$+1)-species ecosystem to those of the original
  $S$-species ecosystem. By construction, this method assumes the existence of a single equilibrium, that is internaly stable with respect to perturbations (such as the addition of one species), and determines the properties of this single equilibrium (its diversity $\phi$, mean abundance $m$, and self-overlap $q_1$) by imposing self-consistency. As a consequence, in principle the cavity method allows one to properly describe only the Unique Fixed Point phase, in which one single uninvadable and internaly stable equilibrium is present. However, as we now show, the cavity method yields meaningful information also within the Multiple Equilibria phase: it describes the properties of a particular family of equilibria, those having diversity $\phi_{\text{Match}}$. When applied to the GLV equations, the cavity method gives an emergent effective variable $(\kappa-\mu m)/\sqrt{\sigma^2 q_1}$ \cite{opper1992phase}, which
  we recognize to coincide (up to a sign) with the parameter $\mathtt{x}_2$ in our annealed formalism; manipulating the standard equations obtained within the cavity method, one can derive a self-consistent equation for this parameter, which reads:
  \begin{equation}\label{eq:SelfConsCav}
    \sigma^2 [w_2(-\mathtt{x}_2) + \gamma w_0(-\mathtt{x}_2)]^2=w_2(-\mathtt{x}_2), \quad \quad w_n(x)= \int_{-\infty}^{x} ds  \frac{e^{-\frac{s^2}{2}}}{\sqrt{2 \pi}}\, (x-s)^n.
  \end{equation}
  From the solution $\mathtt{x}_2^{\rm cav}$ to this equation, one can show that the self-similarity $q_1^{\rm cav}$, average abundance $m^{\rm cav}$ and diversity $\phi^{\rm cav}$ are obtained as (see for instance \cite{royNumericalImplementationDynamical2019}):
  \begin{equation}\label{eq:cavParBun}
    \begin{split}
      &\phi^{\rm cav}=w_0(-\mathtt{x}_2^{\rm cav}),\quad \quad
      q_1^{\rm cav}= \frac{\kappa^2 \, w_2(-\mathtt{x}_2^{\rm cav})}{\grafe{ \mu w_1(-\mathtt{x}_2^{\rm cav})- \mathtt{x}_2^{\rm cav} \sigma^2[w_2(-\mathtt{x}_2^{\rm cav}) + \gamma w_0(-\mathtt{x}_2^{\rm cav})]}^2}, \\
      &  m^{\rm cav}= \frac{\kappa +\mathtt{x}_2^{\rm cav} \sigma \sqrt{q_1^{\rm cav}}}{\mu}.
    \end{split}
  \end{equation}
  The considerations above imply that at $\phi=\phi_{\text{Match}}$, the solutions to the annealed saddle point equations satisfy $\mathtt{x}_1=\mathtt{x}_2$ and $\mathtt{r}e^{\hat{\phi}}=1$. Plugging this into \eqref{eq:xi1y2_ann} and using the explicit expressions of the functions $\omega_n(x)$, we see that at this point $\sigma^2 [\xi_1 y^2]=\sigma^2(1+\mathtt{x}_2^2)-\sigma^2 \omega_2(-\mathtt{x}_2)$, that combined with \eqref{eq:tau2_ann} leads to $\mathtt{r}^2= \sigma^2 \omega_2(-\mathtt{x}_2)$. On the other hand, \eqref{eq:m_ann} with $\mathtt{x}_1=\mathtt{x}_2$ and $\mathtt{r}e^{\hat{\phi}}=1$ implies that
  \begin{equation}
    \begin{split}
      \mathtt{r}&=\gamma \quadre{\frac{\sigma^2}{2} e^{-\frac{\mathtt{x}_2^2}{2}}\tonde{\sqrt{\frac{2}{\pi}} \mathtt{x}_2 - \mathtt{x}_2^2 K(\mathtt{x}_2)}+ \mathtt{r}^2 }+ \mathtt{r}^2\\
      &=\gamma \quadre{\frac{\sigma^2}{2} e^{-\frac{\mathtt{x}_2^2}{2}}\tonde{\sqrt{\frac{2}{\pi}} \mathtt{x}_2 - \mathtt{x}_2^2 K(\mathtt{x}_2)}+ \sigma^2\omega_2(-\mathtt{x}_2)  }+ \mathtt{r}^2\\
      &= \frac{\gamma \sigma^2}{2} \text{Erfc}\tonde{-\frac{\mathtt{x}_2}{\sqrt{2}}} + \mathtt{r}^2= \gamma \sigma^2 \omega_0(-\mathtt{x}_2)+ \mathtt{r}^2, \\
      \mathtt{r}^2&= \sigma^2 \omega_2(-\mathtt{x}_2).
    \end{split}
  \end{equation}
  The square of this equation is equivalent to \eqref{eq:SelfConsCav}. The identities of the order parameters follow immediately.\\

\begin{figure}
    \begin{subfigure}[t]{0.49\textwidth}
      \includegraphics[width = 0.89\textwidth]{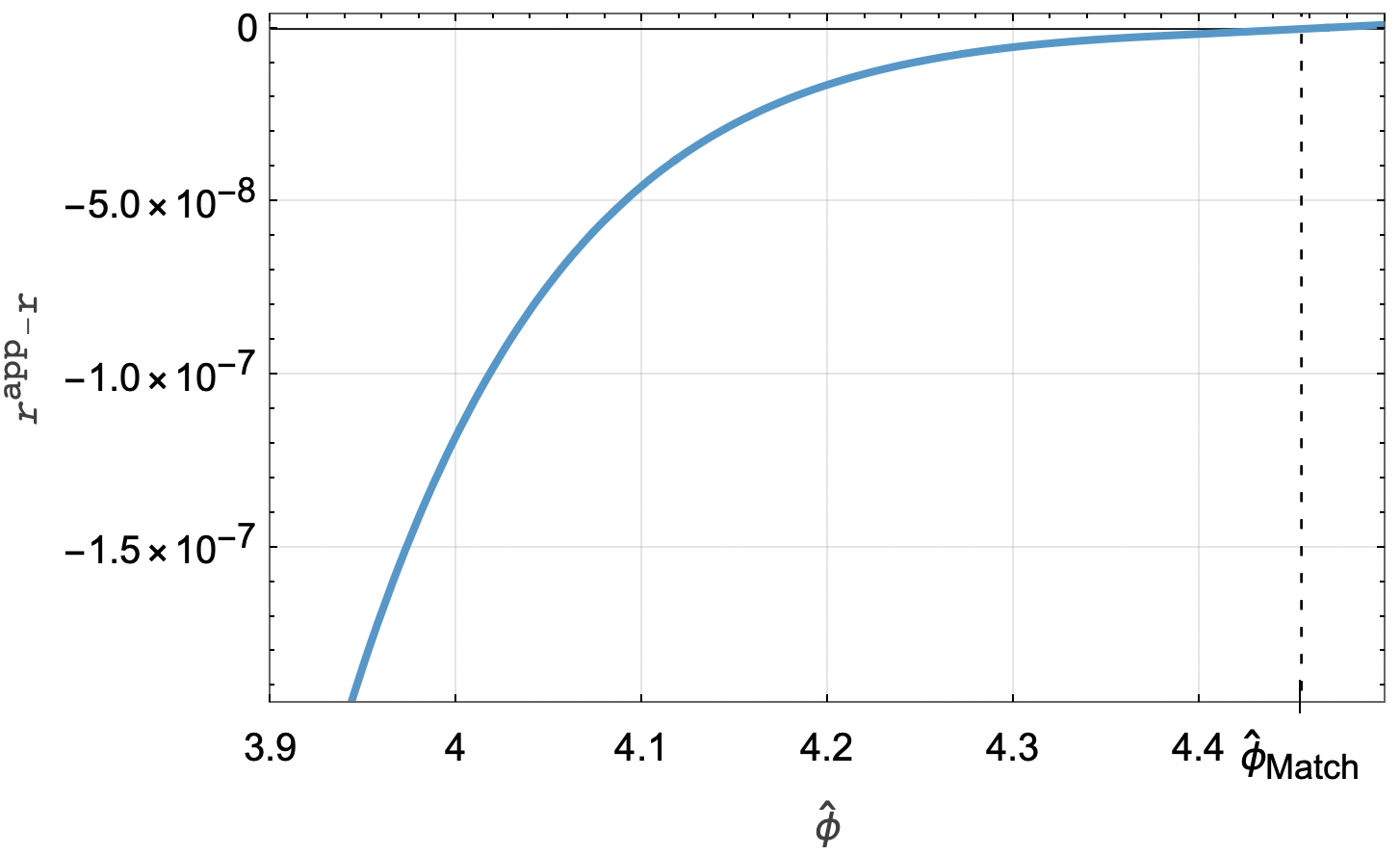}
    \end{subfigure}
    \begin{subfigure}[b]{0.49\textwidth}
      \includegraphics[width = 0.88\textwidth]{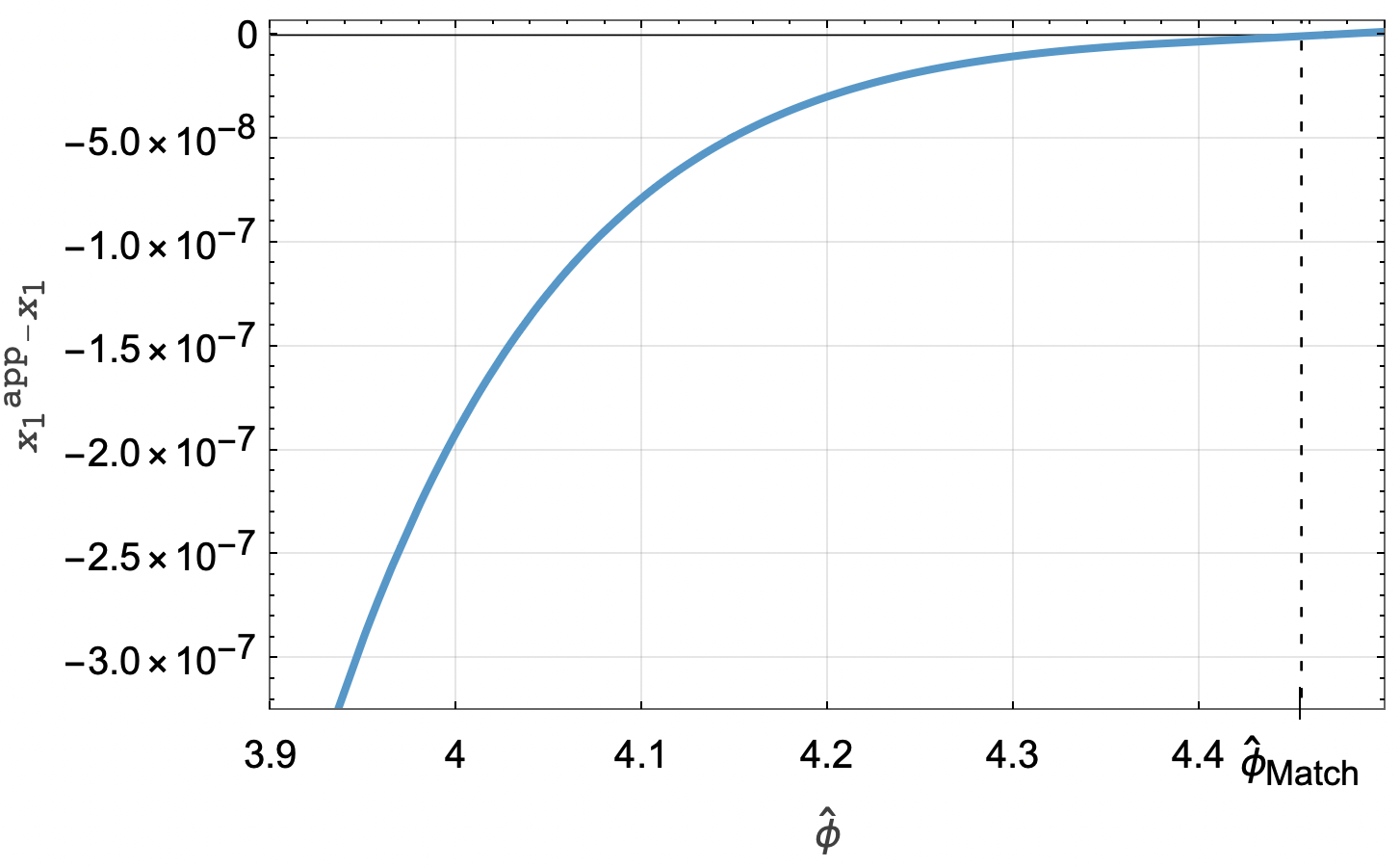}
    \end{subfigure}
    \begin{subfigure}[t]{0.49\textwidth}
      \includegraphics[width = 0.91\textwidth]{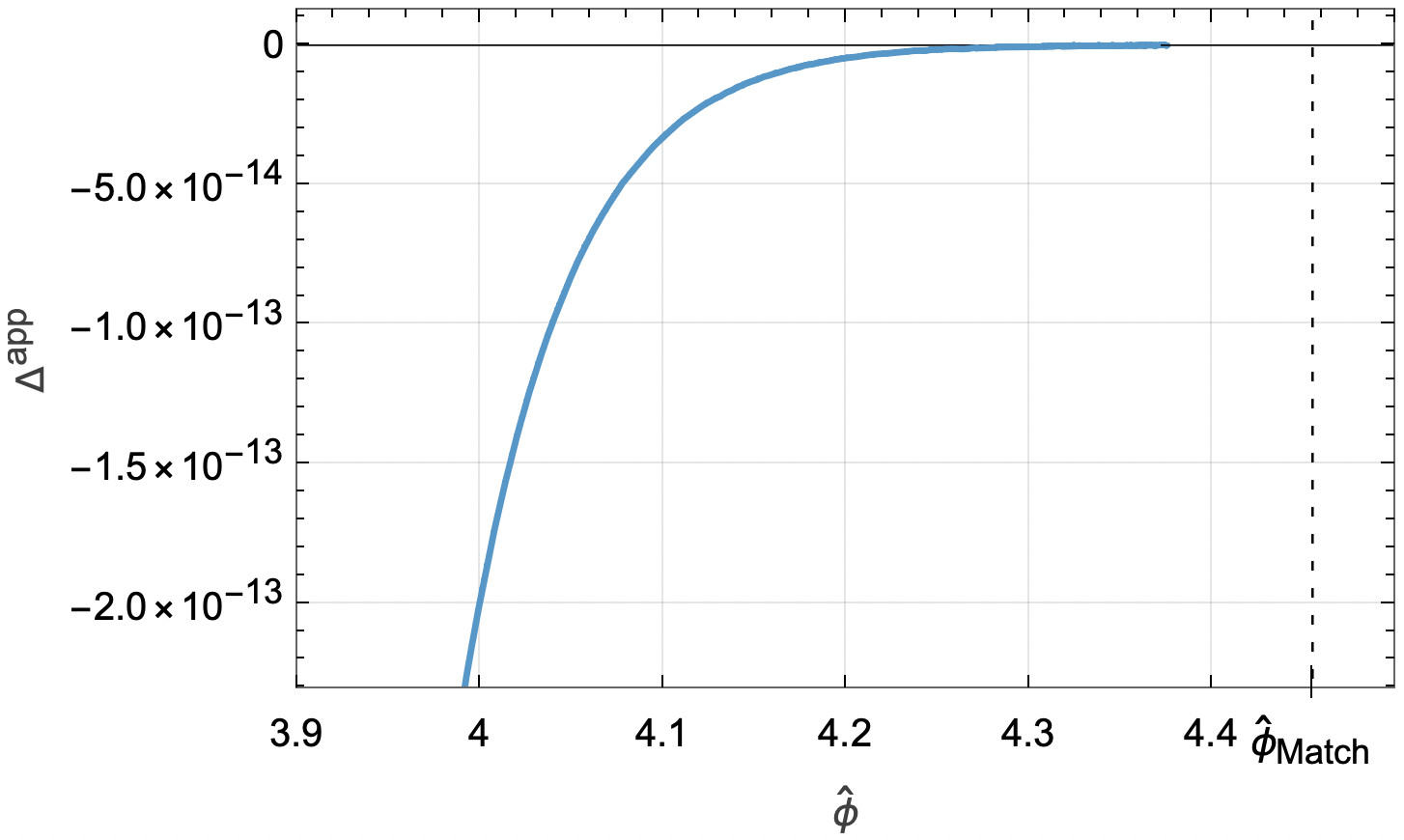}
    \end{subfigure}
    \begin{subfigure}[b]{0.49\textwidth}
      \includegraphics[width = 0.88\textwidth]{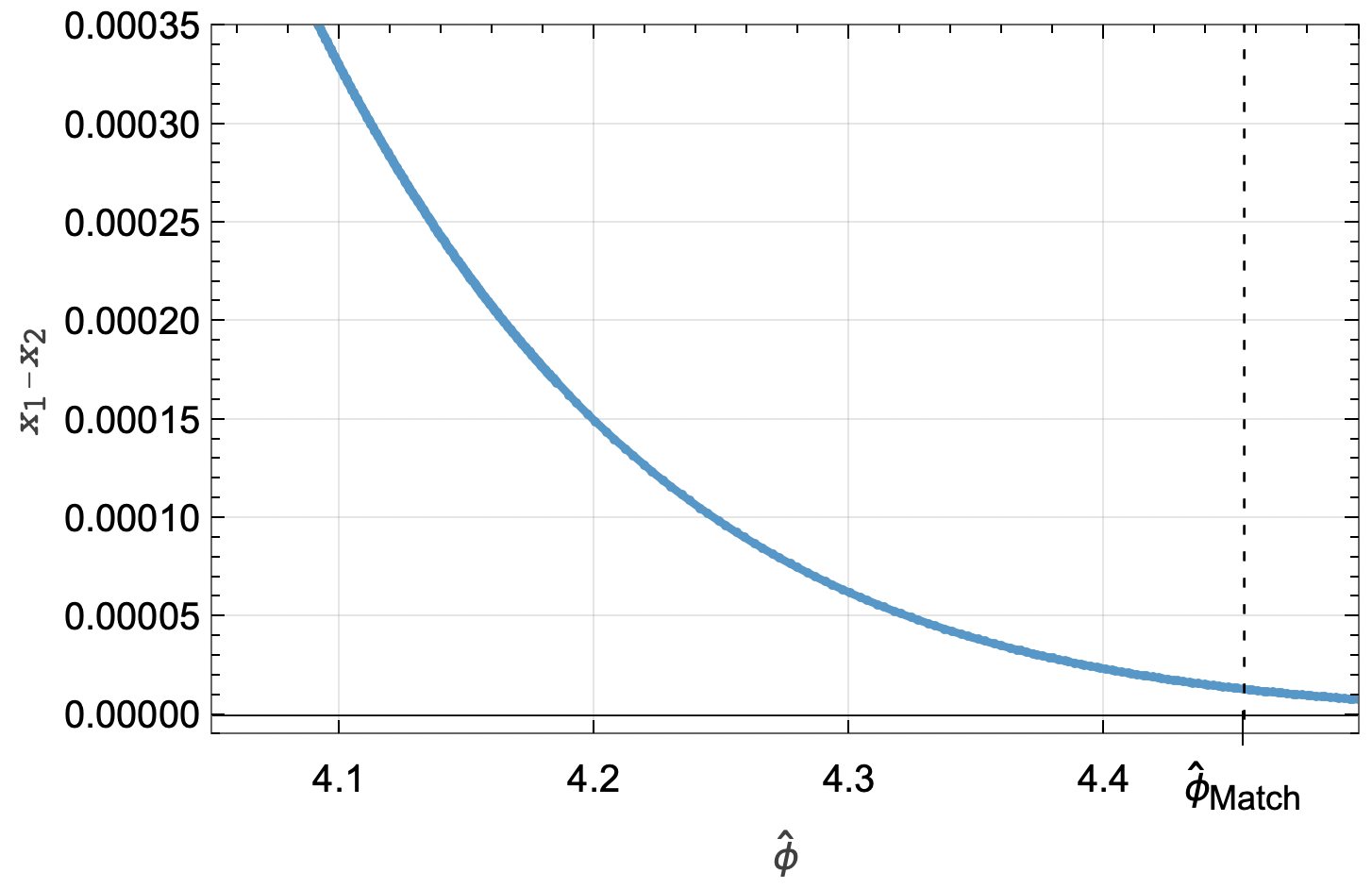}
    \end{subfigure}
    \caption{Plots referring to the total complexity $\alpha=t$, $\sigma=2$ and $\gamma=1/2$. (\emph{Top}) Difference between the right-hand-sides of Eqs.~\ref{eq:IdentGen} evaluated at $x_1^{\rm app}, x_2^{\rm app}$ and $\beta_2^{\rm app}$, and the corresponding parameters $\mathtt{r}, \mathtt{x}_1$ obtained solving the annealed equations. The difference vanishes at the estimated $\hat{\phi}_{\rm{Match}}$. (\emph{Bottom left}) Convergence to zero of $\Delta^{\rm app}= \beta_1^{\rm app} \beta_2^{\rm app}- (\beta_3^{\rm app})^2$ as $\hat{\phi} \to \hat{\phi}_{\rm{Match}}$. (\emph{Bottom right}) Difference between the values of $\mathtt{x}_1, \mathtt{x}_2$ obtained within the annealed calculation. }
    \label{fig:Limit_Ann_ToT}
  \end{figure}

  \emph{Total complexity of equilibria ($\alpha=t$).}
  Also in this case, when \eqref{eq:Delta0}  is attained the  quenched saddle point equations map into the annealed ones, with the identification
  \begin{equation}\label{eq:IdentGen}
    \begin{aligned}
      \mathtt{y} &= \frac{y}{\sqrt{1-\beta_2}},
      & \mathtt{x}_2 &= \frac{x_2}{\sqrt{1-\beta_2}}, \\
      \mathtt{x}_1 &= \frac{x_1}{\sqrt{1+ \tonde{\frac{x_1^2}{x_2^2}-2}\beta_2}},
      & \mathtt{r} &= r \sqrt{\frac{1+ \tonde{\frac{x_1^2}{x_2^2}-2}\beta_2}{1-\beta_2}}.
    \end{aligned}
  \end{equation}
  In particular, by comparing the quenched and annealed sets of self-consistent equations one can show that when \eqref{eq:IdentGen} hold, the values of the order parameters $m,p,q_1, \xi_1$ obtained solving the quenched equations coincide with those obtained solving the annealed equations. In Fig.~\ref{fig:Limit_Ann_ToT}, we show a numerical check that  \eqref{eq:IdentGen}  hold at the value of $\hat{\phi}$ for which $\Delta(\beta_1, \beta_2, \beta_3) =0$. To perform this check, we compute the right-hand-sides of Eq.~\ref{eq:IdentGen} by solving the approximate set of equations obtained placing the representations \eqref{eq:mIntD0}-\eqref{eq:PhiIntD0} into \eqref{eq:VRtop}-\eqref{eq:VRbottom}, and subtract the values obtained within the annealed calculation. Once more, the solutions $x_1^{\rm app}, x_2^{\rm app}, \beta_1^{\rm app},\beta_2^{\rm app},\beta_3^{\rm app}, r^{\rm app}$  of this approximate set of equations coincide with the solutions of \eqref{eq:VRtop}-\eqref{eq:VRbottom} without approximations at the value of $\hat{\phi}$ for which \eqref{eq:Delta0} is satisfied. We find that the second of the limits in \eqref{eq:LimitAnnealed} is again satisfied, while the other two limits are not attained. In particular, at $\hat{\phi}_{\text{Match}}$ it does not hold $\mathtt{x}_1=\mathtt{x}_2$, as we show in Fig.~\ref{fig:Limit_Ann_ToT} (\emph{Bottom right}). This is consistent with the asymmetry between the variables associated to the abundances and those associated to the effective growth rates when $\alpha=t$. Notice that by construction the cavity method allows to characterize only uninvadable equilibria, and thus no comparison can be made between the total complexity (that is contributed by invadable equilibria) and the cavity equations.

  \subsubsection{Beyond the matching point}\label{sec:Beyond}

  At $ \hat{\phi}_{\text{Match}}(\gamma,\sigma)$, the quenched complexity coincides with the annealed one. Beyond that point, i.e., for $ \hat{\phi} \geq \hat{\phi}_{\text{Match}}(\gamma,\sigma)$,  the quenched complexity can either depart from the annealed one, or coincide with it. Discriminating between these two situations numerically is challenging, since it would require to solve the quenched equations for very small values of $\Delta(\beta_1, \beta_2, \beta_2)$. To get insight we assume that for $ \hat{\phi} \geq \hat{\phi}_{\text{Match}}(\gamma,\sigma)$ the conjugate parameters are frozen to the boundary value $\Delta(\beta_1, \beta_2, \beta_2)=0$. We determined a new set of saddle point equations in which $\Delta=0$ is enforced, i.e., $\beta_3$ is not optimized over but set to $\beta_3=-\sqrt{\beta_1 \beta_2}$. This new set of equations is derived in Sec.~\ref{sec:DeltaEnforced}. We solve these equations for $ \hat{\phi} \geq \hat{\phi}_{\text{Match}}(\gamma,\sigma)$, and remark that the resulting complexity is numerically indistinguishable from the annealed in the regime of parameters that we explore, see Fig.~\ref{fig_combined1}~(\emph{left}). Therefore, for $\phi \leq  {\phi}_{\text{Match}}(\gamma,\sigma)$ we make use of the annealed expression for the complexity to derive some of the results reported in the main text. In particular, we remark that the transition between the robust and fragile phase occurs in this regime of parameters.

  \subsection{Details on the numerical solution of the self-consistent equations}{\label{sec:methods}}


  We solved the self-consistent equations numerically using a fixed point (SLSQP) approach. For both the total ($\alpha=t$) and uninvadable ($\alpha=u$) complexities, we
  search a solution at fixed $\gamma$, $\sigma$ and $\hat{\phi}$, the latter fixing the value of $\phi$.  We truncated the integrals to $(u_1,u_2) \in [-20,20]^2$ as the integrands decay rapidly because of the Gaussian weight $\mathcal{G}_{\hat{\bf{x}}}$.
  The integrals were calculated using a quadrature method with $n = 500^2$ points. The results did not change when we increased the number of points or the integration domain. For the initial guess of the fixed point algorithm, we used the results from the uninvadable case in the beginning, then we implemented a step by step continuation method on the grid of parameters $(\sigma,\gamma,a)$, using the previous solution as an initial guess for the next computation.\\
  As discussed above, when
  $\hat{\phi}\approx \hat{\phi}_{\text{Match}}^-(\gamma,\sigma)$ we have the limit $\beta_1\beta_2 - \beta_3^2 \approx 0$ so that the determinant of the Gaussian measure $\mathcal{G}_{\hat{\mathbf{x}}}$ diverges, making the quenched equations difficult to solve. We therefore solve the modified saddle point equations discussed in Sec.~\ref{sec:Beyond} where we enforce  $\Delta= 0$. In this case we compute the integrals sticking to $u\in [-10,10]$. The threshold chosen for the switch between the two sets of equations is $\Delta=\beta_1\beta_2 - \beta_3^2 < 10^{-4}$. We checked that the results do not change when we vary this threshold. We also check that at $\phi=\phi_{\text{Match}}$, the complexities obtained solving the two different sets of equations match perfectly (this is shown by the fact that all curves displayed thereafter are continuous).

  \section{The complexity: additional results}\label{sec:results}

  \begin{figure}
    \begin{subfigure}[t]{0.6\textwidth}
      \centering
      \includegraphics[width = 0.9\textwidth]{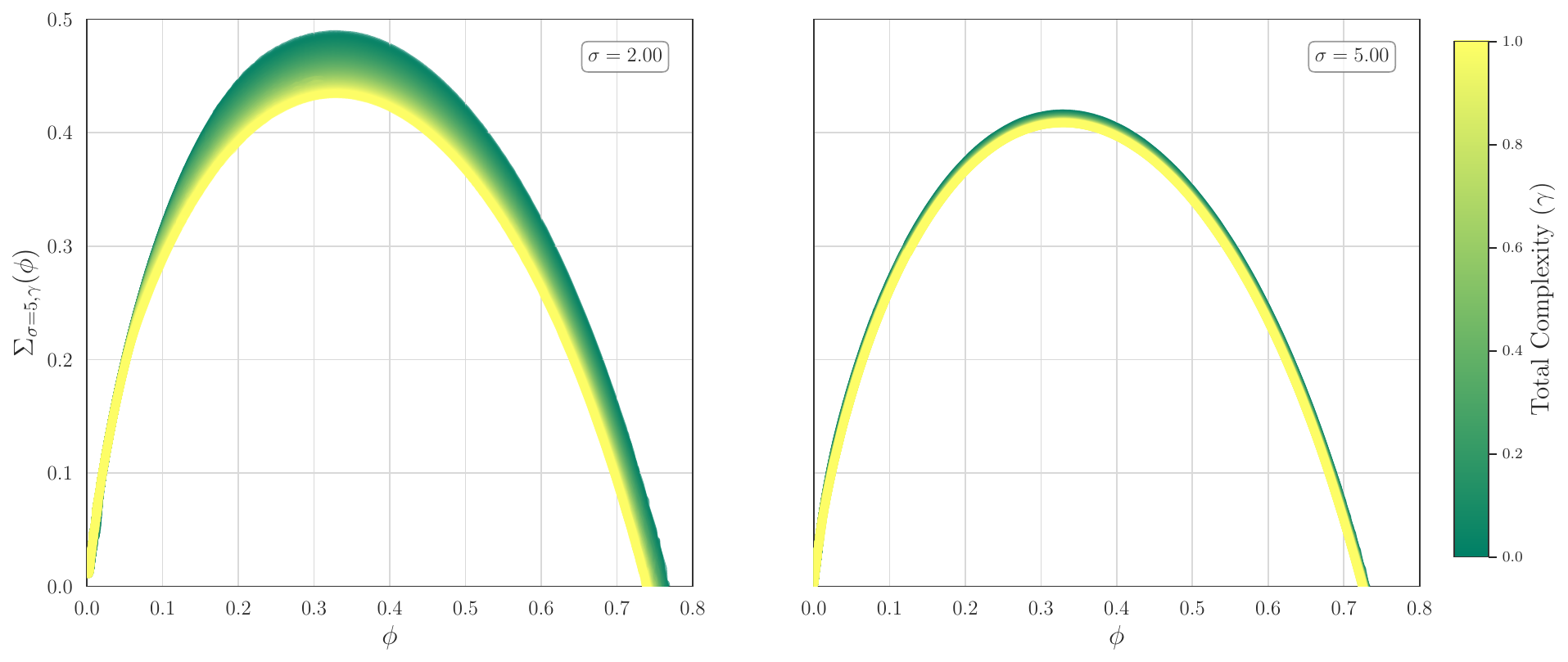}
    \end{subfigure}
    \begin{subfigure}[t]{0.365\textwidth}
      \centering
      \includegraphics[width = 0.9\textwidth]{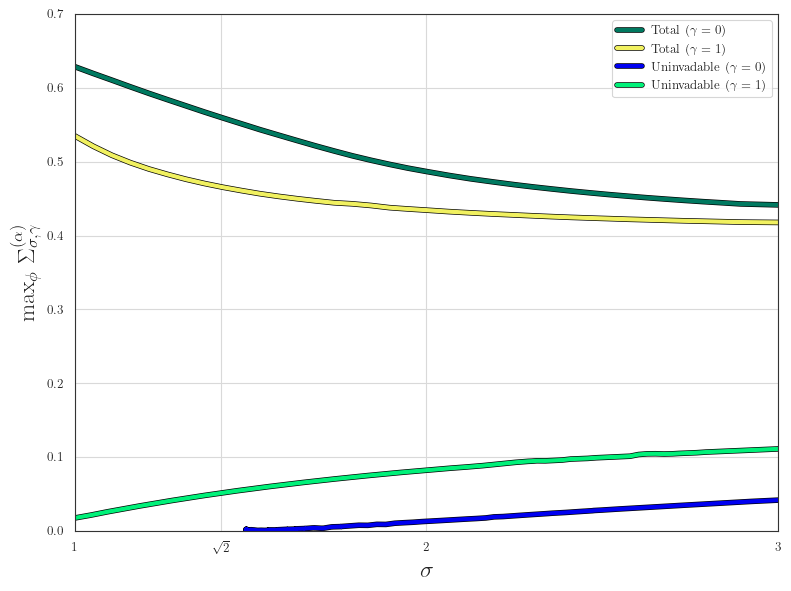}
    \end{subfigure}
    \caption{(\emph{Left}) Total complexities $\Sigma^{(t)}_{\sigma, \gamma}(\phi)$ for $\sigma=2$ and $\sigma=5$. For $\phi<\phi_{\text{match}}$, annealed values are plotted. (\emph{Right}) Maximal complexity $\max_{\phi} \Sigma^{(\alpha)}_{\sigma, \gamma}(\phi)$ as a function of $\sigma$ for  $\gamma \in\{0,1\}$ }
    \label{fig:inv_complex}
  \end{figure}

  \parhead{Total complexity and the Unique Fixed Point phase.} Plots of the total complexity $\Sigma^{(t)}_{\sigma, \gamma}(\phi)$ for two different values of variability $\sigma$ are given in Figure \ref{fig:inv_complex} (\emph{left}).  We make three observations: (i) increasing the reciprocity $\gamma$ of the interactions leads to a smaller number of equilibria, as shown by the fact that the complexity is a decreasing function of $\gamma$ at fixed $\phi,\sigma$. (ii) The complexity is also a decreasing function of $\sigma$ at fixed $\phi,\gamma$. (iii) Decreasing $\sigma$ also increases the spread between the complexity at $\gamma =0$ and $\gamma =1$, especially around the value of diversity where the complexity is maximal. These observations imply that the total complexity remains positive also in the Unique Fixed Point (UFP) phase. This is completely at odds with the uninvadable case, Since $\Sigma^{(u)}_{\sigma, \gamma}(\phi) \to 0$ at $\sigma =\sigma_c(\gamma)= \sqrt{2}/(1+\gamma)$. These behaviors are summarized in Fig.~\ref{fig:inv_complex} (\emph{right}), which shows how  $\max_{\phi} \Sigma^{(\alpha)}_{\sigma, \gamma}(\phi)$ behaves in $\sigma$, for $\gamma \in  \{0,1\}$. We check that indeed, there exists multiple invadable, internally stable equilibria for $\sigma <\sigma_c(\gamma)= \sqrt{2}/(1+\gamma)$. We also verify numerically that the uninvadable complexity is negative in this regime.
  Surprisingly, in the UFP phase the dynamics is not affected by these exponentially-numerous invadable equilibria, as it converges fast to the uninvadable fixed point of the dynamical equations, that is indeed unique. A similar phenomenology is found in \cite{fournier2025non} for a different model. Thus in the UFP phase: (i) There exists a unique, internally stable and uninvadable equilibrium. (ii) There exists exponentially-many other invadable equilibria, some internally stable and some internally unstable. (iii) The dynamics ignores completely the invadable equilibria and converges directly to the unique uninvadable equilibrium.

\begin{figure}[h]
    \begin{subfigure}[c]{0.32\textwidth}
      \centering
      \includegraphics[width = \textwidth]{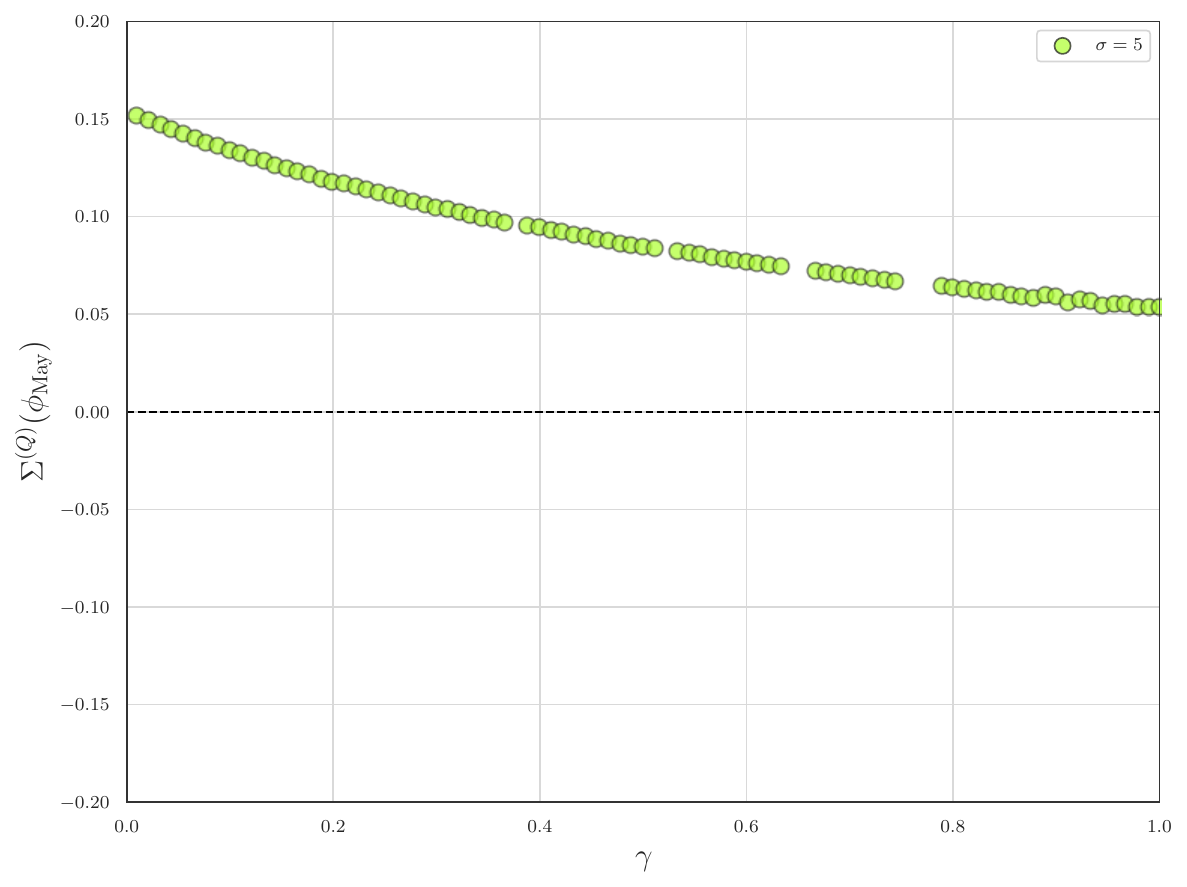}
    \end{subfigure}
    \begin{subfigure}[c]{0.32\textwidth}
      \centering
      \includegraphics[width = \textwidth]{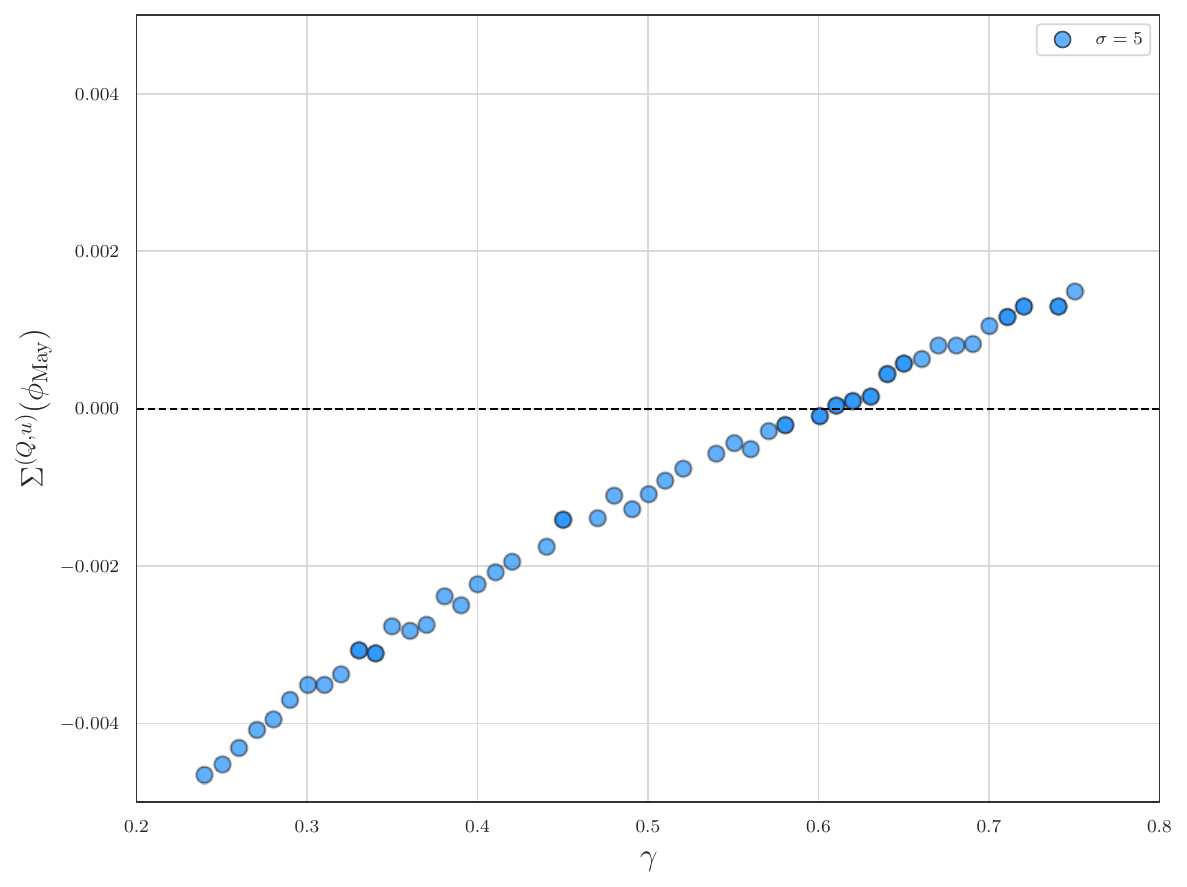}
    \end{subfigure}
    \begin{subfigure}[c]{0.32\textwidth}
      \centering
      \includegraphics[width = \textwidth]{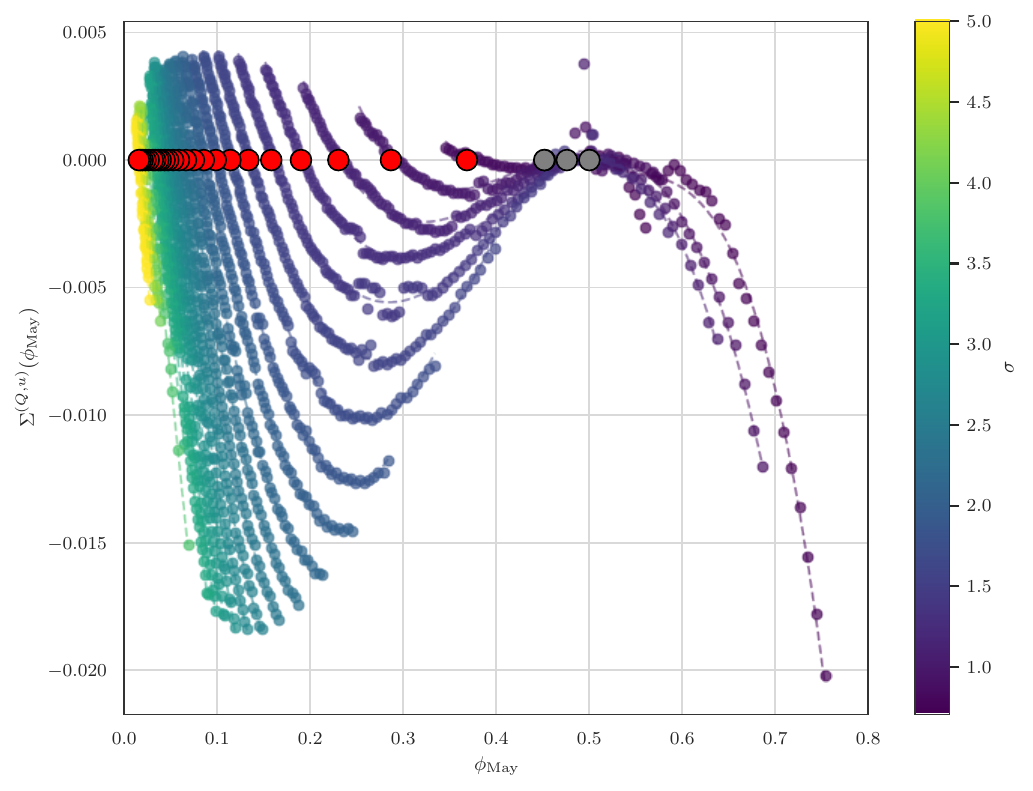}
    \end{subfigure}
    \hfill
    \caption{(\emph{Left}) Total  complexity ($\alpha=t$) evaluated at $\phi_{\mathrm{May}}=[\sigma^2(1+\gamma)^2]^{-1}$ for $\sigma=5$, as a function of $\gamma$. (\emph{Center}) Uninvadable  complexity ($\alpha=u$) evaluated at $\phi_{\mathrm{May}}$ for $\sigma=5$, as a function of $\gamma$. (\emph{Right}) $\Sigma^{(u)}_{\sigma,\gamma} (\phi_{\mathrm{May}})$ evaluated for different $\sigma$ (represented by different colors) and various $\gamma$ (represented by different $\phi_{\text{May}}$). The dashed lines are polynomial fits used to find the critical $\phi_{\mathrm{May,c}}$ such that $\Sigma^{(u)}_{\sigma,\gamma} (\phi_{\mathrm{May,c}})=0$. Red points indicate the $(\phi_{\mathrm{May,c}})$, corresponding to the critical $\gamma_{\text{FR}}$ for each $\sigma$. The grey dots correspond to the UFP phase boundary.}
    \label{fig:phi_may_combined}
  \end{figure}

\begin{figure}
  \centering
  \includegraphics[width=0.6\textwidth]{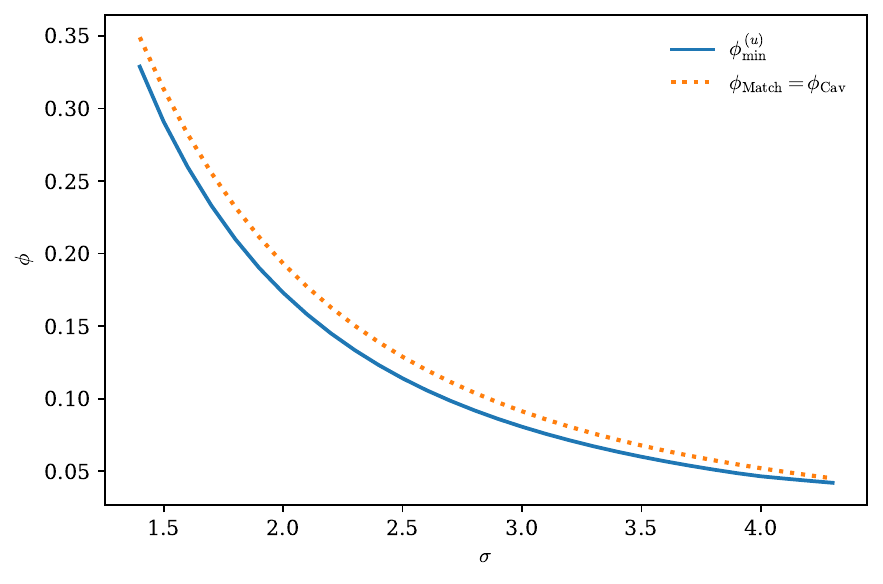}
  \caption{Comparison between $\phi_{\mathrm{min}}^{(u)}(\gamma)$ and $\phi_{\mathrm{Match}}(\gamma)$ for various values of $\sigma$ at $\gamma =0.3$.}
  \label{fig:phi_min_match}
\end{figure}

  \parhead{Uninvadable equilibria, May stability and the fragile-to-robust ME transition.} Let $\left[\phi_{\rm min}^{(u)},\phi_{\rm max}^{(u)}\right]$  denote the support of the quenched complexity, i.e., the range of diversity where the quenched complexity $\Sigma^{(u)}_{\sigma, \gamma}(\phi)$ is positive. A key observation is that while the total complexity remains positive, the uninvadable complexity becomes negative as $\phi\to 0$, since $\phi_{\rm min}^{(u)}>0$. Thus, whether $\phi_{\rm min}^{(u)}$ lies below or above the May threshold $\phi_{\mathrm{May}}=[\sigma^2(1+\gamma)^2]^{-1}$ depends on the values of $(\gamma,\sigma)$, giving rise to the phase boundary between the fragile and robust ME phases discussed in the main text.
  Figure~\ref{fig:phi_may_combined} (\emph{left, center}) shows  $\Sigma^{(\alpha)}_{\sigma, \gamma}(\phi_{\mathrm{May}})$ for $\sigma=5$, as a function of $\gamma$: while the total complexity  $\Sigma^{(t)}_{\sigma, \gamma}(\phi_{\mathrm{May}})$ is always positive (implying that the rGLV equations always admit an exponentially-large number of invadable, marginally stable equilibria), $\Sigma^{(u)}_{\sigma, \gamma}(\phi_{\mathrm{May}})$ vanishes at a  critical value $\gamma_{\text{FR}}$. By continuity and by monotonicity in $\phi$, this implies that for $\gamma < \gamma_{\text{FR}}$ there are no internally stable, uninvadable equilibria as $S \to \infty$, while for $\gamma > \gamma_{\text{FR}}$  they are  exponentially numerous.\\
  To determine $\gamma_{\text{FR}}$ for a fixed $\sigma$, we compute $\Sigma^{(u)}_{\sigma,\gamma} (\phi_{\mathrm{May}})$ for various $\gamma$ and do a polynomial fit of $\phi_{\text{May}} \mapsto \Sigma^{(u)}_{\sigma,\gamma} (\phi_{\text{May}})$ to find the critical $\phi_{\text{May,c}}$ such that $ \Sigma^{(u)}_{\sigma,\gamma} (\phi_{\text{May}}) = 0$, see Fig.~\ref{fig:phi_may_combined} (\emph{right}). We then set  $\phi_{\text{May,c}} = [{\sigma^2(1+\gamma)^2}]$, which gives us $\gamma_{\text{FR}}$.
  The line in the phase diagram in Fig.~1 in the main text is obtained by performing a
  hyperbolic fit to find $\gamma_{\text{FR}} \approx 0.318/\sigma + 0.549$. We remark that in Fig.~\ref{fig:phi_may_combined} (\emph{right}) the complexity at $\phi_{\text{May}}$ increases again with $\phi_{\text{May}}$ for $\gamma$ small. In fact, the complexity becomes 0 again at a $\phi$ that corresponds to $\sigma_c(\gamma) = \sqrt{2}/(1+\gamma)$, which is the boundary of the UFP phase.
Interestingly, we find that $\phi_{\mathrm{min}}^{(u)}(\gamma) < \phi_{\mathrm{Match}}$ for all $\gamma$, so that the transition between the fragile and robust ME phases occurs in the regime of parameters where the quenched complexity coincides with the annealed one. This implies that the transition between the fragile and robust ME phases can be fully characterized by the annealed calculation, which is much simpler to perform than the quenched one. We plot the comparison between these two $\phi$ values in figure \ref{fig:phi_min_match}.

  \section{The replicated Kac-Rice calculation: details on the derivation}\label{sec:appendix}
  \subsection{The derivation of the volume terms}\label{sec:AppVolume}

  The calculation of the asymptotics \eqref{eq:Asy1} and \eqref{eq:Asy2} is identical to that reported in Ref.~\cite{rosQuenchedComplexityEquilibria2023}, and we refer the interested reader to that reference for the details. The calculation of the volume term \eqref{eq:AsyVOl} also follows the steps delineated in Ref.~\cite{rosQuenchedComplexityEquilibria2023}, but it is slightly different depending on whether one considers the total complexity, or the complexity of uninvadable equilibria. For completeness, we show here how this difference arises. Following the same steps as in \cite{rosQuenchedComplexityEquilibria2023}, one shows that
  \eqref{eq:VolumeKR} takes the following form:
  \begin{equation}
    \mathcal{V}^{(\alpha)}_n(\hat{\mathbf{x}}) = \left(
    \prod_{a=1}^n\sum_{\tau^a = 0,1}e^{-\hat{\phi}_a \tau^a}\int d N^a d f^a j^{(\alpha)}_{\hat{\bf x}}(N^a,f^a)\right)^S e^{o(S n)},
  \end{equation}
  \begin{equation}
    \begin{aligned}
      j^{(\alpha)}_{\hat{\bf x}}(N^a,f^a) &= e^{- \sum_{a=1}^n (\hat{m}_a N^a + \hat{p}_a f^a) - \sum_{a \leq b=1}^n (\hat{z}_{ab}N^af^b + \hat{q}_{ab} N^a N^b + \hat{\xi}_{ab}f^af^b)}   \\
      &\quad \times \prod_{a : \tau^a = 1} \theta(N^a) \delta(f^a) \prod_{a : \tau^a = 0} \delta(N^a)\chi^{(\alpha)}(f^a).
    \end{aligned}
  \end{equation}
  Within the replica symmetric assumption, setting $k = \sum_{a}\tau^a$ and defining $g^b = -f^b$, this expression reduces to:
  \begin{equation}\label{eq:104}
    \mathcal{V}^{(\alpha)}_n(\hat{\mathbf{x}})  = e^{S\log\left[\sum_{k=0}^n\binom{n}{k} e^{-k \hat{\phi}}I^{(\alpha)}_k(\hat{\bf x})\right]+ o(S n)}
  \end{equation}
  with
  \begin{equation*}
    \begin{aligned}
      I^{(\alpha)}_k(\hat{\bf x})=   \int_{0}^\infty& \prod_{a=1}^k dN^a \int_{-\infty}^\infty \prod_{b=1}^{n-k + 1 } dg^b \quadre{\delta_{\alpha, t}+ \theta(g^b) \delta_{\alpha, u}}\\
      &\times e^{-\frac{1}{2}\theta_k(N^a,g^b)}  \prod_{a=1}^k e^{-\frac{1}{2}(2\hat{q}_1 - \hat{q}_0)(N^a)^2 - \hat{m} N^a}\prod_{b=1}^{n-k + 1 } e^{-\frac{1}{2}(2\hat{\xi}_1 - \hat{\xi}_0)(g^b)^2 + \hat{p} g^b}
    \end{aligned}
  \end{equation*}
  and
  \begin{equation}\label{eq:thetaConv}
    \theta_k(N^a,g^b) = \hat{q}_0\left(\sum_{a=1}^k N^a \right)^2 + \hat{\xi}_0\left(\sum_{b=1}^{n-k + 1} g^b \right)^2 -2\hat{z} \left(\sum_{a=1}^k N^a\right)\left( \sum_{b=1}^{n-k + 1} g^b\right).
  \end{equation}
  To proceed, we exploit the representation
  \begin{align}\label{eq:IntREpConv}
    e^{-\frac{1}{2}\theta_k(N^a,g^b)} = \int \frac{du_1 du_2}{2\pi \sqrt{\det A}} e^{-\frac{1}{2}(u_1,u_2)A^{-1}(u_1,u_2)^T} \prod_{a=1}^k e^{u_1 N^a} \prod_{b=1}^{n-k+1}e^{-u_2 g^b},
  \end{align}
  where we are assuming that the matrices
  \begin{equation}
    A =
    \begin{pmatrix}
      -\hat{q}_0 & -\hat{z} \\
      -\hat{z} & -\hat{\xi}_0
    \end{pmatrix}, \quad \quad  A^{-1} = \frac{1}{\hat{q}_0 \hat{\xi}_0 - \hat{z}^2}
    \begin{pmatrix}
      -\hat{\xi}_0 & \hat{z} \\
      \hat{z} & -\hat{q}_0
    \end{pmatrix},
  \end{equation}
  are positive definite, namely that $\hat{q}_0 \hat{\xi}_0 - \hat{z}^2>0$. Introducing
  \begin{equation}
    g^{(\alpha)}(x,y)=
    \sqrt{\frac{\pi}{2y}} e^{\frac{x^2}{2y}} \tonde{2 \delta_{\alpha, t} + \delta_{\alpha, u} \text{erfc}\left( \frac{x}{\sqrt{2y}} \right)},
  \end{equation}
  we see that
  \begin{equation}
    \begin{split}
      I^{(\alpha)}_k(\hat{\bf x})&=  \int \frac{du_1 du_2}{2\pi \sqrt{\det A}} e^{-\frac{1}{2}(u_1,u_2)A^{-1}(u_1,u_2)^T}  \left(\int_0^\infty  dN e^{-\frac{1}{2}(2\hat{q}_1 - \hat{q}_0)N^2 -(\hat{m}-u_1) N}  \right)^k \times \\
      &\quad \quad \times\left(\int_{-\infty}^\infty d g \quadre{\delta_{\alpha, t}+ \theta(g) \delta_{\alpha, u}}  e^{-\frac{1}{2}(2\hat{\xi}_1 - \hat{\xi}_0) g^2 + g( \hat{p} - u_2)}  \right)^{n-k+1}\\
      &=  \int \frac{du_1 du_2}{2\pi \sqrt{\det A}} e^{-\frac{1}{2}(u_1,u_2)A^{-1}(u_1,u_2)^T}  \left(  g^{(u)}(\hat{m} - u_1 ; 2\hat{q_1}-\hat{q}_0)\right)^k \left(g^{(\alpha)}(u_2 - \hat{p} ; 2\hat{\xi}_1 - \hat{\xi}_0) \right)^{n-k+1}.
    \end{split}
  \end{equation}
  Performing the sum over $k$ we get
  \begin{equation}\label{lasteq}
    \mathcal{V}^{(\alpha)}_n(\hat{\mathbf{x}})
    =\left[  \int \frac{du_1 du_2}{2\pi \sqrt{\det A}} e^{-\frac{1}{2}(u_1,u_2)A^{-1}(u_1,u_2)^T} \left(e^{-\hat{\phi}}g^{(u)}(\hat{m} - u_1 ; 2\hat{q_1}-\hat{q}_0) +  {g}^{(\alpha)}(u_2 - \hat{p} ; 2\hat{\xi}_1 - \hat{\xi}_0) \right)^n\right]^S,
  \end{equation}
  from which it follows the asymptotics \eqref{eq:AsyVOl}. The quenched expansion \eqref{eq:InQ} then follows straightforwardly from the small $n$ expansions $A^n = 1 + n \log A$  and $\log (1+ A) = A$ when $A \to 0$. The corresponding annealed term \eqref{eq:ValA} is obtained selecting $n=1$ in \eqref{eq:104}. Alternatively, it can be recovered from \eqref{lasteq} setting $n=1$ and performing the integration over the variables $u_1, u_2$, making use of the relation:
  \begin{equation}
    \int \dd{x} \text{Erfc}(a x + b) \cdot \frac{1}{\sqrt{2\pi\sigma^2}} \exp\left( -\frac{(x - \mu)^2}{2\sigma^2} \right)
    = \text{Erfc}\left( \frac{a\mu + b}{\sqrt{1 + 2a^2\sigma^2}} \right).
  \end{equation}

  As stressed above, this derivation assumes that the matrix $A$ is positive definite. For certain relevant values of parameters, however, this assumption is violated, as the matrix $A$ develops a vanishing eigenvalue, leading to a divergence of the Gaussian kernel in \eqref{lasteq}. This occurs whenever the conjugate parameters satisfy the relation $\hat q_0 \hat \xi_0- \hat z^2=0$, corresponding to $\hat z = - \sqrt{\hat q_0 \hat \xi_0}$. When this limit is attained, the integration variables $u_1, u_2$ in \eqref{lasteq} become functionally related. Indeed, when $\hat z = - \sqrt{\hat q_0 \hat \xi_0}$ the term \eqref{eq:thetaConv} reduces to a perfect square, and \eqref{eq:IntREpConv} simplifies into
  \begin{align}\label{eq:IntREpConvDelta0}
    e^{-\frac{1}{2}\theta_k(N^a,g^b)}\; \stackrel{\hat q_0 \hat \xi_0- \hat z^2\to0}{\quad \longrightarrow  }  \;\frac{1}{\sqrt{2\pi}}\int \dd{u}\,  e^{-\frac{1}{2}u^2} e^{u\left(\sqrt{-\hat{q}_0} \sum_{a=1}^k N^a  - \sqrt{ -\hat{\xi}_0 }\sum_{b=1}^{n-k + 1} g^b  \right)}.
  \end{align}
  and therefore
  \begin{equation}
    \begin{split}
      & \lim_{\hat q_0 \hat \xi_0- \hat z^2\to0} I^{(\alpha)}_k(\hat{\bf x})\\
      &\quad =   \int \frac{\dd{u}}{\sqrt{2\pi}} e^{-\frac{u^2}{2}} g^{(u)}(\hat{m} - u\sqrt{-\hat{q}_0} ; 2\hat{q_1}-\hat{q}_0)^k\, {g}^{(\alpha)}( u\sqrt{-\hat{\xi}_0} - \hat{p} ; 2\hat{\xi}_1 - \hat{\xi}_0)^{n-k},
    \end{split}
  \end{equation}
  so that finally
  \begin{align}
    & \lim_{\hat q_0 \hat \xi_0- \hat z^2\to0} \mathcal{J}_n^{(\alpha)}(\hat{\mathbf{x}}) \; \notag \\
    &\quad =\log \left[\int\dd{u} e^{-\frac{ u^2}{2}} \left(e^{-\hat{\phi}} g^{(u)}(\hat{m} - u\sqrt{-\hat{q}_0} ; 2\hat{q_1}-\hat{q}_0) +  {g}^{(\alpha)}( u\sqrt{-\hat{\xi}_0} - \hat{p} ; 2\hat{\xi}_1 - \hat{\xi}_0)\right)^n  \right]
  \end{align}
  and
  \begin{align}\label{eq:VolumeOneInt}
    &\lim_{\hat q_0 \hat \xi_0- \hat z^2\to0}  \bar{\mathcal{J}}^{(\alpha)}(\hat{\mathbf{x}})\;:=\bar{\mathcal{J}}^{(\alpha)}_{\Delta=0}(\hat{\mathbf{x}})\\
    &= \int\dd{u} e^{-\frac{ u^2}{2}} \log \left[e^{-\hat{\phi}} g^{(u)}(\hat{m} - u\sqrt{-\hat{q}_0} ; 2\hat{q_1}-\hat{q}_0) +  {g}^{(\alpha)}( u\sqrt{-\hat{\xi}_0} - \hat{p} ; 2\hat{\xi}_1 - \hat{\xi}_0)  \right].
  \end{align}

  \subsection{Saddle point equations in their original form}\label{supp:OriginalSPE}
  \subsubsection{The quenched equations}
  We report in this section the expression obtained deriving the function \eqref{eq:barA} with respect to the order and conjugate parameters ${\bf x}$ and $\hat{\bf x}$. We begin by noticing that the term that changes with $\alpha=u,t$ (controlling whether the counting is restricted to uninvadable equilibria or not)  is a function of the conjugate parameters only: therefore, the derivatives with respect to the order parameters are equal for both choices of $\alpha$, and equal to those already obtained in \cite{rosQuenchedComplexityEquilibria2023}. They read:
  \begin{align}
    \hat{\xi}_1 &= \frac{q_1 - 2q_0}{2\sigma^2 (q_1 - q_0)^2}, \label{eq:1}\\
    \hat{\xi}_0 &= -\frac{q_0}{\sigma^2 (q_1 - q_0)^2},\label{eq:2} \\
    \hat{p} &= - \frac{\kappa - \mu m}{\sigma^2 (q_1 - q_0)}, \label{eq:3}\\
    \hat{m} &= -\frac{\mu (\kappa - \mu m)}{\sigma^2 (q_1 - q_0)}
    - \frac{(\kappa - \mu m)}{(\gamma + 1) \sigma^2 (q_1 - q_0)}\notag\\
    &\quad + \frac{\mu m}{(\gamma + 1) \sigma^2 (q_1 - q_0)} + \frac{\mu p}{\sigma^2 (q_1 - q_0)}
    - \frac{\gamma}{1 + \gamma} \frac{(\kappa - 2\mu m) z}{\sigma^2 (q_1 - q_0)^2}, \label{eq:4}\\
    \hat{z} &= \frac{\gamma m (\kappa - \mu m)}{(\gamma + 1) \sigma^2 (q_1 - q_0)^2}
    - \frac{\gamma z (q_1 + q_0)}{(\gamma + 1) \sigma^2 (q_1 - q_0)^3}
    - \frac{q_0}{(\gamma + 1) \sigma^2 (q_1 - q_0)^2}, \label{eq:5}\\
    \hat{q}_1 &= - \frac{(\kappa - \mu m) [m + (\gamma + 1)p]}{(\gamma + 1) \sigma^2 (q_1 - q_0)^2}
    + \frac{2(\kappa - \mu m) [m(q_1 - q_0 + \gamma z) + p(\gamma + 1)(q_1 - q_0)]}{(\gamma + 1) \sigma^2 (q_1 - q_0)^3} \nonumber \\
    &\quad + \frac{q_0 (2q_0 z - 2\gamma z^2)}{(\gamma + 1) \sigma^2 (q_1 - q_0)^4}
    - \frac{q_0^2 (3\xi_1 - 2\xi_0)}{2\sigma^2 (q_1 - q_0)^4}
    - \frac{q_1 q_0 (\xi_0 - 2\xi_1)}{\sigma^2 (q_1 - q_0)^4}
    - \frac{q_1 [2q_0 z + \gamma z^2]}{(\gamma + 1) \sigma^2 (q_1 - q_0)^4} \nonumber \\
    &\quad - \frac{\xi_1 q_1^2}{2\sigma^2 (q_1 - q_0)^4}
    - \frac{q_0}{2(q_1 - q_0)^2}
    + \frac{1}{2(q_1 - q_0)}
    - \frac{(\kappa - \mu m)^2}{2\sigma^2 (q_1 - q_0)^2}, \label{eq:6}\\
    \hat{q}_0 &= \frac{2(\kappa - \mu m)(-m - (\gamma + 1)p)}{(\gamma + 1) \sigma^2 (q_1 - q_0)^2}
    + \frac{4(\kappa - \mu m) (m(q_1 - q_0 + \gamma z) + (\gamma + 1)p (q_1 - q_0))}{(\gamma + 1) \sigma^2 (q_1 - q_0)^3} \nonumber \\
    &\quad  + \frac{-2z (q_1^2 + \gamma z (2q_1 + q_0) - q_0^2) +  2(\gamma + 1) \xi_1 q_0 (q_1 - q_0) - (\gamma + 1) \xi_0 (q_1 - q_0)(q_1 + q_0)}{(\gamma + 1) \sigma^2 (q_1 - q_0)^4}
    \nonumber \\
    &\quad - \frac{q_0}{(q_1 - q_0)^2}
    - \frac{(\kappa - \mu m)^2}{\sigma^2 (q_1 - q_0)^2} \label{eq:7}.
  \end{align}

  We now come to the derivatives with respect to the conjugate parameters $\hat{\bf x}$. We introduce the function:
  \begin{align}
    \mathcal{R}^{(\alpha)}_{\hat{\mathbf{x}}}(u_1,u_2) &=  e^{\frac{(\hat{m}-u_1)^2}{2(2\hat{q}_1 - \hat{q}_0)}}\text{Erfc}\left(\frac{\hat{m}-u_1}{\sqrt{2(2\hat{q}_1 - \hat{q}_0)}}\right) \notag\\
    &\quad +  e^{\hat{\phi}}\sqrt{\frac{2\hat{q}_1 - \hat{q}_0}{2\hat{\xi}_1 - \hat{\xi}_0}} e^{\frac{(\hat{p}-u_2)^2}{2(2\hat{\xi}_1 - \hat{\xi}_0)}}\quadre{2 \delta_{\alpha, t} + \text{Erfc}\left(\frac{-\hat{p}+u_2}{\sqrt{2(2\hat{\xi}_1 - \hat{\xi}_0)}}\right) \delta_{\alpha, u}}
  \end{align}
  so that
  \begin{align}
    \bar{\mathcal{J}}^{(\alpha)}(\hat{\mathbf{x}}) &= \int \dd{u_1}\dd{u_2} \mathcal{G}_{\mathbf{\hat{x}}}(u_1,u_2)\log(e^{-\hat{\phi}} \sqrt{\frac{\pi}{2(2\hat{q}_1 - \hat{q}_0)}}\mathcal{R}^{(\alpha)}_{\mathbf{\hat{x}}}(u_1,u_2)).
  \end{align}
  Then, it holds
  \begin{align}
    \phi &= \int \dd{u_1}\dd{u_2} \mathcal{G}_{\hat{\mathbf{x}}}(u_1,u_2) \frac{\partial}{\partial \hat{\phi}} \log(e^{-\hat{\phi}}\mathcal{R}^{\alpha}_{\mathbf{\hat{x}}}(u_1,u_2))\\
    m &= - \int \dd{u_1} \dd{u_2}\frac{\mathcal{G}_{\hat{\mathbf{x}}}(u_1,u_2)}{\mathcal{R}^{\alpha}_{\mathbf{\hat{x}}}(u_1,u_2)} \frac{\partial\mathcal{R}^{\alpha}_{\mathbf{\hat{x}}}(u_1,u_2) }{\partial \hat{m}} \\
    p &= - \int \dd{u_1} \dd{u_2}\frac{\mathcal{G}_{\hat{\mathbf{x}}}(u_1,u_2)}{\mathcal{R}^{\alpha}_{\mathbf{\hat{x}}}(u_1,u_2)} \frac{\partial\mathcal{R}^{\alpha}_{\mathbf{\hat{x}}}(u_1,u_2) }{\partial \hat{p}} \\
    q_1 &= - \int \dd{u_1} \dd{u_2}\frac{ \mathcal{G}_{\hat{\mathbf{x}}}(u_1,u_2)}{\mathcal{R}^{\alpha}_{\mathbf{\hat{x}}}(u_1,u_2)} \sqrt{2\hat{q}_1 - \hat{q}_0}\frac{\partial }{\partial \hat{q}_1}\left( \frac{\mathcal{R}^{\alpha}_{\mathbf{\hat{x}}}(u_1,u_2)}{\sqrt{2\hat{q}_1 - \hat{q}_0}}\right)  \\
    \xi_1 &= - \int \dd{u_1} \dd{u_2}\frac{ \mathcal{G}_{\hat{\mathbf{x}}}(u_1,u_2)}{\mathcal{R}^{\alpha}_{\mathbf{\hat{x}}}(u_1,u_2)} \frac{\partial\mathcal{R}^{\alpha}_{\mathbf{\hat{x}}}(u_1,u_2) }{\partial \hat{\xi}_1} \\
    q_0 &= 2\int \dd{u_1} \dd{u_2}\frac{\partial}{\partial \hat{q}_0}\left[\mathcal{G}_{\hat{\mathbf{x}}}(u_1,u_2)  \log(\frac{\mathcal{R}^{\alpha}_{\mathbf{\hat{x}}}(u_1,u_2)}{\sqrt{2\hat{q}_1 - \hat{q}_0}})\right]\\
    \xi_0 &=2 \int \dd{u_1} \dd{u_2}\frac{\partial}{\partial \hat{\xi}_0}\left[\mathcal{G}_{\hat{\mathbf{x}}}(u_1,u_2)  \log(\mathcal{R}^{\alpha}_{\mathbf{\hat{x}}}(u_1,u_2))\right]\\
    z &= \int \dd{u_1} \dd{u_2}\frac{\partial }{\partial \hat{z}} \left[ \mathcal{G}_{\hat{\mathbf{x}}}(u_1,u_2 ) \log \left(\mathcal{R}^{\alpha}_{\mathbf{\hat{x}}}(u_1,u_2)\right)\right].
  \end{align}
  The first five equations follow straightforwardly from the identity
  \begin{equation}
    \dv{}{x}\left(e^{\frac{x^2}{2}}\text{Erfc}\left(\frac{x}{\sqrt{2}}\right)\right) = x\left(e^{\frac{x^2}{2}}\text{Erfc}\left(\frac{x}{\sqrt{2}}\right)\right) - \sqrt{\frac{2}{\pi}}.
  \end{equation}
  To get a simpler form of the equations for the last three parameters,
  we use the fact that
  \begin{align}
    \frac{\partial\mathcal{G}_{\hat{\mathbf{x}}}(u_1,u_2) }{\partial \hat{q}_0} &= -\frac{1}{2} \frac{\partial^2 \mathcal{G}_{\hat{\mathbf{x}}}(u_1,u_2)}{\partial u_1^2} ,  \\
    \frac{\partial \mathcal{G}_{\hat{\mathbf{x}}}(u_1,u_2)}{\partial \hat{\xi}_0} &= -\frac{1}{2} \frac{\partial^2 \mathcal{G}_{\hat{\mathbf{x}}}(u_1,u_2)}{\partial u_2^2},  \\
    \frac{\partial \mathcal{G}_{\hat{\mathbf{x}}}(u_1,u_2)}{ \partial \hat{z}}  &=-\frac{\partial^2 \mathcal{G}_{\hat{\mathbf{x}}}(u_1,u_2)}{\partial u_1 \partial u_2},
  \end{align}
  so that, integrating by parts:
  \begin{align}
    q_0 &= 2\int du_1 du_2 \mathcal{G}_{\hat{\mathbf{x}}}(u_1,u_2) \left[  \frac{\partial}{\partial \hat{q}_0 } - \frac{1}{2} \frac{\partial^2 }{\partial u_1^2}\right]  \log(\frac{\mathcal{R}^{(\alpha)}_{\mathbf{\hat{x}}}(u_1,u_2)}{\sqrt{2\hat{q}_1 - \hat{q}_0}})\\
    \xi_0 &= 2\int du_1 du_2 \mathcal{G}_{\hat{\mathbf{x}}}(u_1,u_2) \left[  \frac{\partial}{\partial \hat{\xi}_0 } - \frac{1}{2} \frac{\partial^2 }{\partial u_2^2}\right]  \log(\mathcal{R}^{(\alpha)}_{\mathbf{\hat{x}}}(u_1,u_2))\\
    z &= \int du_1 du_2 \mathcal{G}_{\hat{\mathbf{x}}}(u_1,u_2 )\left[\frac{\partial }{\partial \hat{z}} - \frac{\partial^2}{\partial u_1 \partial u_2} \right] \log(\mathcal{R}^{(\alpha)}_{\mathbf{\hat{x}}}(u_1,u_2)).
  \end{align}
  Performing the derivatives and setting
  \begin{equation}
    \mathpzc{d}{\bf u}= \dd{u_1} \dd{u_2} \mathcal{G}_{\hat{\mathbf{x}}}(u_1,u_2)
  \end{equation}
  we finally get:
  \begin{align}
    m &= \int   \mathpzc{d}{\bf u} \frac{1}{\sqrt{2\hat{q}_1 - \hat{q}_0}}\frac{\sqrt{\frac{2}{\pi}} - \frac{(\hat{m}-u_1)}{\sqrt{2\hat{q}_1 - \hat{q}_0}}e^{\frac{(\hat{m}-u_1)^2}{2(2\hat{q}_1 - \hat{q}_0)}}\text{Erfc}\left(\frac{\hat{m}-u_1}{\sqrt{2(2\hat{q}_1 - \hat{q}_0)}}\right)}{\mathcal{R}^{(\alpha)}_{\mathbf{\hat{x}}}(u_1,u_2)}\label{eqm1}\\
    p &=  \int  \mathpzc{d}{\bf u}  \frac{e^{\hat{\phi}}\sqrt{\frac{2\hat{q}_1 - \hat{q}_0}{2\hat{\xi}_1 - \hat{\xi}_0}}}{\sqrt{2\hat{\xi}_1 - \hat{\xi}_0}} \quadre{\frac{\frac{u_2-\hat{p}}{\sqrt{2\hat{\xi}_1 -\hat{\xi}_0}} e^{\frac{(\hat{p}-u_2)^2}{2(2\hat{\xi}_1 - \hat{\xi}_0)}} \tonde{2 \delta_{\alpha, t}+ \delta_{\alpha,u}\text{Erfc}\left(\frac{u_2-\hat{p}}{\sqrt{2(2\hat{\xi}_1 - \hat{\xi}_0)}}\right)} -\delta_{\alpha,u}\sqrt{\frac{2}{\pi}}}{\mathcal{R}^{(\alpha)}_{\mathbf{\hat{x}}}(u_1,u_2)}}\\
    q_1 &= \int  \mathpzc{d}{\bf u} \frac{1}{2\hat{q}_1 - \hat{q}_0} \frac{\left(1 + \frac{(\hat{m}-u_1)^2}{2\hat{q}_1 - \hat{q}_0} \right)e^{\frac{(\hat{m}-u_1)^2}{2(2\hat{q}_1 - \hat{q}_0)}}\text{Erfc}\left(\frac{(\hat{m}-u_1)}{\sqrt{2(2\hat{q}_1 - \hat{q}_0)}}\right) - \frac{\hat{m}-u_1}{\sqrt{2\hat{q}_1 - \hat{q}_0}}\sqrt{\frac{2}{\pi}}  }{\mathcal{R}^{(\alpha)}_{\mathbf{\hat{x}}}(u_1,u_2)} \\
    \xi_1 &= \int \mathpzc{d}{\bf u} \; \frac{e^{\hat{\phi}}\sqrt{\frac{2\hat{q}_1 - \hat{q}_0}{2\hat{\xi}_1 - \hat{\xi}_0}}}{2\hat{\xi}_1 - \hat{\xi}_0}\\
    &\quad \times\quadre{
    \frac{\left(1 + \frac{(u_2-\hat{p})^2}{2\hat{\xi}_1 - \hat{\xi}_0}\right)e^{\frac{(u_2-\hat{p})^2}{2(2\hat{\xi}_1 - \hat{\xi}_0)}} \tonde{2 \delta_{\alpha, t}+ \delta_{\alpha,u}\text{Erfc}\left(\frac{u_2-\hat{p}}{\sqrt{2(2\hat{\xi}_1 - \hat{\xi}_0)}}\right)} - \delta_{\alpha,u}\sqrt{\frac{2}{\pi}}\frac{u_2-\hat{p}}{\sqrt{2\hat{\xi}_1 - \hat{\xi}_0}} }{\mathcal{R}^{(\alpha)}_{\mathbf{\hat{x}}}(u_1,u_2)}}\label{eqm4}\\
    q_0 &=  \int  \mathpzc{d}{\bf u} \; \frac{1}{2\hat{q}_1 - \hat{q}_0}\left[\frac{\sqrt{\frac{2}{\pi}} - \frac{\hat{m} - u_1}{\sqrt{2\hat{q}_1 - \hat{q}_0}}e^{\frac{(\hat{m}-u_1)^2}{2(2\hat{q}_1 - \hat{q}_0)}} \text{Erfc}\left(\frac{\hat{m}-u_1}{\sqrt{2(2\hat{q}_1 - \hat{q}_0)}}\right)}{\mathcal{R}^{(\alpha)}_{\mathbf{\hat{x}}}(u_1,u_2)} \right]^2 \label{eqm5}\\
    \xi_0 &= \int  \mathpzc{d}{\bf u} \;  \frac{e^{2\hat{\phi}} \tonde{\frac{2\hat{q}_1 - \hat{q}_0}{2\hat{\xi}_1 - \hat{\xi}_0} }}{2\hat{\xi}_1 - \hat{\xi}_0} \quadre{\frac{\frac{u_2-\hat{p}}{\sqrt{2\hat{\xi}_1 -\hat{\xi}_0}} e^{\frac{(\hat{p}-u_2)^2}{2(2\hat{\xi}_1 - \hat{\xi}_0)}} \tonde{2 \delta_{\alpha, t}+ \delta_{\alpha,u}\text{Erfc}\left(\frac{u_2-\hat{p}}{\sqrt{2(2\hat{\xi}_1 - \hat{\xi}_0)}}\right)} -\delta_{\alpha,u}\sqrt{\frac{2}{\pi}}}{\mathcal{R}^{(\alpha)}_{\mathbf{\hat{x}}}(u_1,u_2)}}^2\label{eqm6}\\
    z &= \int\mathpzc{d}{\bf u} \; \frac{e^{\hat{\phi}}}{2\hat{\xi}_1 - \hat{\xi}_0} \quadre{\frac{\sqrt{\frac{2}{\pi}} - \frac{(\hat{m}-u_1)}{\sqrt{2\hat{q}_1 - \hat{q}_0}}e^{\frac{(\hat{m}-u_1)^2}{2(2\hat{q}_1 - \hat{q}_0)}}\text{Erfc}\left(\frac{\hat{m}-u_1}{\sqrt{2(2\hat{q}_1 - \hat{q}_0)}}\right)}{\mathcal{R}^{(\alpha)}_{\mathbf{\hat{x}}}(u_1,u_2)}}
    \notag \\
    &\hspace{1in} \times \quadre{\frac{\frac{u_2-\hat{p}}{\sqrt{2\hat{\xi}_1 -\hat{\xi}_0}} e^{\frac{(\hat{p}-u_2)^2}{2(2\hat{\xi}_1 - \hat{\xi}_0)}} \tonde{2 \delta_{\alpha, t}+ \delta_{\alpha,u}\text{Erfc}\left(\frac{u_2-\hat{p}}{\sqrt{2(2\hat{\xi}_1 - \hat{\xi}_0)}}\right)} -\delta_{\alpha,u}\sqrt{\frac{2}{\pi}}}{\mathcal{R}^{(\alpha)}_{\mathbf{\hat{x}}}(u_1,u_2)}}\label{eqm7}\\
    \phi &= \int\mathpzc{d}{\bf u} \;  \frac{\exp\left(\frac{(\hat{m}-u_1)^2}{2(2\hat{q}_1 - \hat{q}_0)}\right) \text{Erfc}\left(\frac{\hat{m}-u_1}{\sqrt{2(2\hat{q}_1 - \hat{q}_0)}}\right)}{\mathcal{R}^{(\alpha)}_{\mathbf{\hat{x}}}(u_1,u_2)}\label{eqm11}.
  \end{align}
  For $\alpha=u$, we recover the equations already reported in Ref.~\cite{rosQuenchedComplexityEquilibria2023}.

  \subsubsection{The annealed equations}
  We report the expression obtained deriving the function \eqref{eq:Aann} with respect to the order and conjugate parameters ${\bf x}$ and $\hat{\bf x}$. As for the quenched case, the term that changes with $\alpha$ is a function of the conjugate parameters only, and the derivatives with respect to the order parameters are equal for both choices of $\alpha$, and equal to those already obtained in \cite{rosQuenchedComplexityEquilibria2023}:
  \begin{align}
    &\hat{p}=- \frac{(\kappa-\mu m)}{\sigma^2 \; q_1}\label{eq_an12}\\
    &\hat{\xi_1}= \frac{1}{2\sigma^2 \; q_1}\\
    &\hat{m} =\frac{\mu p}{\sigma^2 q_1} + \frac{\mu m}{\sigma^2 q_1(1+\gamma)} - \frac{\mu(\kappa-\mu m)}{\sigma^2 q_1} + \frac{ \gamma m (\kappa-\mu m) \quadre{\mu m - (\kappa-\mu m)} }{\sigma^2 (1+\gamma) q_1^2}-\frac{(\kappa-\mu m)}{ \sigma^2 q_1 (1+ \gamma)} \\
    &\hat{q}_1 =-\frac{\xi_1}{ 2 \sigma^2 q_1^2} + \frac{2(\kappa-\mu m) [(\gamma+1) p+m]}{2 \sigma^2(1+\gamma)q_1^2}-\frac{(\kappa-\mu m)^2}{2 \sigma^2 q_1^2}+ 2 \frac{\gamma}{1+\gamma} \frac{m^2 (\kappa-\mu m)^2}{2 \sigma^2 q_1^3} + \frac{1}{2 q_1}\label{eq_an122}.
  \end{align}
  We now consider the derivatives with respect to the conjugate parameters. The derivative with respect to $\hat \phi$ reads
  \begin{equation}\label{eq:phia1}
    \phi= \frac{e^{\frac{\hat{m}^2}{4 \hat{q}_1}}\text{Erfc} \tonde{\frac{\hat{m}}{2 \sqrt{\hat{q}_1}}}} {e^{\frac{\hat{m}^2}{4 \hat{q}_1}} \text{Erfc} \tonde{\frac{\hat{m}}{2 \sqrt{\hat{q}_1}}}+ e^{ \hat{\phi}}\sqrt{\frac{\hat{q}_1}{\hat{\xi}_1}} e^{\frac{\hat{p}^2}{4 \hat{\xi}_1}} \tonde{2 \delta_{\alpha, t} + \delta_{\alpha, u} \text{Erfc} \tonde{\frac{-\hat{p}}{2 \sqrt{\hat{\xi}_1}}}}}.
  \end{equation}
  Differentiating with respect to the conjugate parameters $\hat{\bf{x}}$ we obtain:
  \begin{align}
    m&= \frac{\sqrt{\frac{2}{\pi}}-\frac{\hat{m}}{
      \sqrt{2 \hat{q}_1}} e^{\frac{\hat{m}^2}{4\hat{q}_1}}
    \text{Erfc}\left(\frac{\hat{m}}{2 \sqrt{\hat{q}_1}}\right)}{ \sqrt{2 \hat{q}_1} \left(e^{\frac{\hat{m}^2}{4\hat{q}_1}} \text{Erfc}\left(\frac{\hat{m}}{2
        \sqrt{\hat{q}_1}}\right)+e^{\hat \phi}
        \sqrt{\frac{\hat{q}_1}{\hat{\xi}_1}}
        e^{\frac{\hat{p}^2}{4 \hat{\xi}_1}}
    \quadre{2 \delta_{\alpha, t} + \delta_{\alpha,u}\text{Erfc}\left(\frac{-\hat{p}}{2 \sqrt{\hat{\xi}_1}}\right)}\right)}, \label{eq:annm1}\\
    q_1&=\frac{ e^{\frac{\hat{m}^2}{4 \hat{q}_1}} \left(\hat{m}^2+2
      \hat{q}_1\right) \text{Erfc}\left(\frac{\hat{m}}{2 \sqrt{\hat{q}_1}}\right)-\sqrt{\frac{2}{\pi}}
    \hat{m} \sqrt{2\hat{q}_1}}{4 \hat{q}_1^2 \left(e^{\frac{\hat{m}^2}{4\hat{q}_1}} \text{Erfc}\left(\frac{\hat{m}}{2
        \sqrt{\hat{q}_1}}\right)+e^{\hat \phi}
        \sqrt{\frac{\hat{q}_1}{\hat{\xi}_1}}
        e^{\frac{\hat{p}^2}{4 \hat{\xi}_1}}
    \quadre{2 \delta_{\alpha, t} + \delta_{\alpha,u}\text{Erfc}\left(\frac{-\hat{p}}{2 \sqrt{\hat{\xi}_1}}\right)}\right)},\\
    p&=\frac{\sqrt{2\hat{q}_1} e^{\hat{\phi}} \left(\delta_{\alpha, u}\sqrt{\frac{2}{\pi
        }}+\frac{\hat{p}
    e^{\frac{\hat{p}^2}{4 \hat{\xi}_1}}}{\sqrt{2\hat{\xi}_1}}\quadre{2 \delta_{\alpha, t} + \delta_{\alpha,u}\text{Erfc}\left(\frac{-\hat{p}}{2 \sqrt{\hat{\xi}_1}}\right)}\right)}{2 \hat{\xi}_1 \left(e^{\frac{\hat{m}^2}{4\hat{q}_1}} \text{Erfc}\left(\frac{\hat{m}}{2
        \sqrt{\hat{q}_1}}\right)+e^{\hat \phi}
        \sqrt{\frac{\hat{q}_1}{\hat{\xi}_1}}
        e^{\frac{\hat{p}^2}{4 \hat{\xi}_1}}
    \quadre{2 \delta_{\alpha, t} + \delta_{\alpha,u}\text{Erfc}\left(\frac{-\hat{p}}{2 \sqrt{\hat{\xi}_1}}\right)}\right)},\\
    \xi_1&= -\frac{\sqrt{2\hat{q}_1} e^{\hat{\phi}} \left(\delta_{\alpha, u}\sqrt{\frac{2}{\pi}} \hat{p}+\frac{e^{\frac{\hat{p}^2}{4
        \hat{\xi}_1}} \left(2 \hat{\xi}_1+\hat{p}^2\right)}{
    \sqrt{2 \hat{\xi}_1}}\quadre{2 \delta_{\alpha, t} + \delta_{\alpha,u}\text{Erfc}\left(\frac{-\hat{p}}{2 \sqrt{\hat{\xi}_1}}\right)}\right)}{4\hat{\xi}_1^2 \left(e^{\frac{\hat{m}^2}{4\hat{q}_1}} \text{Erfc}\left(\frac{\hat{m}}{2
        \sqrt{\hat{q}_1}}\right)+e^{\hat \phi}
        \sqrt{\frac{\hat{q}_1}{\hat{\xi}_1}}
        e^{\frac{\hat{p}^2}{4 \hat{\xi}_1}}
    \quadre{2 \delta_{\alpha, t} + \delta_{\alpha,u}\text{Erfc}\left(\frac{-\hat{p}}{2 \sqrt{\hat{\xi}_1}}\right)}\right)}\label{eq:annxi1}
  \end{align}
  Notice that, exploiting the equation for $\phi$, these equations can be re-written in the form:
  \begin{align}
    m&= -\phi \frac{\hat{m}}{2 \hat{q}_1}+ \frac{\phi}{\sqrt{\pi \; \hat{q}_1 }}\frac{e^{-\frac{\hat{m}^2}{4 \hat{q}_1}}}{ \text{Erfc} \tonde{\frac{\hat{m}}{2 \sqrt{\hat{q}_1}}}}\\
    q_1&=\frac{ \phi}{2 \hat{q}_1}+\frac{\phi \hat{m}^2}{ 4\; \hat{q}_1^2} -\frac{\phi \hat{m}}{2 \sqrt{\pi} \hat{q}_1^{\frac{3}{2}}}\frac{ e^{-\frac{\hat{m}^2}{4 \hat{q}_1}}}{ \text{Erfc} \tonde{\frac{\hat{m}}{2 \sqrt{\hat{q}_1}}}}\\
    p&= -(1- \phi)
    \quadre{\frac{\hat{p}}{ 2\hat{\xi}_1} + \delta_{\alpha, u} \frac{1}{\sqrt{\pi \hat \xi_1}} \frac{e^{-\frac{\hat{p}^2}{ 4\hat{\xi}_1}}}{1+ \text{Erf} \tonde{\frac{\hat{p}}{2 \sqrt{\hat{\xi}_1}}} }}\\
    \xi_1&= (1- \phi) \quadre{\frac{1}{2 \hat{\xi}_1}+\frac{\hat{p}^2}{ 4\; \hat{\xi}_1^2} + \delta_{\alpha, u} \frac{\hat{p}}{ 2 \sqrt{\pi}\; \hat{\xi}_1^{\frac{3}{2}}} \frac{e^{-\frac{\hat{p}^2}{ 4\hat{\xi}_1}}}{1+ \text{Erf} \tonde{\frac{\hat{p}}{2 \sqrt{\hat{\xi}_1}}} }}.
  \end{align}
  This form shows a decoupling between the two equations for the parameters $\hat{m}, \hat{q}_1$ associated to the abundances, and those for the parameters $\hat{p}, \hat{\xi}_1$ associated to the effective growth rates. This decoupling is natural within the annealed approximation. Again, for $\alpha=u$ we recover the results of \cite{rosQuenchedComplexityEquilibria2023}.

  \subsection{Self-consistent equations for the conjugate parameters in the rescaled variables}
  \subsubsection{The quenched self-consistent equations}
  The above equations are more conveniently expressed in terms of the rescaled conjugate parameters \eqref{eq:ROP}. It is straightforward to show that the integral representations \eqref{eqm1}-\eqref{eqm11} are equivalent to \eqref{eq:mInt}-\eqref{eq:PhiInt}. We now aim at re-writing Eqs. \eqref{eq:1}-\eqref{eq:7} in terms of these rescaled conjugate parameters \eqref{eq:ROP}.
  Combining \eqref{eq:1} and \eqref{eq:2} we immediately obtain:
  \begin{equation}
    y=\sqrt{2 \hat \xi_1 - \hat \xi_0}=\frac{1}{\sqrt{\sigma^2 (q_1 - q_0)}},
  \end{equation}
  which allows us to re-write the denominators in Eqs. \eqref{eq:1}-\eqref{eq:7} in terms of powers of $y$. The first three equations \eqref{eq:1}, \eqref{eq:2} and \eqref{eq:3} take therefore the form:
  \begin{align}
    1=&\sigma^2\tonde{[q_1 y^2]- [q_0 y^2]}\label{eq:UnoNos}\\
    \beta_2 =&- \sigma^2  [ q_0y^2]\\
    x_2=&-\kappa y + \mu [m y]\label{eqy},
  \end{align}
  where the quantities in brackets can be replaced by the integral representations \eqref{eq:mInt}-\eqref{eq:PhiInt}. Using these three equations, the remaining Eqs. \eqref{eq:4}-\eqref{eq:7} take the following form:
  \begin{align}
    r x_1&=
    x_2\left(\frac{1}{(\gamma + 1)} + \mu
    + \frac{\gamma}{1 + \gamma} \sigma^2[ zy^2]\right)
    +\mu \frac{[my]}{\gamma + 1}  + \mu [py]
    + \mu \frac{\gamma}{1 + \gamma} \sigma^2 [zy^2] [my],\label{eq:p}\\
    \beta_3 &=-\frac{\gamma}{\gamma+1}\sigma^2 x_2 [my]- \frac{\gamma}{\gamma +1}  \sigma^2[z y^2]+ \beta_2\left(\frac{1 + 2\gamma\sigma^2[zy^2]  }{\gamma+1} \right),\\
    r^2 &= \sigma^2 \left([\xi_0 y^2] - [\xi_1 y^2] +\frac{2[zy^2]}{\gamma +1}\right) +\sigma^2 + 2\frac{\gamma}{\gamma+1}\sigma^4[zy^2]^2,\\
    \beta_1&=  -2 x_2 \sigma^2 \left( \frac{[m y]}{1+\gamma} + [p y] \right)- \frac{4\,\sigma^4 x_2 \gamma}{1+\gamma}\, [z y^2]\, [my] - \sigma^2 x_2^2
    +\beta_2 \sigma^2\tonde{1-2 [\xi_1 y^2]}\\
    &\quad- \frac{2\gamma \sigma^4}{1+\gamma}\,(1-\beta_2) [z y^2]^2 \notag  \\
    &\; +
    \sigma^2 (2 \beta_2-1) \tonde{ \frac{2}{1+\gamma} [z y^2] + \frac{2\gamma}{1+\gamma}\,\sigma^2\, [z y^2]^2 +[\xi_0 y^2]}\label{eq:UnoNos2},
  \end{align}
  where once more the quantities in brackets can be expressed in terms of the integral representations \eqref{eq:mInt}-\eqref{eq:PhiInt}. Since the latter are functions of the rescaled conjugate parameters only, these equations allow to determine such parameters self consistently. From these equations one sees that introducing the parameter $y$ through the rescaling \eqref{eq:ROP} is convenient, since the integral representations corresponding to the terms in brackets do not depend explicitly on $y$, and therefore instead of solving 7 coupled self-consistent equations, we can solve 6 of them (all except \eqref{eqy}) and then determine $y$ using \eqref{eqy}.

  Despite these simplifications, solving the system of coupled equations in the form given above is still suboptimal, since the equations contain multiple terms that are expressed in terms of integral convolutions. We now discuss how to reach the simpler form of the equations that we have reported in Sec.~\ref{sec:SCE_q}. As a first step, one can notice that the integral representations \eqref{eq:mInt}-\eqref{eq:PhiInt} satisfy the following relations, obtained through partial integration:
  \begin{align}\label{pr}
    rx_1 [my] &=  -(\beta_1 + r^2)[q_1 y^2] +\beta_1 [q_0 y^2] + \phi +\beta_3 [z y^2]\\
    x_2[py] &= \beta_3 [zy^2] +\beta_2[\xi_0 y^2]   -(\beta_2+1) [\xi_1 y^2] + (1-\phi )\label{pr2}.
  \end{align}
  These are obtained making use of the identities:
  \begin{align}
    & \beta_1 {\partial}_{u_1}\mathpzc{g}_{\hat{\bf x}}(u_1,u_2) + \beta_3 r \, \partial_{u_2} \mathpzc{g}_{\hat{\bf x}}(u_1,u_2)  = - r^2(x_1 - u_1) \mathpzc{g}_{\hat{\bf x}}(u_1,u_2) +r^2x_1 \mathpzc{g}_{\hat{\bf x}}(u_1,u_2), \\
    & \beta_3 \partial_{u_1} \mathpzc{g}_{\hat{\bf x}}(u_1,u_2) + \beta_2 r \partial_{u_2} \mathpzc{g}_{\hat{\bf x}}(u_1,u_2) = - r(x_2 - u_2) \mathpzc{g}_{\hat{\bf x}}(u_1,u_2) +rx_2 \, \mathpzc{g}_{\hat{\bf x}}(u_1,u_2),
  \end{align}
  and
  \begin{align}
    \dv{}{u_1}K(x_1-u_1) &= -(x_1-u_1) K(x_1-u_1) + \sqrt{\frac{2}{\pi}},\\
    \dv{}{u_2}K(u_2-x_2) &= -(x_2-u_2) K(u_2-x_2) - \sqrt{\frac{2}{\pi}},
  \end{align}
  where we remind that $ K(x) = e^{x^2/2}\text{Erfc}\left(x/\sqrt{2}\right)$. Summing the two equations \eqref{pr} and \eqref{pr2} we get:
  \begin{align}
    rx_1 [my] + x_2[py] &= -r^2[q_1 y^2] +\beta_1([q_0y^2] - [q_1y^2])\\
    &\quad + \beta_2([\xi_0y^2] - [\xi_1y^2]) - [\xi_1 y^2]  + 2\beta_3[zy^2]+1.
    \label{eq:sumipp}
  \end{align}
  Plugging the equations for $\beta_1$ and $r^2$ we obtain the identity
  \begin{align}
    rx_1 [my] + x_2[py] &=  - \left(1 + 2\frac{1 + \gamma \sigma^2 [zy^2]}{\mu(\gamma+1)}\right)x_2^2 + \frac{2}{\mu}rx_1x_2.
  \end{align}
  This identity can be exploited in combination with Eq.~\eqref{eq:p}, to show that the latter equation is equivalent to the following two relations (assuming $y>0$):
  \begin{align}
    &rx_1 -\frac{x_2}{\gamma +1} \left( 1 + \gamma \sigma^2[zy^2]\right)=0,\label{eq:newp1}\\
    & x_2 + [py] + \frac{[my]}{\gamma + 1} + \frac{\gamma}{\gamma +1 }\sigma^2 [zy^2][my] = 0.\label{eq:newp2}
  \end{align}
  Notice that for $\gamma=0$, \eqref{eq:newp1} allows to eliminate the variable $x_2$ and reduce the number of coupled self-consistent equations to five.  If $\gamma \neq 0$, these identities are still useful: first, they imply that the set of six self-consistent equations does not depend on the parameter $\mu$, only the equation used to determine $y$ does; moreover, the integral $[z y^2]$ can be eliminated from all the equations using:
  \begin{equation}
    \sigma^2 [z y^2]=\frac{1+ \gamma}{\gamma} \tonde{\frac{r x_1}{x_2}-\frac{1}{1+ \gamma}}, \quad \quad (\text{for} \quad \gamma \neq 0).
  \end{equation}
  With this substitution, we are left with six self-consistent equations, containing six integral representations. Substituting one equation into the other, the system of equations is reduced to the form \eqref{eq:VRtop}-\eqref{eq:VRbottom}, in which each equation contains only one integral representation, that is particularly convenient for the numerical solution.

  \subsubsection{The annealed self-consistent equations}
  As in the quenched case, also in the annealed case the saddle point equations are more compactly written in terms of the rescaled parameters \eqref{eq:NewParametersA}. Again, one can easily check that the equations \eqref{eq:phia1}-\eqref{eq:annxi1} are equivalent to \eqref{eq:my_ann}-\eqref{eq:phi_ann2}. Eqs.~\eqref{eq_an12}-\eqref{eq_an122} take the following form:
  \begin{align}
    &\sigma^2[ q_1\mathtt{y}^2]=1,\\
    &\mathtt{x}_2=-\kappa\mathtt{y} + \mu [m\mathtt{y}],\\
    &\mathtt{x}_1\mathtt{r}=\mu \tonde{[p\mathtt{y}]+\frac{[m\mathtt{y}]}{1+\gamma}+\mathtt{x}_2-\frac{\gamma}{1+\gamma}\sigma^2\mathtt{x}_2[m\mathtt{y}]^2}-\frac{\gamma}{1+\gamma}\sigma^2\mathtt{x}_2^2[m\mathtt{y}]+\frac{\mathtt{x}_2}{1+\gamma},\label{eq:elimin}\\
    &\mathtt{r}^2=- \sigma^2[\xi_1\mathtt{y}^2]- 2\mathtt{x}_2 \sigma^2\tonde{[p\mathtt{y}]+\frac{[m\mathtt{y}]}{1+\gamma}}-\sigma^2\mathtt{x}_2^2+ 2 \frac{\gamma}{1+\gamma}\sigma^4 \mathtt{x}_2^2 [m\mathtt{y}]^2+ \sigma^2\label{eq:anneqextended},
  \end{align}
  where the quantities in brackets $[\cdot]$ denote the representations \eqref{eq:my_ann}-\eqref{eq:phi_ann2}.
  Similarly to the quenched case, one can derive the following identity satisfied by these representations:
  \begin{equation}\label{eq:ineqa}
    \mathtt{r}^2 [q_1 \mathtt{y}^2] + [\xi_1\mathtt{y}^2] = -\mathtt{r}\mathtt{x}_1 [m\mathtt{y}] -\mathtt{x}_2 [p\mathtt{y}] + 1 \Leftrightarrow\mathtt{r}^2  + \sigma^2[\xi_1 \mathtt{y}^2]- \sigma^2  = -\sigma^2\mathtt{r}\mathtt{x}_1 [m \mathtt{y}] - \sigma^2\mathtt{x}_2 [p \mathtt{y}].
  \end{equation}
  Eq.\eqref{eq:anneqextended} implies
  \begin{equation}
    1-\frac{\mathtt{r}^2}{\sigma^2}  - [\xi_1 \mathtt{y}^2] = -\mathtt{x}_2 ^2 +\frac{2}{\mu}\mathtt{x}_2 \left(\mathtt{r}\mathtt{x}_1 - \frac{\mathtt{x}_2}{\gamma +1}\right)  + 2\frac{\gamma}{\mu(\gamma +1)} \sigma^2 [m\mathtt{y}]\mathtt{x}_2^3 =\mathtt{r}\mathtt{x}_1 [m \mathtt{y}] +\mathtt{x}_2 [p \mathtt{y}].
  \end{equation}
  Using \eqref{eq:elimin} to eliminate $[p\mathtt{y}]$, we find that the right-hand side also equals to:
  \begin{align}
    &\quad \mathtt{r}\mathtt{x}_1 [m \mathtt{y}] +\mathtt{x}_2 [p \mathtt{y}]\notag \\
    &= \mathtt{r}\mathtt{x}_1 [m\mathtt{y}] +\frac{\mathtt{x}_2}{\mu} \left(\mathtt{r}\mathtt{x}_1  -\mathtt{x}_2\left( \frac{1}{\gamma + 1} + \mu \right) - \frac{\mu [m\mathtt{y}]}{\gamma + 1} + \frac{\gamma}{1+\gamma} \sigma^2\mathtt{x}_2 [m\mathtt{y}] (\mu [m\mathtt{y}] +\mathtt{x}_2)  \right)
  \end{align}
  which plugged into \eqref{eq:ineqa} gives
  \begin{align}
    \left(\mathtt{r}\mathtt{x}_1 - \frac{\mathtt{x}_2}{\gamma +1} +\sigma^2\frac{\gamma }{\gamma +1} [m\mathtt{y}]\mathtt{x}_2^2   \right) \frac{\kappa\mathtt{y}}{\mu} = 0
  \end{align}
  so again for $\mathtt{y} \neq 0$ we get the relation:
  \begin{equation}
    \mathtt{r}\mathtt{x}_1 = \frac{\mathtt{x}_2}{\gamma +1}\left(1 - \gamma \sigma^2 [m\mathtt{y}]\mathtt{x}_2 \right).
  \end{equation}
  This relation, plugged into \eqref{eq:elimin}, shows that that equation is equivalent to the two relations
  \begin{align}
    &\mathtt{x}_1\mathtt{r}=-\frac{\gamma}{1+\gamma}\sigma^2\mathtt{x}_2^2[m\mathtt{y}]+\frac{\mathtt{x}_2}{1+\gamma},\\
    &[p\mathtt{y}]=-\frac{[m\mathtt{y}]}{1+\gamma}-\mathtt{x}_2+\frac{\gamma}{1+\gamma}\sigma^2\mathtt{x}_2[m\mathtt{y}]^2.
  \end{align}
  Once more, for $\gamma=0$ these imply $\mathtt{x}_1\mathtt{r}=\mathtt{x}_2$, allowing to eliminate one conjugate parameter. For $\gamma \neq 0$, one has:
  \begin{align}
    \sigma^2[m\mathtt{y}]=\frac{1+\gamma}{ \gamma }\tonde{\frac{\mathtt{1}}{\mathtt{x}_2(1+\gamma)} -\frac{\mathtt{x}_1\mathtt{r}}{\mathtt{x}_2^2}}, \quad \quad (\text{for} \quad \gamma \neq 0).
  \end{align}
  Plugging these into \eqref{eq:anneqextended}, one recovers the system of equations presented in Sec.~\ref{supp:AnnShort}.

  \subsection{The complexity as a function of the rescaled variables}\label{supp:COmpXhat}

  \subsubsection{Quenched complexity}
  We here account for the derivation of the expressions \eqref{eq:QuenchedSigma}. We begin with the expression \eqref{eq:barA}: evaluated at the solutions of the saddle point equations, this equals to the quenched complexity. The identity \eqref{eq:sumipp} reads $ \hat{m}m + \hat{p}p= 2(\hat{z}z-\hat{q}_1q_1 -\hat{\xi}_1\xi_1 )+ \hat{q}_0q_0 + \hat{\xi}_0\xi_0 + 1$, so that
  \begin{equation}
    \Sigma^{(\alpha)}_{\sigma,\gamma}(\phi) = \overline{p}(x) + \mathpzc{d}(\phi) +\frac{1}{2}\left( \hat{m} m + \hat{p} p +1 \right) +\hat{\phi} \phi +  \bar{\mathcal{J}}^{(\alpha)}(\hat{x}).
  \end{equation}
  Written in terms of the rescaled variables:
  \begin{equation}
    \frac{1}{2}(m\hat{m} + p\hat{p}) = \frac{1}{2}\left( rx_1[my] + x_2 [py] \right)
  \end{equation}
  and
  \begin{equation}
    \begin{aligned}
      \bar{p}(\mathbf{x}) &= -rx_1 [my]
      - x_2 [py]
      + \beta_3 [zy^2]+ \frac{\gamma}{(\gamma + 1)}\sigma^2 [my] + \sigma^2\frac{\gamma}{2(\gamma +1)}[zy^2]^2 - \frac{1}{2\sigma^2(\gamma + 1)} \\
      &\quad - \frac{1}{2}[\xi_1 y^2] + \frac{\beta_2}{2}([\xi_0y^2] - [\xi_1y^2]) - \frac{1}{2} x_2^2 - \frac{\log[\frac{2\pi}{y^2}]}{2}+\frac{\beta_2}{2}\left(1  - 2\frac{\gamma\sigma^2 [zy^2]^2}{\gamma + 1}\right)
    \end{aligned}
  \end{equation}
  so that, using again \eqref{eq:sumipp} and the equation
  \begin{equation}
    \begin{aligned}
      \frac{\beta_1}{2\sigma^2} &= \frac{\beta_2}{2}(2\frac{r^2}{\sigma^2} - 1) -\frac{1}{2} (\xi_1y^2)+ \frac{1}{2} - \frac{1}{2\sigma^2}r^2 \\
      &\quad + \frac{1}{2} x_2^2 -\sigma^2 \frac{\gamma}{\gamma+1}\left([my][zy^2]x_2 + (1-\beta_2)[zy^2]^2  \right),
    \end{aligned}
  \end{equation}
  which is equation \ref{eq:UnoNos2} where we plugged equation \ref{eq:newp2}, we have
  \begin{align}
    \frac{1}{2}(m\hat{m} + p\hat{p} + 1)  + \bar{p} &= -\frac{1}{2}\left(-r^2[q_1 y^2] +\beta_1([q_0y^2] - [q_1y^2]) \right) \notag\\
    &\quad -\frac{1}{2}\left( \beta_2([\xi_0y^2]  - [\xi_1y^2]) - [\xi_1 y^2]  + 2\beta_3[zy^2]+1\right)  \notag\\
    &\quad  +\frac{1}{2}+ \beta_3 [zy^2]+ \frac{\gamma}{(\gamma + 1)}\sigma^2 [my] + \sigma^2\frac{\gamma}{2(\gamma +1)}[zy^2]^2 - \frac{1}{2\sigma^2(\gamma + 1)}  \notag\\
    &\quad - \frac{1}{2}[\xi_1 y^2] + \frac{\beta_2}{2}([\xi_0y^2] - [\xi_1y^2])\notag \\
    &\quad - \frac{1}{2} x_2^2 - \frac{\log\left(\frac{2\pi}{y^2}\right)}{2}+\frac{\beta_2}{2}\left(1  - 2\frac{\gamma\sigma^2 [zy^2]^2}{\gamma + 1}\right)\\
    &= \frac{1}{2}\left(1 - r^2[q_0 y^2] - [\xi_1 y^2] - \frac{1}{\sigma^2(\gamma + 1)}   \right)\notag \\
    &\quad  - \sigma^2 \frac{\gamma}{2(\gamma + 1)}[zy^2]^2 - \frac{1}{2}\log\left(\frac{2\pi}{y^2}\right).
  \end{align}
  Eq. \eqref{eq:QuenchedSigma} follows from noticing that

  \begin{align}
    \bar{\mathcal{J}}^{(\alpha)}(\hat{x}) &= \int \dd{u_1} \dd{u_2}\mathpzc{g}_{\mathbf{\hat{x}}}(u_1,u_2) \log\left(\frac{K(x_1 - u_1) + e^{\hat\phi}r  [\delta_{\alpha, t} 2 e^{\frac{(u_2-x_2)^2}{2}}+ \delta_{\alpha, u}K(u_2-x_2)]}{2 r }\right)\notag\\
    &\quad+\frac{1}{2} \log\left(\frac{2\pi}{y^2}\right) -\hat{\phi}.
  \end{align}

  \subsubsection{Annealed Complexity}
  We here account for the derivation of \eqref{eq:AnnealedSigma}.
  The expression \eqref{eq:Aann}, evaluated at the solutions of the saddle point equations, gives the annealed complexity. The identity \eqref{eq:ineqa} is equivalent to $ \hat{q}_1 q_1 + \hat{\xi}_1 \xi_1 + \hat{m} m + \hat{p} p = \frac{1}{2}( \hat{m} m + \hat{p} p + 1)$, so that:
  \begin{equation}
    \Sigma^{(\alpha, A)}_{\sigma,\gamma}(\phi) = {p}_1({\bf x}) + \mathpzc{d}(\phi) +\frac{1}{2}\left( \hat{m} m + \hat{p} p +1 \right) +\hat{\phi} \phi +  \mathcal{J}^{(\alpha)}_1(\hat{\bf x}).
  \end{equation}
  In terms of the rescaled variables:
  \begin{equation}
    \frac{1}{2}( \hat{m} m + \hat{p} p) = \frac{1}{2}\left(\mathtt{r}\mathtt{x}_1 [m\mathtt{y}] +\mathtt{x}_2 [p \mathtt{y}] \right) = \frac{1}{2}\left(1 - \frac{\mathtt{r}^2}{\sigma^2} - [\xi_1\mathtt{y}^2] \right)
  \end{equation}
  and
  \begin{align*}
    p_1(\mathbf{x})&= -\frac{1}{2 } \bigg[
      \mathtt{x}_2^2 \left( 1 -\sigma^2 \frac{\gamma [m \mathtt{y}]^2}{1 + \gamma} \right)
      + 2\mathtt{x}_2 \left( [p \mathtt{y}] + \frac{[m \mathtt{y}]}{1 + \gamma} \right)
    + [\xi_1 \mathtt{y}^2]  \bigg] - \frac{1}{2} \log(\frac{2\pi}{\mathtt{y}^2})\notag \\
    &\quad - \frac{1}{2 \sigma^2 (1 + \gamma)}\\
    & =  -\frac{1}{2 } \bigg[
    -\frac{\mathtt{r}^2}{\sigma^2} + 1  +\sigma^2\frac{\gamma}{\gamma+1} [m\mathtt{y}]^2\mathtt{x}_2^2  \bigg] - \frac{1}{2} \log(\frac{2\pi}{\mathtt{y}^2})
    - \frac{1}{2 \sigma^2 (1 + \gamma)},
  \end{align*}
  where we used \eqref{eq:anneqextended} in the second line. Combining everything, we recover \eqref{eq:AnnealedSigma} using that:
  \begin{align}
    \mathcal{J}_1^{(\alpha)}(\hat{\mathbf{x}}) = \frac{1}{2} \log\left(\frac{2\pi}{\mathtt{y}^2}\right) - \hat{\phi} + \log\left(K(\mathtt{x}_1)+ e^{\hat{\phi}}\mathtt{r} [\delta_{\alpha, t} 2e^{\frac{\mathtt{x}_2^2}{2}} + \delta_{\alpha, u} K(-\mathtt{x}_2)]\right)- \log (2\mathtt{r}).
  \end{align}

  \subsection{Self-consistent equations enforcing $\Delta=0$}\label{sec:DeltaEnforced}
  Enforcing $\hat{z} = -\sqrt{\hat{q}_0 \hat{\xi}_0 }$ into \eqref{eq:barA} one finds the modified function:
  \begin{align}\label{eq:NewAct0}
    \bar{\mathcal{A}}^{(\alpha)}_{\Delta=0}(\mathbf{x}, \hat{\mathbf{x}}, \phi) &= \bar{p}(\mathbf{x}) + \ell(\phi) + \hat{q}_1 q_1 + \hat{\xi}_1 \xi_1 + \hat{m} m + \hat{p} p + \hat{\phi} \phi \nonumber \\
    &\quad - \frac{1}{2} \left( \hat{q}_0 q_0 + \hat{\xi}_0 \xi_0 \right) + \sqrt{\hat{q}_0 \hat{\xi}_0 }z + \bar{\mathcal{J}}^{(\alpha)}_{\Delta = 0}(\hat{\mathbf{x}}),
  \end{align}
  with $\bar{\mathcal{J}}^{(\alpha)}_{\Delta = 0}(\hat{\mathbf{x}})$ given in \eqref{eq:VolumeOneInt}.
  The modified saddle point equations are obtained taking the derivatives of $\bar{\mathcal{A}}^{(\alpha)}_{\Delta=0}(\mathbf{x}, \hat{\mathbf{x}}, \phi)$  with respect to the parameters ${\bf x}$ and $\hat{\bf x}$. Notice that this leads to one less equation, since $\hat{z}$ does not have to be determined through the saddle point.\\
  The derivatives with respect to the order parameters ${\bf x}$ are identical to \eqref{eq:1}-\eqref{eq:7}, where $\hat{z}$ in Eq.~\eqref{eq:5} has to be replaced with $\hat{z}\to -\sqrt{\hat{q}_0 \hat{\xi}_0 }$. When expressed in terms of the rescaled conjugate parameters, this set of equations maps into Eqs.~\eqref{eq:UnoNos}-\eqref{eq:UnoNos2}, with $\beta_3\to -\sqrt{\beta_1 \beta_2}$, and where the expression for the order parameters $m y,p y,q_1 y^2, \xi_1 y^2, q_0 y^2, \xi_0 y^2$ have to be determined taking the derivatives of
  $\bar{\mathcal{A}}^{(\alpha)}_{\Delta=0}(\mathbf{x}, \hat{\mathbf{x}}, \phi)$ with respect to $\hat{\bf x}$. Notice that $z y^2$ is fixed from Eq.~\eqref{eq:5}, as:
  \begin{equation}
    z y^2= -\frac{\gamma+1}{\sigma^2\gamma} \frac{\sqrt{\beta_1\beta_2}}{2\beta_2 -1}  - \frac{1}{2\beta_2 -1} \left( \frac{\beta_2}{\gamma \sigma^2} - x_2 m y\right).
  \end{equation}
  When computing these derivatives we recover Eqs.~\eqref{eq:mIntD0}-\eqref{eq:xi1IntD0}, and Eq.~\eqref{eq:PhiIntD0}. The equations for the remaining parameters $q_0, \xi_0$ take the modified form:
  \begin{equation}
    \begin{aligned}
      q_0 y^2 &\to \int \dd{u}\,    \mathpzc{g}_{\hat{\bf x}}(u)\,  \frac{1}{r^2} \left[ \frac{\sqrt{\frac{2}{\pi}} - (x_1 - u) K(x_1- u) }{K(x_1- u) + e^{\hat\phi}r [\delta_{\alpha, t} 2 e^{\frac{1}{2}\tonde{\lambda u-x_2}^2}+ \delta_{\alpha, u}K\tonde{\lambda u-x_2}]} \right]^2  \\
      &\quad - \sqrt{\frac{\beta_2}{\beta_1}} zy^2  \\
      &\quad + \sqrt{\frac{\beta_2}{\beta_1}}  \int \dd{u}\,    \mathpzc{g}_{\hat{\bf x}}(u)\, r e^{\hat\phi}\frac{\sqrt{\frac{2}{\pi}} - (x_1 - u) K(x_1- u)}{K(x_1- u) + e^{\hat\phi}r [\delta_{\alpha, t} 2 e^{\frac{1}{2}\tonde{\lambda u-x_2}^2}+ \delta_{\alpha, u}K\tonde{\lambda u-x_2}]} \\
      &\quad \hspace{1in} \times r e^{\hat\phi}\frac{ -\delta_{\alpha, u}\sqrt{\frac{2}{\pi}} + (\lambda u  x_2 )  [\delta_{\alpha, t} 2 e^{\frac{1}{2}\tonde{\lambda u-x_2}^2}+ \delta_{\alpha, u}K\tonde{\lambda u-x_2}] }{K(x_1- u) + e^{\hat\phi}r [\delta_{\alpha, t} 2 e^{\frac{1}{2}\tonde{\lambda u-x_2}^2}+ \delta_{\alpha, u}K\tonde{\lambda u-x_2}]} ,
    \end{aligned}
  \end{equation}

  and
  \begin{equation}
    \begin{aligned}
      \xi_0 y^2 &\to \int \dd{u}\,    \mathpzc{g}_{\hat{\bf x}}(u)\,  r^2e^{2\hat\phi} \left[ \frac{-\delta_{\alpha, u}\sqrt{\frac{2}{\pi}} + (\lambda u -x_2) [\delta_{\alpha, t} 2 e^{\frac{1}{2}\tonde{\lambda u-x_2}^2}+ \delta_{\alpha, u}K\tonde{\lambda u-x_2}] }{K(x_1- u) + e^{\hat\phi}r [\delta_{\alpha, t} 2 e^{\frac{1}{2}\tonde{\lambda u-x_2}^2}+ \delta_{\alpha, u}K\tonde{\lambda u-x_2}]}\right]^2\\
      &\quad  - \sqrt{\frac{\beta_1}{\beta_2}} zy^2  \\
      &\quad + \sqrt{\frac{\beta_1}{\beta_2}}  \int \dd{u}\,    \mathpzc{g}_{\hat{\bf x}}(u)\,  r e^{\hat\phi}\frac{\sqrt{\frac{2}{\pi}} - (x_1 - u) K(x_1- u) }{K(x_1- u) + e^{\hat\phi}r [\delta_{\alpha, t} 2 e^{\frac{1}{2}\tonde{\lambda u-x_2}^2}+ \delta_{\alpha, u}K\tonde{\lambda u-x_2}]}  \\
      &\quad \hspace{1in} \times \frac{-\delta_{\alpha, u}\sqrt{\frac{2}{\pi}} + (\lambda u  x_2 )  [\delta_{\alpha, t} 2 e^{\frac{1}{2}\tonde{\lambda u-x_2}^2}+ \delta_{\alpha, u}K\tonde{\lambda u-x_2}]  }{K(x_1- u) + e^{\hat\phi}r [\delta_{\alpha, t} 2 e^{\frac{1}{2}\tonde{\lambda u-x_2}^2}+ \delta_{\alpha, u}K\tonde{\lambda u-x_2}]} .
    \end{aligned}
  \end{equation}

  Plugging these expressions back into Eqs. \eqref{eq:1}-\eqref{eq:7}, one gets the modified saddle point equations that we discussed in Sec.~\ref{sec:delta_zero}.

\end{appendix}

\end{document}